\documentclass[a4paper,fleqn,usenatbib]{mnras}

\usepackage[T1]{fontenc}
\usepackage{ae,aecompl}

\usepackage{graphicx}	
\usepackage{amsmath}	
\usepackage{amssymb}	
\usepackage{pdflscape}

\def\sersic{{\tt S\'ersic}}
\def\devauc{{\tt deVauc}}
\def\sedisc{{\tt SeDisc}}
\def\sedibar{{\tt SeDiBar}}

\title[Photometric analysis of Abell~1689]{A photometric analysis of
  Abell~1689: two-dimensional multi-structure decomposition,
  morphological classification, and the Fundamental Plane}
\author[E. Dalla Bont\`a et al.]{E. Dalla
  Bont\`a$^{1,2}$\thanks{E-mail: elena.dallabonta@unipd.it},
  R.~L. Davies$^{3}$, R.~C.~W. Houghton$^{3}$, F. D'Eugenio$^{4,5}$,
  \newauthor and J. M\'endez-Abreu$^{6,7,8}$\\$^{1}$Dipartimento di
  Fisica e Astronomia ``G. Galilei'', Universit\`a degli Studi di
  Padova, Vicolo dell'Osservatorio 3, I-35122, Padova,
  Italy.\\$^{2}$INAF Osservatorio Astronomico di Padova, Vicolo
  dell'Osservatorio 5, I-35122, Padova, Italy\\$^{3}$Physics
  Department, University of Oxford, Denys Wilkinson Building, Keble
  Road, Oxford OX1 3RH, UK\\$^{4}$Research School of Astronomy and
  Astrophysics, Australian National University, Canberra, ACT 2611,
  Australia\\$^{5}$ARC Centre of Excellence for All-Sky Astrophysics
  (CAASTRO)\\$^{6}$School of Physics and Astronomy, University of St
  Andrews, SUPA, North Haugh, St Andrews, KY16 9SS,
  UK\\$^{7}$Instituto de Astrof\'isica de Canarias, 38200 La Laguna,
  Tenerife, Spain\\$^{8}$Universidad de La Laguna,
  Dept. Astrof\'isica, 38206 La Laguna, Tenerife, Spain}

\date{Accepted 2017 September 22. Received 2017 September 22; in original form 2017 March 16}

\pubyear{2017}

\begin{document}
\label{firstpage}
\pagerange{\pageref{firstpage}--\pageref{lastpage}}
\maketitle

\begin{abstract}
We present a photometric analysis of 65 galaxies in the rich cluster
Abell~1689 at $z=0.183$, using the {\em Hubble Space Telescope}
Advanced Camera for Surveys archive images in the rest-frame
$V$-band. We perform two-dimensional multi-component photometric
decomposition of each galaxy adopting different models of the
surface-brightness distribution.  We present an accurate morphological
classification for each of the sample galaxies. For 50 early-type
galaxies, we fit both a de Vaucouleurs and S\'ersic law; S0s are
modelled by also including a disc component described by an
exponential law.  Bars of SB0s are described by the profile of a
Ferrers ellipsoid. For the 15 spirals, we model a S\'ersic bulge,
exponential disc, and, when required, a Ferrers bar component. We
derive the Fundamental Plane by fitting 40 early-type galaxies in the
sample, using different surface-brightness distributions. We find that
the tightest plane is that derived by S\'ersic bulges. We find that
bulges of spirals lie on the same relation. The Fundamental Plane is
better defined by the bulges alone rather than the entire
galaxies. Comparison with local samples shows both an offset and
rotation in the Fundamental Plane of Abell~1689.
\end{abstract}




\section{Introduction}
\label{sec:intro}

\begin{figure*}
\includegraphics[scale=.6,angle=0.]{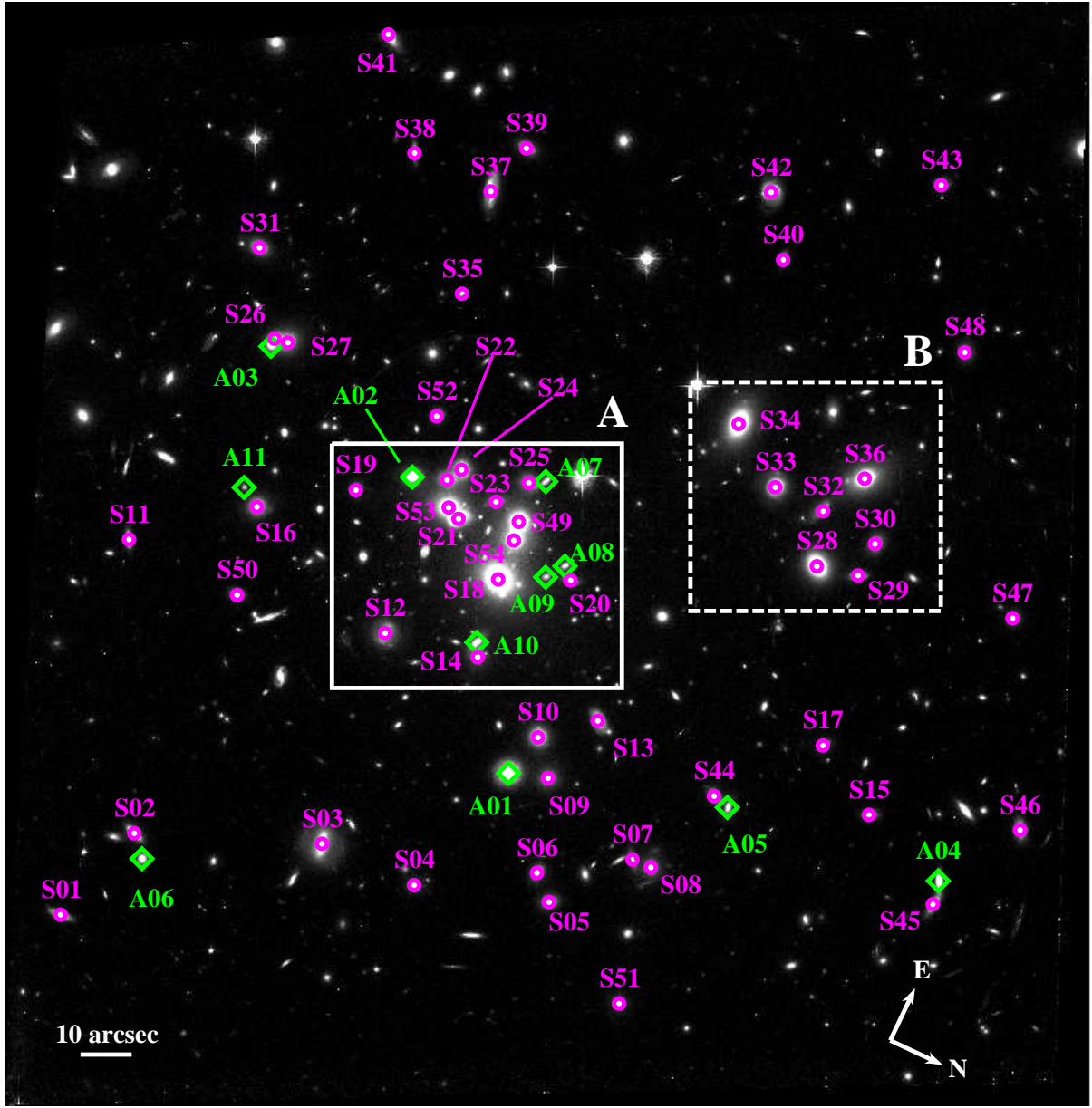}
\caption{ACS/WFC/F625W image of Abell~1689. The white continuous and
  dashed frames highlight the regions A and B where the photometric
  decomposition was particularly challenging due to the light
  contamination of the crowded galaxies. Magenta circles mark the
  spectroscopic sample and green diamonds mark the ancillary
  sample. The linear scale and orientation are shown.}
\label{fig:FoV}
\end{figure*}

The discovery of the FP three decades ago \citep{djo87,dre87,fab87}
constituted an important milestone on understanding galaxy evolution.
Stellar velocity dispersion, $\sigma_{\star}$, effective radius
$R_{\rm e}$, and average surface brightness within $R_{\rm e}$,
$\left<I\right>_{\rm e}$, of ETGs define a remarkably tight plane in
the form $R_{\rm e} \,\alpha\, \sigma_{\star}^b \left<I\right>_{\rm
  e}^c$. Under the assumptions of structural homology and uniform
mass-to-light ratio, the virial theorem predicts $b=2$ and $c=-1$;
because the best-fit values of $b$ and $c$ deviate from this
prediction, the FP is said to be ``tilted'' \citep{bur97,tru04}. The
FP remains a potentially powerful tool to investigate galaxy mass
assembly and luminosity evolution with redshift, by comparing the
values of the FP coefficients over time. However, there is little
uniformity in the details of how the observables are measured, which
makes direct comparisons difficult or inappropriate.  Differences in
the derived FP coefficients can be due to the algorithm used, whether
the fit is direct or orthogonal, choice of the dependent variable
\citep[e.g.,][]{she12}, passband \citep[e.g.,][]{ber03}, and sample
selection \citep[e.g.,][]{nig08}.  In this contribution, we will focus
on the photometric parameters which enter the FP and investigate
whether or not possible discrepancies can arise with the change of
photometric models. Indeed, originally $R_{\rm e}$ and
$\left<I\right>_{\rm e}$ of galaxies were measured by fitting a de
Vaucouleurs law to the growth curve (see Paper I for a
description). Later on S\'ersic profile to the growth curve was
adopted \citep[e.g.,][]{cao93,lab02}. A de Vaucouleurs bulge plus
exponential disc decomposition has also been used
\citep[e.g.,][]{sag97,fri05}, as have two-dimensional
surface-brightness decompositions
\citep[e.g.,][]{fri09,sim02,tra03,fer11}.

Abell 1689 \citep{Abell58} is a richness class 4 cluster at redshift
$z=0.183$ \citep{strub99}.  As a Coma cluster analogue, it provides an
opportunity to study the evolution of galaxies in dense environments
over the last 2.26 Gyr. It is a dynamically active, merging system
with discrete mass components as revealed by substructure in X-ray,
lensing, and near-infrared maps of this cluster \citep{hai10}. The
galaxy alignment appears to be stronger towards the centre and is
mostly present among the fainter galaxies, whereas bright galaxies are
unaligned \citep{hun10}. The luminosity function shows a steep red
faint end upturn, suggesting that the least massive galaxies are just
being quenched at this epoch \citep{ban10}. Moreover, the cluster
population shows two distinct populations: two-thirds are unremarkable
blue, late-type spirals; the remainder, found only in the cluster
outskirts, are dusty red sequence galaxies whose star formation is
heavily obscured. There is also an excess of 100 $\mu$m-selected
galaxies that extend $\sim6$ Mpc in length along an axis that runs
NE-SW through the cluster center \citep{bal02,hai10}.

This is the third paper in a series on Abell~1689, and a fourth one is
in preparation.  \citet[][hereafter Paper I]{hought12} presents
imaging and spectroscopy of the cluster and analyses the
Faber-Jackson, Kormendy, and colour-magnitude relations, based on data
from the Advanced Camera for Surveys (ACS) on the {\em Hubble Space
  Telescope (HST)} and the Gemini Multi-Object Spectrograph on the
Gemini North telescope (GMOS-N).  \citet[][hereafter Paper II]{deug13}
presents integral field spectroscopy of a sample of galaxies observed
with the Fibre Large Array Multi Element Spectrograph (FLAMES) at the
Very Large Telescope, European Southern Observatory (ESO), and
investigates their internal kinematics.  In this paper, we analyse the
photometry of 65 galaxies at the centre of Abell~1689, perform
two-dimensional multi-component surface brightness decompositions,
provide a morphological classification, and derive the Fundamental
Plane (FP) relationship for the early-type galaxies (ETGs; i.e.,
ellipticals or lenticulars) using different photometric models. In
Paper IV (in preparation), we will provide a deep interpretation of
the FP by measuring accurate dynamical masses of the sample galaxies
observed with FLAMES and ACS.

This work is organised as follows. The sample selection is presented
in Sect. \ref{sec:sample}. The photometric analysis is described in
Sect.  \ref{sec:2Dparam_fit}. The morphological classification is
discussed in Sect. \ref{sec:morph}. In Sect. \ref{sec:FP} the FP is
derived and the results are compared to the local FP. In
Sect. \ref{sec:end} we draw our conclusions.  We assume $H_0=71$ km
s$^{-1}$ Mpc$^{-1}$, $\Omega_m=0.27$, and $\Omega_{\Lambda}=0.73$,
following the seven-year {\it Wilkinson Microwave Anisotropy Probe}
(WMAP7) cosmology \citep{kom11}, as in Paper I.

\section{Sample}
\label{sec:sample}

In this investigation, we used images from {\it HST} ACS.  We
downloaded the data, originally obtained for program GO-9289 (PI:
H.\ Ford), from MAST\footnote{Mikulsi Archive for Space Telescopes at
  the Space Telescope Science Institute}. The images are from the Wide
Field Channel (WFC) with the F625W filter, which approximates the
Sloan Digital Sky Survey $r$ filter and is nearly equivalent to
rest-frame $V$-band at the redshift of Abell~1689 ($z=0.183$). Our
data reduction procedures are described in Paper I.

We performed a photometric analysis of 65 galaxies, i.e., 54 galaxies
from the spectroscopic sample and 11 from the ancillary sample, as
described below.
  
The primary sample we selected are galaxies from Paper I that were
observed with GMOS-N plus those that were observed with FLAMES from
Paper II. The field of view of the ACS/WFC/F625W image contains 43
galaxies from Paper I and 29 galaxies from Paper II. The two samples
have 18 galaxies in common, so our entire spectroscopic sample
consists of 54 individual galaxies.

We performed a two-dimensional photometric decomposition of the
spectroscopic sample.  This also required photometric analysis of 11
additional galaxies that affect the surface-brightness distribution of
some of the spectroscopic sample galaxies on account of their
proximity (Sect. \ref{sec:fitproc}). These 11 galaxies were therefore
modelled with the aim of subtracting their two-dimensional
surface-brightness distributions to improve the fits for the primary
sample. We provide the derived parameters of our photometric
decomposition as ancillary data.

Visual inspection of the images of the spectroscopy sample reveals
that 41 are ETGs and 13 are late-type galaxies (LTGs; i.e.,
spirals). We list galaxy names, coordinates, morphological
classification, and central stellar velocity dispersions
$\sigma_{\star}$ (see Sect.\ \ref{sec:sigma_star}) of the
spectroscopic sample in Table \ref{tab:spec_sample}.  Our visual
inspection of the contaminating galaxies forming the ancillary sample
reveals that nine galaxies are ETGs and two are LTGs (Table
\ref{tab:ancill_sample}).

\begin{table*}
\begin{minipage}{126mm}
\caption{\small{Spectroscopic sample.}}
\label{tab:spec_sample}
\begin{scriptsize}
\begin{center}
\begin{tabular}{lcrlllrc}
\hline
\multicolumn{2}{c}{Galaxy} &
\multicolumn{1}{c}{RA} &
\multicolumn{1}{c}{DEC} &
\multicolumn{1}{c}{Type} &
\multicolumn{1}{c}{Data} &
\multicolumn{1}{c}{$\sigma_{\star}$}&
\multicolumn{1}{c}{FP sample} \\
\multicolumn{2}{c}{ID} &
\multicolumn{1}{c}{(h m sec)} &
\multicolumn{1}{c}{($\degr$ $\prime$ $\prime \prime$)} &
\multicolumn{1}{c}{} &
\multicolumn{1}{c}{} &
\multicolumn{1}{c}{(km s$^{-1}$)} &
\multicolumn{1}{c}{} \\
\multicolumn{1}{c}{(1)} &
\multicolumn{1}{c}{(2)} &
\multicolumn{1}{c}{(3)} &
\multicolumn{1}{c}{(4)} &
\multicolumn{1}{c}{(5)} &
\multicolumn{1}{c}{(6)} &
\multicolumn{1}{c}{(7)} &
\multicolumn{1}{c}{(8)} \\
\hline
S01 &  286,---   & 13 11 23.09 & $-$1 21 17.1    & Late   & G   &  150.7 $\pm$   2.9  &  No   \\  
S02 &  341, 28   & 13 11 24.47 & $-$1 21 10.9    & Early  & G,F &  190.5 $\pm$   3.6  &  Yes  \\           
S03 &  368,---   & 13 11 25.39 & $-$1 20 36.8    & Late   & G   &  167.1 $\pm$   3.3  &  No   \\  
S04 &  371,---   & 13 11 25.41 & $-$1 20 17.0    & Early  & G   &  166.9 $\pm$   6.5  &  Yes  \\             
S05 &  390, 30   & 13 11 25.96 & $-$1 19 51.7    & Early  & G,F &  171.5 $\pm$   3.3  &  Yes  \\      
S06 &  398, 14   & 13 11 26.24 & $-$1 19 56.3    & Early  & G,F &  280.2 $\pm$   4.7  &  Yes  \\      
S07 &  433,---   & 13 11 26.93 & $-$1 19 40.5    & Early  & G   &   67.7 $\pm$  11.8  &  Yes  \\               
S08 &  435, 16   & 13 11 26.94 & $-$1 19 36.6    & Early  & G,F &  151.7 $\pm$   4.1  &  Yes  \\               
S09 &  463, 13   & 13 11 27.43 & $-$1 20  2.3    & Early  & G,F &  182.7 $\pm$   3.6  &  Yes  \\               
S10 &  476,---   & 13 11 27.86 & $-$1 20  7.5    & Early  & G   &  260.4 $\pm$   5.1  &  Yes  \\        
S11 &  481,---   & 13 11 27.94 & $-$1 21 36.5    & Early  & G   &  143.8 $\pm$   4.1  &  Yes  \\        
S12 &  501,---   & 13 11 28.25 & $-$1 20 43.3    & Late   & G   &  149.1 $\pm$   4.8  &  No   \\     
S13 &  508,---   & 13 11 28.39 & $-$1 19 58.3    & Late   & G   &  126.9 $\pm$   5.5  &  No   \\          
S14 &  514, 29   & 13 11 28.48 & $-$1 20 24.9    & Early  & G,F &  179.5 $\pm$   2.3  &  Yes  \\             
S15 &  531,---   & 13 11 28.78 & $-$1 19  2.4    & Late   & G   &  101.5 $\pm$   6.8  &  No   \\     
S16 &  549, 10   & 13 11 29.04 & $-$1 21 16.6    & Early  & G,F &  220.3 $\pm$   2.9  &  Yes  \\             
S17 &  567, 17   & 13 11 29.35 & $-$1 19 16.4    & Early  & G,F &  250.4 $\pm$   3.8  &  No   \\         
S18 &  584, 12   & 13 11 29.52 & $-$1 20 27.8    & Early  & G,F &  270.2 $\pm$   5.0  &  Yes  \\       
S19 &  593,---   & 13 11 29.79 & $-$1 21  0.5    & Early  & G   &  131.9 $\pm$   3.2  &  Yes  \\        
S20 &  601,---   & 13 11 29.91 & $-$1 20 14.9    & Early  & G   &  109.3 $\pm$   6.7  &  Yes  \\        
S21 &  610, 9    & 13 11 30.02 & $-$1 20 39.9    & Late   & G,F &  122.1 $\pm$   2.3  &  No   \\     
S22 &  635,---   & 13 11 30.42 & $-$1 20 45.2    & Early  & G   &  255.0 $\pm$   3.6  &  Yes  \\               
S23 &  636,---   & 13 11 30.43 & $-$1 20 34.7    & Early  & G   &  152.6 $\pm$   5.5  &  Yes  \\               
S24 &  645,---   & 13 11 30.62 & $-$1 20 43.5    & Early  & G   &  175.9 $\pm$   4.5  &  Yes  \\               
S25 &  655,---   & 13 11 30.84 & $-$1 20 30.5    & Early  & G   &  151.5 $\pm$   3.4  &  Yes  \\              
S26 &  670, 25   & 13 11 31.14 & $-$1 21 27.6    & Early  & G,F &  240.9 $\pm$   4.4  &  Yes  \\               
S27 &  677,---   & 13 11 31.17 & $-$1 21 24.9    & Early  & G   &  185.8 $\pm$   5.9  &  Yes  \\                
S28 &  690, 5    & 13 11 31.45 & $-$1 19 32.5    & Early  & G,F &  285.0 $\pm$   2.4  &  Yes  \\          
S29 &  698,---   & 13 11 31.57 & $-$1 19 24.4    & Late   & G   &   94.0 $\pm$   5.6  &  No   \\             
S30 &  717, 18   & 13 11 32.04 & $-$1 19 24.1    & Early  & G,F &  182.5 $\pm$   3.9  &  Yes  \\             
S31 &  723, 24   & 13 11 32.14 & $-$1 21 37.9    & Early  & G,F &  183.4 $\pm$   3.8  &  Yes  \\        
S32 &  724,---   & 13 11 32.14 & $-$1 19 36.0    & Late   & G   &   37.2 $\pm$  13.1  &  No   \\     
S33 &  726,---   & 13 11 32.16 & $-$1 19 46.5    & Late   & G   &  218.0 $\pm$   2.6  &  No   \\          
S34 &  753, 6    & 13 11 32.71 & $-$1 19 58.3    & Early  & G,F &  312.6 $\pm$   2.6  &  Yes  \\             
S35 &  755,---   & 13 11 32.72 & $-$1 20 58.2    & Early  & G   &   92.8 $\pm$   6.5  &  Yes  \\        
S36 &  756, 19   & 13 11 32.76 & $-$1 19 31.4    & Early  & G,F &  266.9 $\pm$   3.1  &  Yes  \\               
S37 &  814,---   & 13 11 34.10 & $-$1 21  1.7    & Late   & G   &  130.1 $\pm$   3.9  &  No   \\     
S38 &  816,---   & 13 11 34.13 & $-$1 21 18.4    & Early  & G   &  114.8 $\pm$   3.4  &  Yes  \\              
S39 &  848, 22   & 13 11 34.81 & $-$1 20 59.0    & Early  & G,F &  185.7 $\pm$   3.4  &  Yes  \\        
S40 &  852, 21   & 13 11 34.91 & $-$1 20  4.2    & Early  & G,F &  116.7 $\pm$   3.1  &  Yes  \\               
S41 &  874,---   & 13 11 35.40 & $-$1 21 33.0    & Early  & G   &  205.9 $\pm$   2.7  &  Yes  \\         
S42 &  883,---   & 13 11 35.65 & $-$1 20 12.0    & Late   & G   &   49.2 $\pm$   8.2  &  No   \\                
S43 &  906,---   & 13 11 36.68 & $-$1 19 42.5    & Late   & G   &   37.2 $\pm$   9.3  &  No   \\             
S44 &  ---, 1    & 13 11 28.14 & $-$1 19 31.4    & Early  & F   &  236.8 $\pm$   6.6  &  Yes  \\   
S45 &  ---, 2    & 13 11 28.07 & $-$1 18 43.6    & Early  & F   &   93.8 $\pm$   4.8  &  Yes  \\   
S46 &  ---, 3    & 13 11 29.44 & $-$1 18 34.4    & Early  & F   &  198.9 $\pm$   5.6  &  Yes  \\   
S47 &  ---, 4    & 13 11 31.92 & $-$1 18 53.5    & Early  & F   &  101.5 $\pm$   6.3  &  Yes  \\   
S48 &  ---, 7    & 13 11 34.82 & $-$1 19 24.3    & Early  & F   &  129.1 $\pm$   4.3  &  Yes  \\   
S49 &  ---, 8    & 13 11 30.32 & $-$1 20 29.0    & Early  & F   &  223.7 $\pm$   3.3  &  Yes  \\   
S50 &  ---, 11   & 13 11 27.88 & $-$1 21 12.7    & Late   & F   &  112.3 $\pm$   4.0  &  No   \\   
S51 &  ---, 15   & 13 11 25.14 & $-$1 19 30.8    & Early  & F   &  161.1 $\pm$   5.9  &  Yes  \\   
S52 &  ---, 23   & 13 11 31.12 & $-$1 20 52.4    & Early  & F   &  181.4 $\pm$   5.6  &  Yes  \\   
S53 &  ---, 26   & 13 11 30.10 & $-$1 20 42.6    & Early  & F   &  250.0 $\pm$   5.9  &  Yes  \\   
S54 &  ---, 27   & 13 11 30.07 & $-$1 20 28.3    & Early  & F   &  231.7 $\pm$   6.1  &  Yes  \\   
\hline
\end{tabular}
\end{center}
{\em Note.} Col. (1): galaxy ID from this paper.
Col. (2): galaxy ID from paper I and/or from paper II.
Col. (3): right ascension (J2000.0). 
Col. (4): declination (J2000.0). 
Col. (5): Early/late type classification.  
Col. (6): spectroscopic data available, GMOS-N (G) and/or FLAMES
(F). 
Col. (7): central stellar velocity dispersion and its 1$\sigma$ error.
Col. (8): sample adopted in FP analysis.
\end{scriptsize}
\end{minipage}
\end{table*}

\section{Two-dimensional surface-brightness fits}
\label{sec:2Dparam_fit}
To perform a photometric decomposition of each galaxy, we used the
code {\tt GASP2D}, which is described in detail by
\cite{mend08,mend14}.  Briefly, {\tt GASP2D} assumes that the
surface-brightness distribution of elliptical galaxies consists of a
single bulge component, and that disc galaxies are the sum of a bulge,
a disc and, if necessary, a bar component. Each structure has
elliptical and concentric isophotes with constant ellipticity,
$\epsilon=1-q$, and constant position angle (PA).  This algorithm has
been used successfully to model ellipticals and brightest cluster
galaxies \citep[e.g.,][]{asc11}, unbarred and barred disc galaxies
\citep[e.g.,][]{mor12}, active galaxies with an unresolved component
\citep[e.g.,][]{ben13}, high-$z$ galaxies \citep{zan16}, and, more
recently, the large sample of galaxies from the Calar Alto Legacy
Integral Field Area data release 3 \citep[CALIFA-DR3,][]{mend17}.

\subsection{Photometric model}
\label{sec:photmod}
For ellipticals and bulge components, we
adopt the \cite{Sersic1963} law, i.e., 
\begin{equation} 
I_{\rm b}(r)=I_{\rm e}\,{\rm e}^{-b_n\left[\left(r/r_{\rm e} 
\right)^{1/n}-1\right]},
\label{eqn:bulge_surfbright} 
\end{equation} 
%
where $r_{\rm e}$, $I_{\rm e}$, and $n$ are the effective (or
half-light) radius, the surface brightness at $r_{\rm e}$, and a shape
parameter describing the curvature of the surface-brightness profile,
respectively. The value of $b_n$ is coupled to $n$ so that half of the
total luminosity of the bulge is within $r_{\rm e}$ and can be
approximated as $b_n = 2n -0.324$ \citep{cio91}. 
The total luminosity of the bulge is
%
\begin{equation} 
L_{\rm bulge} = 2 \pi I_{\rm 0,bulge}  \,n 
\,r_{\rm e}^{2}\frac{\Gamma(2n)}{b_{n}^{2n}} \,q_{\rm bulge}, 
\label{eq:total_bulge}
\end{equation} 
%
where $I_{\rm 0,bulge}=I_{\rm e}\,10^{b_n}$ is the central surface
brightness of the bulge, $q_{\rm bulge}$ is the bulge axial ratio, and
$\Gamma$ is the Euler gamma function.

We consider as a special case the \cite{deV48} law,
which is essentially Eq. \ref{eqn:bulge_surfbright}
with a fixed value of the S\'{e}rsic index $n=4$.

We describe the surface brightness of the disc component by an
exponential law \citep{Freeman70},
\begin{equation} 
I_{\rm d}(r) = I_{\rm 0,disc}\,e^{-r/h}, 
\label{eqn:disc_surfbright} 
\end{equation} 
%
where $I_{\rm 0,disc}$ and $h$ are the central surface brightness and scale-length 
of the disc, respectively. 
The total luminosity of the disc is
%
\begin{equation} 
L_{\rm disc} = 2 \pi I_{\rm 0,disc}\,h^2\,q_{\rm disc}, 
\label{eq:total_disc}
\end{equation} 
where $q_{\rm disc}$ is the disc axial ratio.

%
We adopt the radial  surface-brightness  profile   of  a  
\cite{Ferrers1877} ellipsoid
 to describe  bar components,
%
\begin{equation} 
I_{\rm bar}(r) = \left\{ 
\begin{array}{ll}
  I_{\rm 0,bar} \left[1- \left(\frac{r}{r_{\rm bar}} \right)^2 
    \right]^{n_{\rm bar}+0.5} & \mbox{$r \le r_{\rm bar}$} \\
  0 & \mbox{$r > r_{\rm bar}$} ,
\end{array} 
\right.
\label{eq:ferrers}
\end{equation} 
%
where $I_{\rm 0,bar}$, $r_{\rm bar}$, and $n_{\rm bar}$ are the
central surface brightness, length, and shape parameter of the
surface-brightness profile of the bar, respectively. The total
luminosity of the bar is
%
\begin{equation} 
L_{\rm bar} = 2 \pi I_{\rm 0, bar} r_{\rm bar}^{4}
  \int_{0}^{\infty}r (r_{\rm bar}^{2}-r^{2})^{n_{\rm bar}+0.5} dr.
\label{eq:total_ferrers_n}
\end{equation} 

 We chose to fix the $n_{\rm bar}$ parameter at  $n_{\rm bar}=2$,
 following \cite{Lau05}. The total luminosity of the bar
 for $n_{\rm bar}=2$ is
%
\begin{equation} 
L_{\rm bar}= \pi I_{\rm 0,bar} (1-\epsilon_{\rm bar}) 
  r_{\rm bar}^{2}\frac{\Gamma(7/2)}{\Gamma(9/2)}. 
\label{eq:total_ferrers_2}
\end{equation} 

\subsection{Fitting procedure}
\label{sec:fitproc}
We performed multiple fits of the sky-subtracted images of the 
galaxies. Specifically, each ETG was fitted
\begin{enumerate}
\item as a single bulge component following a de Vaucouleurs profile
  (fits hereafter referred to as \devauc);
\item as a single bulge component following a S\'ersic profile,
   (hereafter \sersic);
\item as a sum of a bulge following a S\'ersic profile, and a disc
  component (hereafter \sedisc); or
\item when a bar is present, also as a sum of a bulge following a
  S\'ersic profile, a disc, and a bar component (hereafter \sedibar).
\end{enumerate}

Each LTG was fitted with a \sedisc\ model, or a \sedibar\ model in cases where a bar was
detected. 

Since {\tt GASP2D} accounts for seeing effects, for each 
galaxy we used an appropriate PSF, whose details are given in Paper I.

\begin{table}
\caption{\small{Ancillary sample.}}
\label{tab:ancill_sample}
\begin{scriptsize}
\begin{center}
\begin{tabular}{crrl}
\hline
\multicolumn{1}{c}{Galaxy} &
\multicolumn{1}{c}{RA} &
\multicolumn{1}{c}{Dec.} &
\multicolumn{1}{c}{Type} \\
\multicolumn{1}{c}{ID} &
\multicolumn{1}{c}{(h m sec)} &
\multicolumn{1}{c}{($\degr$ arcmin arcsec)} &
\multicolumn{1}{c}{} \\
\multicolumn{1}{c}{(1)} &
\multicolumn{1}{c}{(2)} &
\multicolumn{1}{c}{(3)} &
\multicolumn{1}{c}{(4)} \\
\hline
A01  &  13 11 27.27 & $-$1 20 09.7   & Late    \\ 
A02  &  13 11 30.26 & $-$1 20 51.6   & Early   \\ 
A03  &  13 11 31.03 & $-$1 21 27.6   & Early   \\ 
A04  &  13 11 28.38 & $-$1 18 44.6   & Early   \\ 
A05  &  13 11 28.08 & $-$1 19 28.1   & Early   \\ 
A06  &  13 11 24.21 & $-$1 21 07.4   & Late    \\ 
A07  &  13 11 30.95 & $-$1 20 27.6   & Early   \\ 
A08  &  13 11 30.05 & $-$1 20 17.1   & Early   \\ 
A09  &  13 11 29.81 & $-$1 20 19.6   & Early   \\ 
A10  &  13 11 28.65 & $-$1 20 26.3   & Early   \\ 
A11  &  13 11 29.20 & $-$1 21 20.5   & Early   \\ 
\hline
\end{tabular}
\end{center}
{\em Note.} Col. (1): galaxy ID (Fig.\ \ref{fig:FoV}). Col. (2): right
ascension (J2000.0). Col. (3): declination (J200.0). Col. (4):
Early/Late type classification.
\end{scriptsize}
\end{table}

The choice of the region in which we perform the $\chi^2$ minimization
(see \citealt{mend08} for details on the minimization algorithm) is a
crucial issue.  After extensive testing with mock galaxies, we
concluded that the most-suitable maximum fitting radius, $r_{\rm
  max}$, is where $I(r_{\rm max})=1.5 \,\sigma_{\rm sky}$.  Indeed we
created artificial galaxies as described in
Sect. \ref{sec:err_photfit}, and performed photometric decompositions
to a limit surface brightness of $0.1\,\sigma_{\rm sky}$,
$0.25\,\sigma_{\rm sky}$, $0.50\,\sigma_{\rm sky}$, $0.75\,\sigma_{\rm
  sky}$, $1.0\,\sigma_{\rm sky}$,..., and $4.5\,\sigma_{\rm sky}$. We
then analysed the distribution of the errors on the parameters (as in
Sect. \ref{sec:err_photfit}).  Extending the fitting area to pixels
where the sky noise dominates over the surface-brightness of the
galaxy leads to significant systematic errors in the fitted
photometric parameters. In particular, it leads to an overestimate of
$R_e$ and S\'ersic index $n$, if a single S\'ersic component is
fitted, and an overestimate of $R_e$, $n$, and also the scale length
$h$, if a sum of S\'ersic and exponential components are fitted. In
both cases, the size of the galaxy is overestimated. On the other
hand, if the fit is performed within a region that is too restricted,
the size of the galaxy derived from the photometric decomposition is
underestimated. We find that thorough testing to identify the optimal
maximum fitting radius is essential to avoiding potentially severe
systematic errors in scaling relations involving galaxy sizes.

\begin{figure*}
\includegraphics[scale=.9,angle=0.]{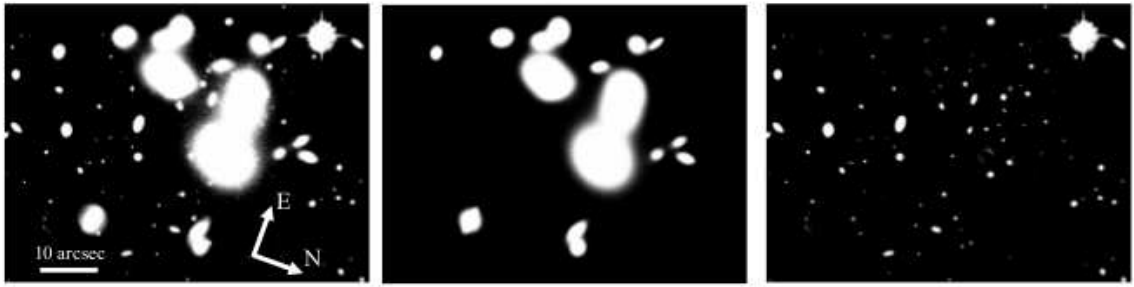}
\caption{Left: centre of Abell~1689, corresponding to region A of
  Fig. \ref{fig:FoV}. Centre: synthetic image of region A, resulting
  from the composition of the modelled surface-brightness distribution
  of the eighteen brightest galaxies. Right: residual (observed $-$
  modelled) image of region A.  The gray-scale, linear scale and
  orientation used for the panels are kept the same. Linear scale and
  orientation are shown on the left panel.}
\label{fig:synth_centre}
\end{figure*}

It is challenging to fit the surface brightness distributions of
galaxies that overlap.  Each of these galaxies consequently has an
underlying surface brightness gradient that is due to its neighbours,
and cannot be neglected. It must be treated as extra background light
that must be removed. Although {\tt GASP2D} is able to fit multiple
galaxies simultaneously, dealing with more than two galaxies at once
leads to degeneracy in the fit parameters. Therefore, for each galaxy,
we took into account contamination due to the neighbours by
subtracting their surface brightness models in an iterative way.

The proximity problem is particularly onerous in two dense regions of
the cluster, marked ``A'' and ``B'' in Fig.\ \ref{fig:FoV}. Region A,
which is the centre of the cluster with a surface area of $\sim 0.5$
arcmin$^2$, includes thirteen of our galaxies.  As noted in
Sect. \ref{sec:sample}, we fitted eleven ancillary galaxies in the
field of view whose surface brightness affects the sample galaxies and
whose photometric decomposition is presented in this paper. Five of
these additional galaxies are in region A. We thus modelled the
central eighteen galaxies in the following iterative fashion:
\begin{enumerate}
\item Fit the central cD galaxy (galaxy S18) and subtract its model;
\item Fit the outer less-contaminated galaxies and subtract their models;
\item Proceeding inward, fit the less-contaminated galaxies and
  subtract their models\footnote{For each galaxy, we fitted a \sersic,
    \sedisc, and, in cases where a bar is present, \sedibar, and then
    chose the model that best describes the surface brightness
    distribution of the galaxy, particularly in its outer regions, by
    visual inspection of the residuals images.};
\item Repeat the previous step until the sample is complete; 
\item Repeat steps (i)--(iv) for each galaxy. Each fit
  is performed on the observed image from which the 
  models of the surrounding galaxies from the most recent iteration are subtracted, 
  leaving a final image that contains only the galaxy currently being modelled.
  \end{enumerate}

\noindent Step (v) is repeated until consistent values of fitted
parameters for the whole central sample are obtained. For region A, we
performed step (v) six times to obtain convergence.  The comparison
between the observed surface-brightness distributions of the eighteen
galaxies in the centre of Abell~1689 and their models is shown in
Fig. \ref{fig:synth_centre}.

We then subtracted from the whole observed ACS/WFC/F625W image the
models of the eighteen galaxies and used the resulting image to fit
the seven galaxies in region B. We used the same iterative method
described above, starting with the most extended galaxy (galaxy S34).

Finally, we subtracted from the original observed image the models of
the twenty-five galaxies fitted in regions A and B and used the
resulting image to fit the rest of the sample galaxies. We adopted the
iterative method described above for a few sub-groups of three or four
galaxies.  When the iterative process converged, we cut a frame for
each sample galaxy and used it for the final fits. In all cases, the
individual frames were large enough to include the entire region
defined by $r_{\rm max}$.

We were able to fit all the galaxies with the exception of S17. In
this case, inspection reveals the presence of an edge-on disk, for
which a thick-disk model is required. {\tt GASP2D} is not yet able to
fit a thick-disk model, so only \devauc\ and \sersic\ fits of S17 were
performed.

\subsection{Error estimates}
\label{sec:err_photfit}
To estimate the errors on the fitted parameters, we ran a series of
Monte Carlo simulations. For every fit type --- \devauc, \sersic,
\sedisc, and \sedibar\ --- we created 250 artificial galaxies
characterised by parameters appropriate to the specific
model. Simulations were carried out in one-magnitude bins, and five
bins were required to cover the luminosity range of our sample. Thus,
for each fit type, about 1250 artificial galaxies were created. Each
parameter $p_i$ was randomly chosen in the range $p_{\rm min}- 0.3
p_{\rm min}< p_i <p_{\rm max}+0.3\,p_{\rm max}$, where $p_{\rm min}$
and $p_{\rm max}$ are the minimum and maximum values of the fitted
parameter on the real images in that particular magnitude bin.

The size of each artificial frame is $700\times 700$ pixel$^2$,
equivalent to $21\times21$ arcsec$^2$ (pixel scale $=0.03$ arcsec
pixel$^{-1}$). This is large enough to enclose $r_{\rm max}$ for all
fits.  We separately produced 250 mock galaxies in frames of
$1600\times 1600$ pixel$^2$, equivalent to $48\times48$ arcsec$^2$, to
run simulations for the central cD galaxy. All the synthetic galaxies
were convolved with a PSF that was randomly chosen from those produced
for the fits to the observed image.  The pixel scale, CCD gain, and
read-out-noise of the artificial images match those of the real
HST/ACS/F625W image. In addition, we added photon noise in order to
obtain a signal-to-noise ratio consistent with that of the original
image.

We then ran the {\tt GASP2D} two-dimensional parametric decomposition
as described above to analyze the images of the mock galaxies. We
studied the distribution of the relative errors on the parameters as
$(p_{\rm output}/p_{\rm input}-1)$. For position angles and axis
ratios we derived the absolute errors, $(p_{\rm output}-p_{\rm
  input})$.  All the distributions appear to be nearly Gaussian.  We
measured the median and absolute deviation of each distribution and
applied $5\sigma$-clipping to reject outliers. Median values were used
to detect the possible presence of systematic errors and the absolute
deviations were used to derive the errors on the single parameters. We
did not identify any systematic errors, as all median values are
consistent with zero.

In Table \ref{tab:phot_par}, we present the best-fit observed
parameters with their errors for the whole sample, adopting \sersic,
\sedisc, and \sedibar\ models according to the morphological
classification presented in Sect.\ \ref{sec:morph}. In Figures
\ref{fig:fit_S01}-\ref{fig:fit_A11} we show the corresponding {\tt
  GASP2D} fits.  We give the results of the photometric decomposition
of the ETGs of the spectroscopic sample with \devauc\ and
\sersic\ models in Tables \ref{tab:devauc_par} and
\ref{tab:sersic_par}, respectively.

\begin{figure*}
\centering
\includegraphics[scale=0.7,angle=90.]{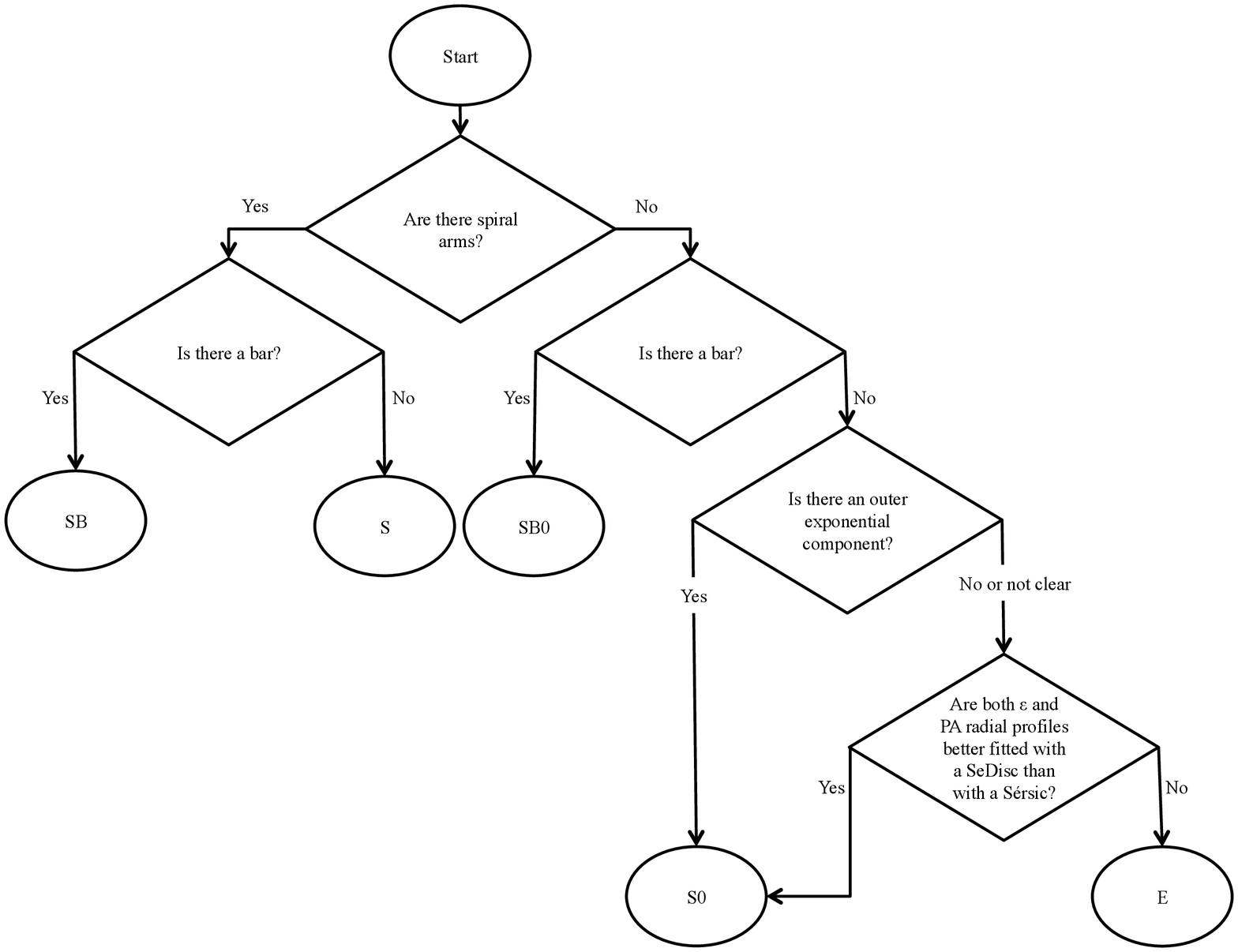}
\caption{Flowchart describing the method applied to classify the
  galaxies.}
\label{fig:flowchart}
\end{figure*}

\section{Morphological Classification}
\label{sec:morph}
We were able to distinguish between ETGs and LTGs by visual
inspection, as mentioned in Sect. \ref{sec:sample}, because the
presence of spiral arms is clearly detectable given the high
signal-to-noise ratio and spatial resolution of the data.

Nevertheless, on the basis of visual inspection alone, it is not
always possible to distinguish among ellipticals (E), unbarred
lenticulars (S0), and possibly barred lenticulars (SB0), or to
distinguish between spirals (S) and barred spirals (SB). This
necessitates a more sophisticated and quantitative approach.  We
therefore made use of the multi-component photometric decompositions
and use the isophotal parameters derived in our fits to check for
signatures of bars and discs.  Barred galaxies are characterised by
the presence of a local maximum in the ellipticity radial profile and
constant PA in the bar region \citep[e.g.,][]{ague09}. A disc
component is characterised by an exponential surface-brightness radial
profile with constant ellipticity and PA.

By the combination of visual inspection and analysis of the isophotal
parameters, we are able to detect with confidence the presence of a
bar and therefore classify a galaxy as spiral (S), barred spiral (SB),
or barred lenticular (SB0).

A more difficult problem arises when we need to distinguish between an
E and S0, i.e., detect the presence of a disc.  We note that for an E
or S0, a \sersic\ model is always a poorer fit than a \sedisc\ model,
as the former has seven free parameters (i.e., $I_e$, $r_e,n$,
$\epsilon_b$, PA$_b$, and the centre $x_0$, $y_0$) and the latter has
four more ($I_0$, $h$, $\epsilon_d$, PA$_d$).  We therefore
conservatively classify a galaxy as an S0 only if we can associate the
fitted exponential component to a real structure of the galaxy and not
use it just as a mathematical expedient \citep[see
    also][]{fri05,mend17}.

For each candidate E or S0 galaxy, we visually examined the
ellipse-averaged radial profile of the surface brightness,
ellipticity, and position angle. We also compared the \sersic\ and
\sedisc\ fits, and closely inspected the modelled and residual images,
as the latter are particularly useful for detection of any structured
residual of the galaxy. If an outer exponential component is present,
the galaxy is classified as S0.  If no outer exponential is detected
or the result is ambiguous, the galaxy is classified as S0 if both the
ellipticity and PA radial profiles are better fitted with a \sedisc,
otherwise the galaxy is classified as E.  No additional spiral
galaxies were detected from the analysis of the residuals of the
photometric decomposition.

The method used to classify the galaxies is shown in the flowchart in
Fig.\ \ref{fig:flowchart}. We further subclassified the ellipticals as
E$n$, where $n$ is the integer approximating the value
$10\times(1-q_{\rm bulge})$ and $0<n<6$, following the \cite{vdb76}
classification. For unbarred and barred S0s and spirals, we also used
the subclasses ``a, b, c'' \citep{vdb76} on the basis of the
disc-to-bulge luminosity ratio \citep[][Kormendy, private
    communication]{kor12}.

Galaxy S18 is a cD, a giant elliptical with a typical extended
envelope which is very well fit by an exponential component. Thus, its
total surface-brightness distribution is best fit by a \sedisc\ model.

The morphological classification of the galaxies and the features that
allow us to discriminate among the different classes are shown in
Table \ref{tab:morph_type}. Es are better fit by a \sersic\ model, S0s
and Ss by a \sedisc\ model, and SB0s and SBs by a \sedibar\ model. The
\devauc\ model provides poorer fits of our ETGs than the
\sersic\ model, given that typically, S\'ersic indices $n\neq4$.


\begin{table*}
\begin{minipage}{126mm}
\caption{\small{Morphological classification of the sample galaxies.}}
\label{tab:morph_type}
\begin{scriptsize}
\begin{center}
\begin{tabular}{clccccc}
\hline
\multicolumn{1}{c}{Galaxy} &
\multicolumn{1}{c}{Type} &
\multicolumn{1}{c}{Spiral arms} &
\multicolumn{1}{c}{Bar} &
\multicolumn{1}{c}{Exp. component} &
\multicolumn{1}{c}{$\varepsilon$} &
\multicolumn{1}{c}{PA} \\
\multicolumn{1}{c}{ID} &
\multicolumn{1}{c}{} &
\multicolumn{1}{c}{} &
\multicolumn{1}{c}{} &
\multicolumn{1}{c}{} &
\multicolumn{1}{c}{} &
\multicolumn{1}{c}{} \\
\multicolumn{1}{c}{(1)} &
\multicolumn{1}{c}{(2)} &
\multicolumn{1}{c}{(3)} &
\multicolumn{1}{c}{(4)} &
\multicolumn{1}{c}{(5)} &
\multicolumn{1}{c}{(6)} &
\multicolumn{1}{c}{(7)} \\
\hline
     &                &   & Spectroscopic sample \\
S01  & Sb          & Y & N &           &   &   \\    
S02  & S0b         & N & N & Y         &   &   \\    
S03  & SBbc        & Y & Y &           &   &   \\    
S04  & E2          & N & N & N         & N & N \\    
S05  & E2          & N & N & N         & N & N \\    
S06  & E4          & N & N & N         & Y & N \\    
S07  & S0b         & N & N & Y         &   &   \\    
S08  & S0ab        & N & N & not clear & Y & Y \\    
S09  & S0ab        & N & N & Y         &   &   \\    
S10  & S0ab        & N & N & Y         &   &   \\    
S11  & S0b         & N & N & Y         &   &   \\    
S12  & SBb         & Y & Y &           &   &   \\    
S13  & Sb          & Y & N &           &   &   \\    
S14  & S0ab        & N & N & Y         &   &   \\    
S15  & SBbc        & Y & Y &           &   &   \\    
S16  & S0ab        & N & N & Y         &   &   \\    
S17  & S0          & N & N & Y         &   &   \\    
S18  & cD          & N & N & Y         &   &   \\    
S19  & S0ab        & N & N & Y         &   &   \\    
S20  & SB0bc       & N & Y &           &   &   \\    
S21  & SBbc        & Y & Y &           &   &   \\    
S22  & S0ab        & N & N & Y         &   &   \\    
S23  & S0ab        & N & N & not clear & Y & Y \\    
S24  & S0b         & N & N & Y         &   &   \\    
S25  & SB0ab       & N & Y &           &   &   \\    
S26  & S0ab        & N & N & Y         &   &   \\    
S27  & E1          & N & N & not clear & Y & N \\    
S28  & E2          & N & N & N         & N & N \\    
S29  & Sbc         & Y & N &           &   &   \\    
S30  & S0a         & N & N & Y         &   &   \\    
S31  & E3          & N & N & N         & N & N \\    
S32  & Sb          & Y & N &           &   &   \\    
S33  & Sb          & Y & N &           &   &   \\    
S34  & S0ab        & N & N & Y         &   &   \\    
S35  & S0ab        & N & N & Y         &   &   \\    
S36  & S0ab        & N & N & Y         &   &   \\    
S37  & SBc         & Y & Y &           &   &   \\    
S38  & S0ab        & N & N & Y         &   &   \\    
S39  & SB0ab       & N & Y &           &   &   \\    
S40  & S0b         & N & N & Y         &   &   \\    
S41  & S0b         & N & N & Y         &   &   \\    
S42  & Sc          & Y & N &           &   &   \\    
S43  & SBc         & Y & Y &           &   &   \\    
S44  & S0b         & N & N & Y         &   &   \\    
S45  & S0b         & N & N & Y         &   &   \\    
S46  & S0ab        & N & N & Y         &   &   \\    
S47  & S0ab        & N & N & Y         &   &   \\    
S48  & SB0ab       & N & Y &           &   &   \\    
S49  & S0ab        & N & N & not clear & Y & Y \\    
S50  & SBb         & Y & Y &           &   &   \\    
S51  & S0a         & N & N & not clear & Y & Y \\    
S52  & S0b         & N & N & Y         &   &   \\    
S53  & S0b         & N & N & Y         &   &   \\    
S54  & S0b         & N & N & not clear & Y & Y \\    
\hline                                              
     &                &   & Ancillary sample \\     
A01  & Sc          & Y & N &           &   &   \\    
A02  & S0b         & N & N & Y         &   &   \\    
A03  & S0ab        & N & N & not clear & Y & Y \\    
A04  & S0bc        & N & N & Y         &   &   \\    
A05  & E5          & N & N & N         & N & N \\    
A06  & Sab         & Y & N &           &   &   \\    
A07  & S0b         & N & N & Y         &   &   \\    
A08  & S0ab        & N & N & Y         &   &   \\    
A09  & E2          & N & N & not clear & N & N \\    
A10  & S0ab        & N & N & Y         &   &   \\    
A11  & S0bc        & N & N & Y         &   &   \\    
\hline                                              
\end{tabular}
\end{center}
{\em Note.} Col. (1): galaxy ID (Fig.\ \ref{fig:FoV}).  Col. (2): morphological type.
Col. (3): presence of spiral arms. Col. (4): presence of a
bar. Col. (5): presence of an outer exponential component. Col. (6):
$\epsilon$ radial profile better fitted with a \sedisc\ rather than a
\sersic. Col. (7): PA radial profile better fitted with a
\sedisc\ rather than a \sersic.
\end{scriptsize}
\end{minipage}
\end{table*}

\section{FP Analysis}
\label{sec:FP}

The sample analysed to determine the FP coefficients for Abell 1689 is
composed of the ETGs of the spectroscopic sample with successful
photometric decomposition.  Only galaxy S17 is excluded from this
analysis on account of its edge-on disk component
(Sect. \ref{sec:fitproc}), so the total sample used in the FP analysis
consists of the 40 galaxies listed in Table \ref{tab:spec_sample} .

\subsection{Central stellar velocity dispersions}
\label{sec:sigma_star}

\begin{figure}
\centering
\includegraphics[scale=0.4,angle=0.]{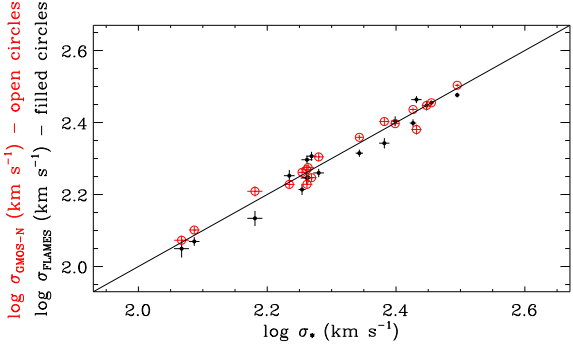}
\caption{Central stellar velocity dispersions from GMOS-N (red open
  circles) and FLAMES (black filled circles) versus the adopted
  $\sigma_{\star}$ values. The continuous line defines the one-to-one
  relation.}
\label{fig:sigma_comp}
\end{figure}

We use central stellar velocity dispersions $\sigma_{\star}$ from
Paper I, which are already corrected to a standard projected aperture
of 1.62 kpc, equivalent to 3.4 arcsec at the distance of the Coma
  galaxy cluster \citep{jor95b}. Stellar velocity dispersions of the
sample galaxies from Paper II were re-extracted for this study from
the FLAMES/GIRAFFE spectra using a synthetic circular aperture that
projects to $1.62\,$ kpc and adjusting the seeing to that of the
GMOS-N data (${\rm FWHM} \approx 1$\, arcsec).  For the 18 galaxies
for which we have both GMOS-N and re-extracted FLAMES measurements, we
take $\sigma_{\star}$ to be the weighted mean of the two values. The
comparison between GMOS-N, re-extracted FLAMES, and mean velocity
dispersions is shown in Fig. \ref{fig:sigma_comp}. The average
  difference between GMOS-N and FLAMES stellar velocity dispersion
  values, $\left<\sigma_{{\rm
      GMOS-N},i}-\sigma_{{\rm FLAMES},i}\right>=4.3$ km\,s$^{-1}$, that is
  within the mean $1\sigma$ error in the velocity dispersion
  ($\left<1\sigma_{\rm FLAMES}\right>=5.9$ km\,s$^{-1}$ and
  $\left<1\sigma_{\rm GMOS-N}\right>=4.4$ km\,s$^{-1}$). For this
  reason we conclude that the two sets of data are consistent. The
values of $\sigma_{\star}$ adopted for this analysis are given in
Table \ref{tab:spec_sample}.

\subsection{FP fits of Abell~1689 ETGs}
\label{sec:fp_fits}

We use the fitting algorithm {\tt LTS\_PLANEFIT} described by
\citet{cap13}, which combines the robust Least Trimmed Squares
technique of \citet{rou06} with a least-squares fitting algorithm that
allows for errors in all variables as well as intrinsic scatter.  The
best-fitting plane is defined as $z=a+b(x-x_0)+c(y-y_0)$, where $x_0$
and $y_0$ are the median of the measured values $x_j$ and $y_j$,
respectively. The intrinsic scatter, $\epsilon_z$, is in the
  $z$-coordinate and defined in Sect. 3.2.1 of \citet[][Eq. 7 and
    following paragraph]{cap13}. The observed scatter, $\Delta$, is
  defined as the standard deviation of
  $[a+b(x_j-x_0)+c(y_j-y_0)-z_j]$, where $x_j$, $y_j$, and $z_j$ are
  the fitted data values. In all our fits, we set the clipping
parameter to $5\sigma$, which results in no rejections of galaxies.
Our choice of a large clipping parameter is driven by two
considerations: (i) we have carefully checked each individual galaxy
while performing the photometric decomposition and find no physical
reason to exclude any galaxy, and (ii) for a direct comparison of the
FP fits for different photometric models, we want the sample of
galaxies to be the same in each case.  The central cD galaxy S18 could
be considered an ``outlier'' for its peculiar surface brightness
distribution, but we find consistent results regardless of whether or
not S18 is included in the sample.

\begin{figure}
\includegraphics[scale=.6,angle=0.]{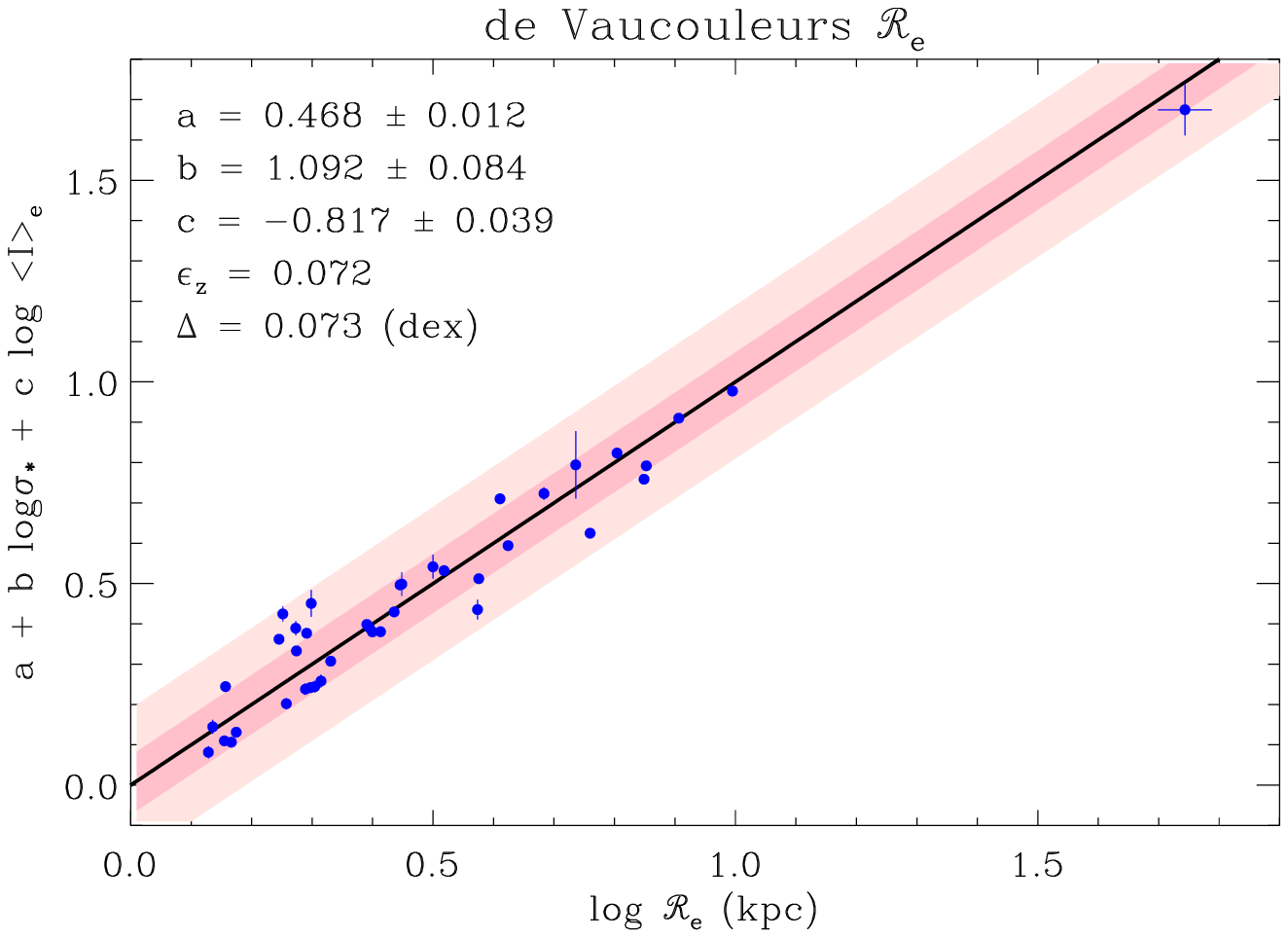}
\includegraphics[scale=.6,angle=0.]{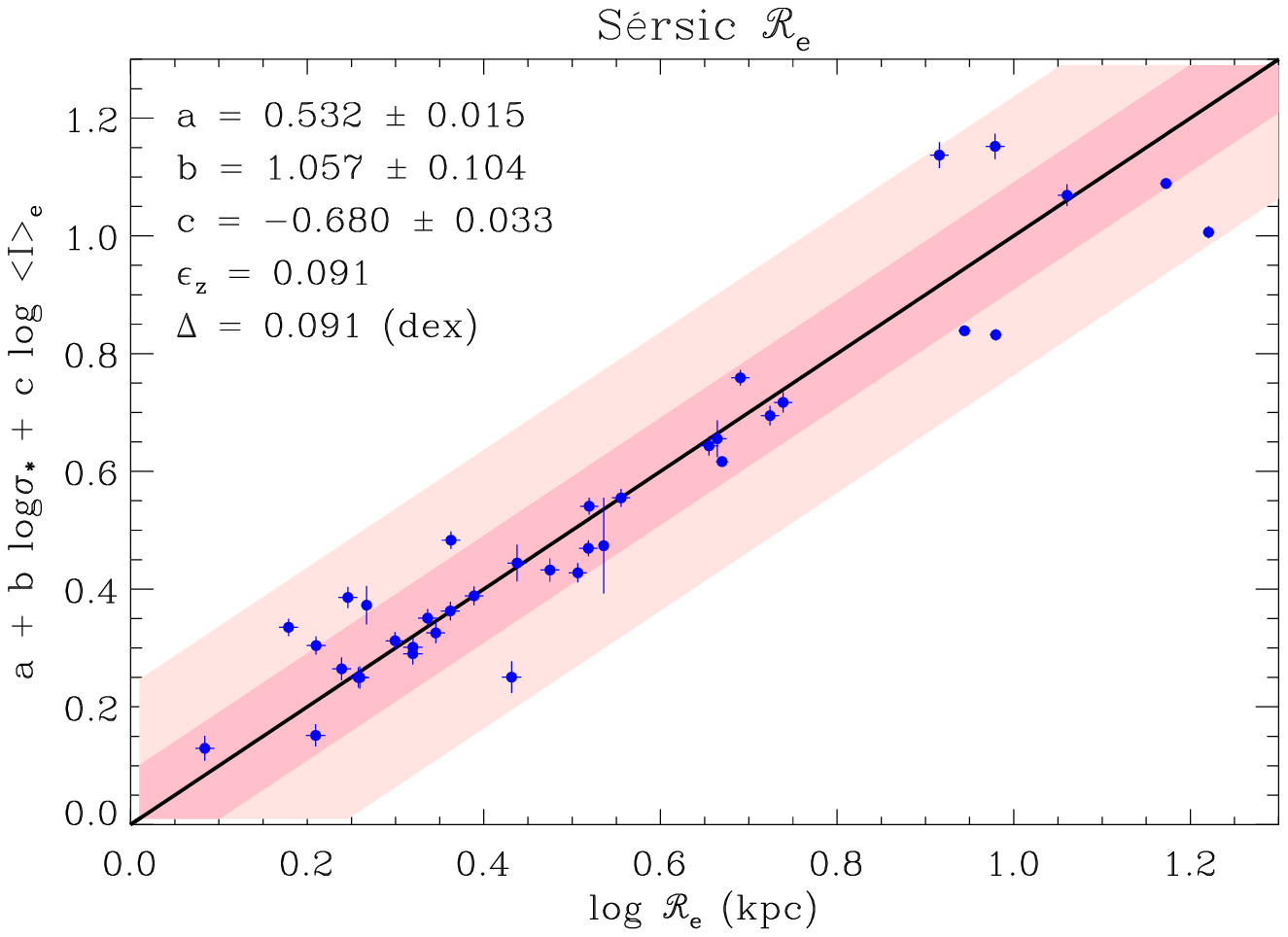}
\includegraphics[scale=.6,angle=0.]{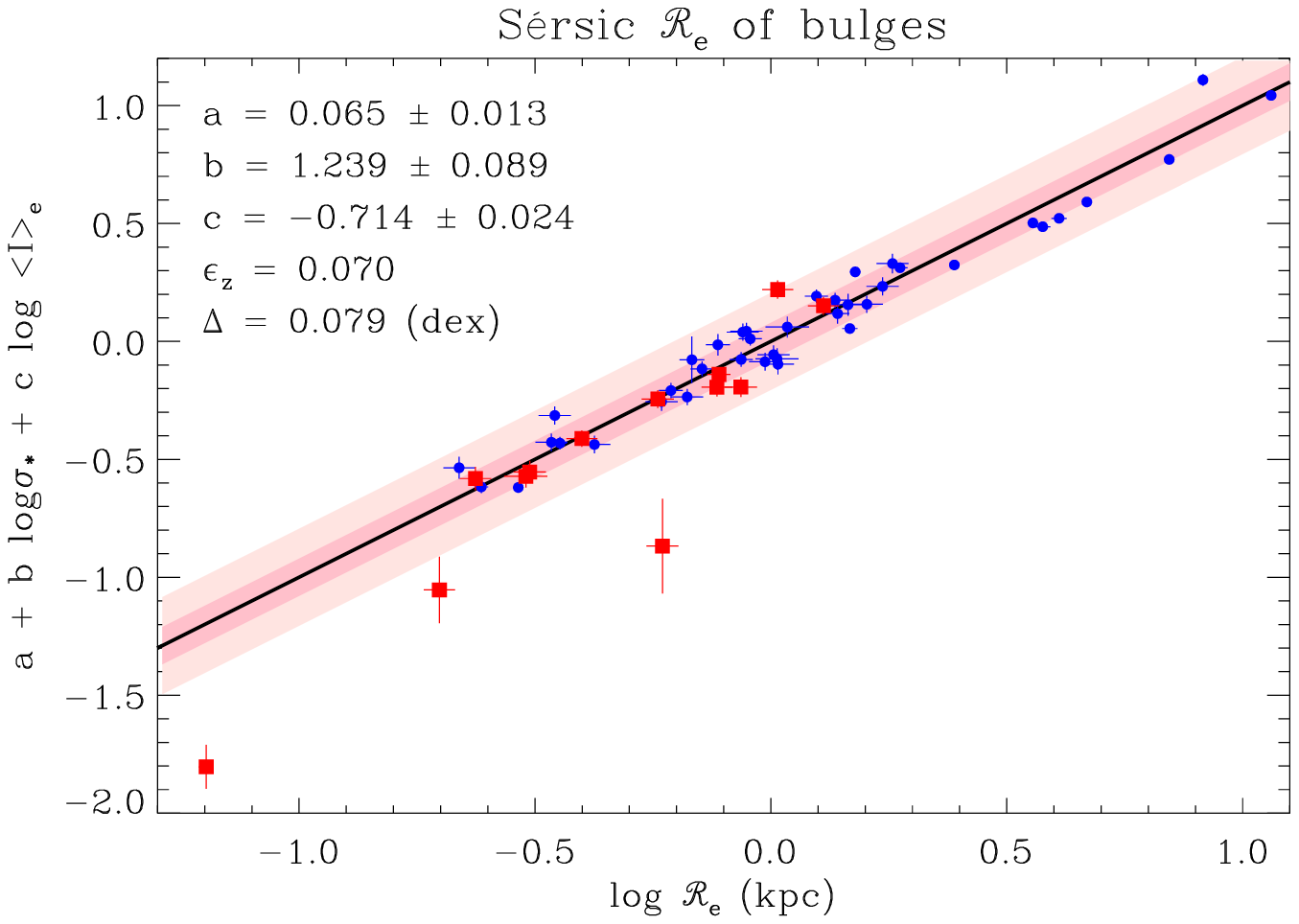}
\caption{Edge-on view of the FP with $\log\mathcal{R}_{\rm e}$ as
  dependent variable and using $\mathcal{R}_{\rm e}$ and $\left<I
  \right>_{\rm e}$ from \devauc\ photometric model (top panel);
  \sersic\ model (middle panel), and \sersic\ model of bulges (bottom
  panel), as described in the text. Blue filled circles: ETG sample;
  red filled squares: bulges of the LTG sample. The LTG sample is not
  used for the fit and plotted to show how it lies on the FP. The
  dark- and light-pink shaded regions enclose the $1\sigma$
  (equivalent to 68 per cent of the values for a Gaussian
  distribution) and $2.6\sigma$ (99 per cent) observed scatter,
  respectively.}
\label{fig:fp_re}
\end{figure}

\subsubsection{$\log\mathcal{R}_{\rm e}$ as the dependent variable}
\label{sec:fp_re}
We first fitted the FP in the classical form \citep{djo87},
\begin{equation}
\log \mathcal{R}_{\rm e}=a+b\,\log\sigma_{\star}+
c\,\log \left<I \right>_{\rm e}
\label{eqn:fp_re}
\end{equation}
where $\mathcal{R}_{\rm e}=r_{\rm e}\left(q_{\rm bulge}\right)^{1/2}$
is the circularised effective radius in kpc, $\sigma_{\star}$ is the
central stellar velocity dispersion in km\,s$^{-1}$
(Sect.\ \ref{sec:sigma_star}), and $\left<I \right>_{\rm e}=I_{\rm
  e}\,{\rm exp}(b_n)\,n\,\Gamma(2n)\,b_n^{-2n}$ is the average surface
brightness within the effective radius, in $L_\odot$ pc$^{-2}$.  The
conversion to $L_\odot$ pc$^{-2}$ is obtained from
$I=10^{-0.4(\mu-\mu_\odot)}$, where $\mu_\odot=26.222$ mag
arcsec$^{-2}$ is a constant depending on the absolute magnitude of the
Sun in the observed passband. Each magnitude and surface brightness is
corrected for Galactic extinction following \citet{sch98}, adopting an
absorption $A=0.073$ mag for the coordinates of Abell~1689 in the
SDSS-$r$ band. In each case, the surface brightness is also corrected
for cosmological $(1+z)^4$ dimming \citep{tol30}.
  
With the aim of comparing the FP coefficients derived by using
different fits for the surface-brightness distributions of the galaxies,
we perform the following fits, in which
$\mathcal{R}_{\rm e}$ and $\left<I \right>_{\rm e}$ are derived from
\begin{description}
\item 1) a \devauc\ model for all the galaxies;
\item 2) a \sersic\ model for all the galaxies;
\item 3) a S\'ersic model for all galactic bulges, i.e., taken from a
  \sersic\ model for Es, a \sedisc\ model for S0s, and a
  \sedibar\ model for SB0s.
 \end{description}

We present FP coefficients along with intrinsic and observed scatter
for the three fits in Table \ref{tab:fp_results}, and the
corresponding plots are shown in Fig. \ref{fig:fp_re}.  We note that
the FP coefficients for different photometric models are not
consistent.

We specify that we derived the FP corresponding to a \devauc\ model
because it is usually done in literature, but with the warning that
the \devauc\ model is not a good representation of the ETGs and does
not provide very reliable values of $\mathcal{R}_{\rm e}$ and $\left<I
\right>_{\rm e}$. The FP derived by using a S\'ersic model for all
galactic bulges is the tightest, having a smaller intrinsic and
observed scatter than the FP derived by adopting a
\sersic\ model\footnote{We exclude from this comparison the FP derived
  by using a \devauc\ model, for the reasons explained above.}. From
this, we conclude that the FP is defined by the bulges alone, rather
than by the entire galaxies. This conclusion is strengthened by adding
the bulges of the LTG sample; they all lie on the FP, with the
exception of three galaxies (namely S32, S42, and S43) out of
thirteen. These outliers are the galaxies with the lowest value of
$\sigma_{\star} \sim 40$ km s$^{-1}$. According to \citet{kor04}, they
could be pseudo-bulges, which are similar to small discs (and
therefore rotation supported) and made by slow evolution internal to
galaxy discs. Indeed, the FP relation for elliptical and classical
bulges holds till very low values of velocity dispersion \citep{cos17}
and refers to pressure supported systems.

\subsubsection{$\log \,\sigma_{\star}$ as the dependent variable}
\label{sec:fp_sigma}

\begin{figure}
\includegraphics[scale=.6,angle=0.]{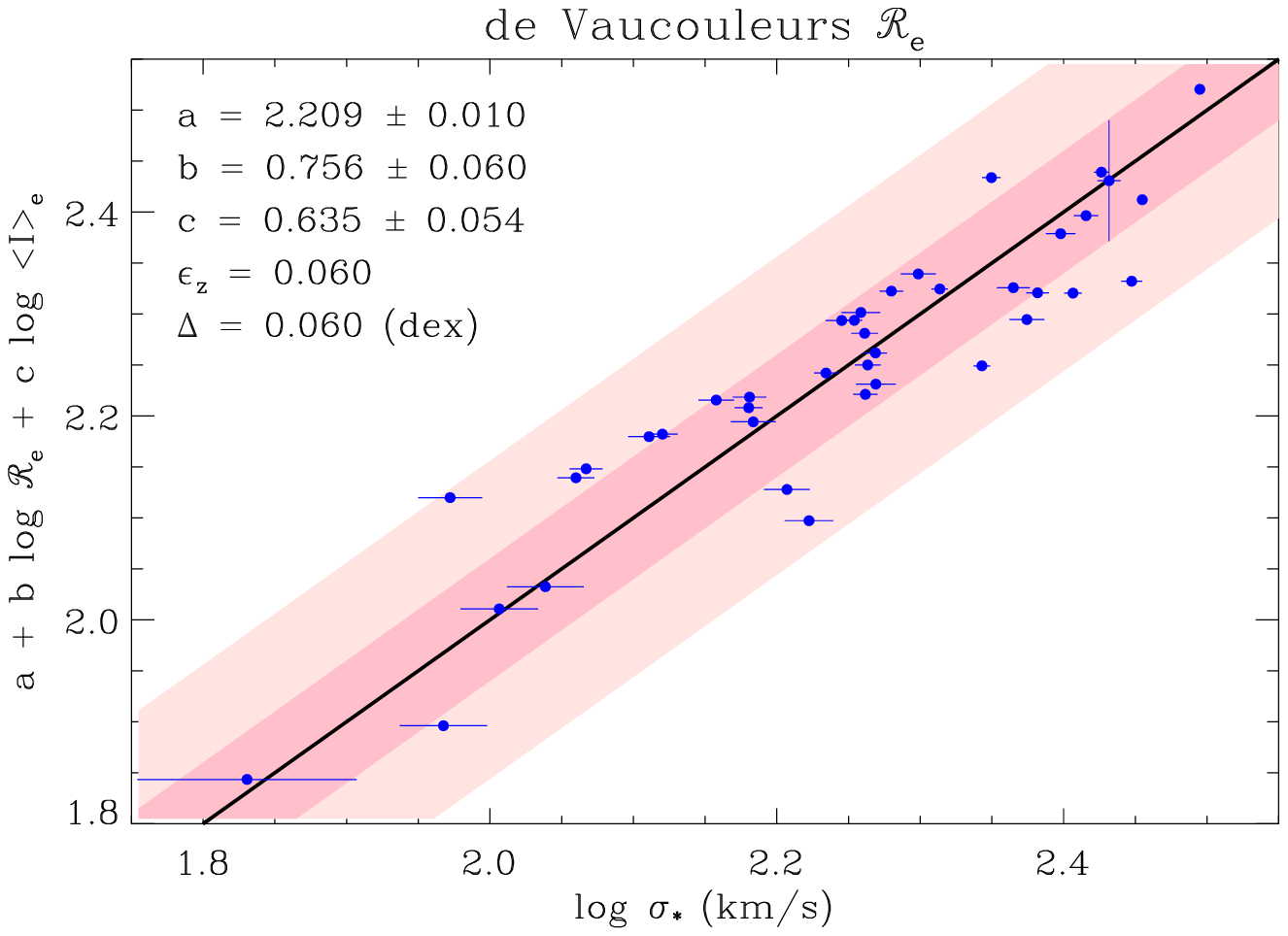}
\includegraphics[scale=.6,angle=0.]{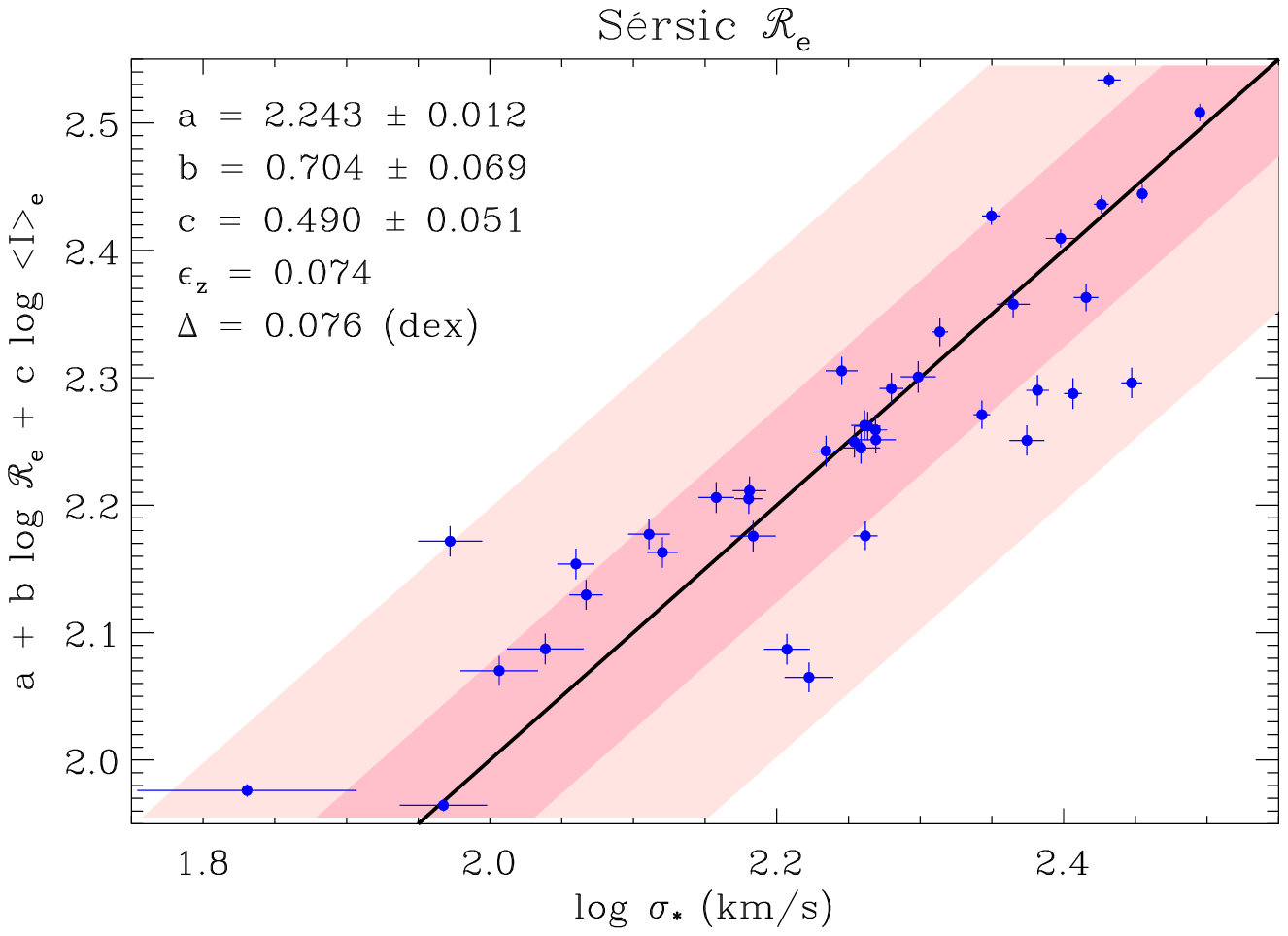}
\includegraphics[scale=.6,angle=0.]{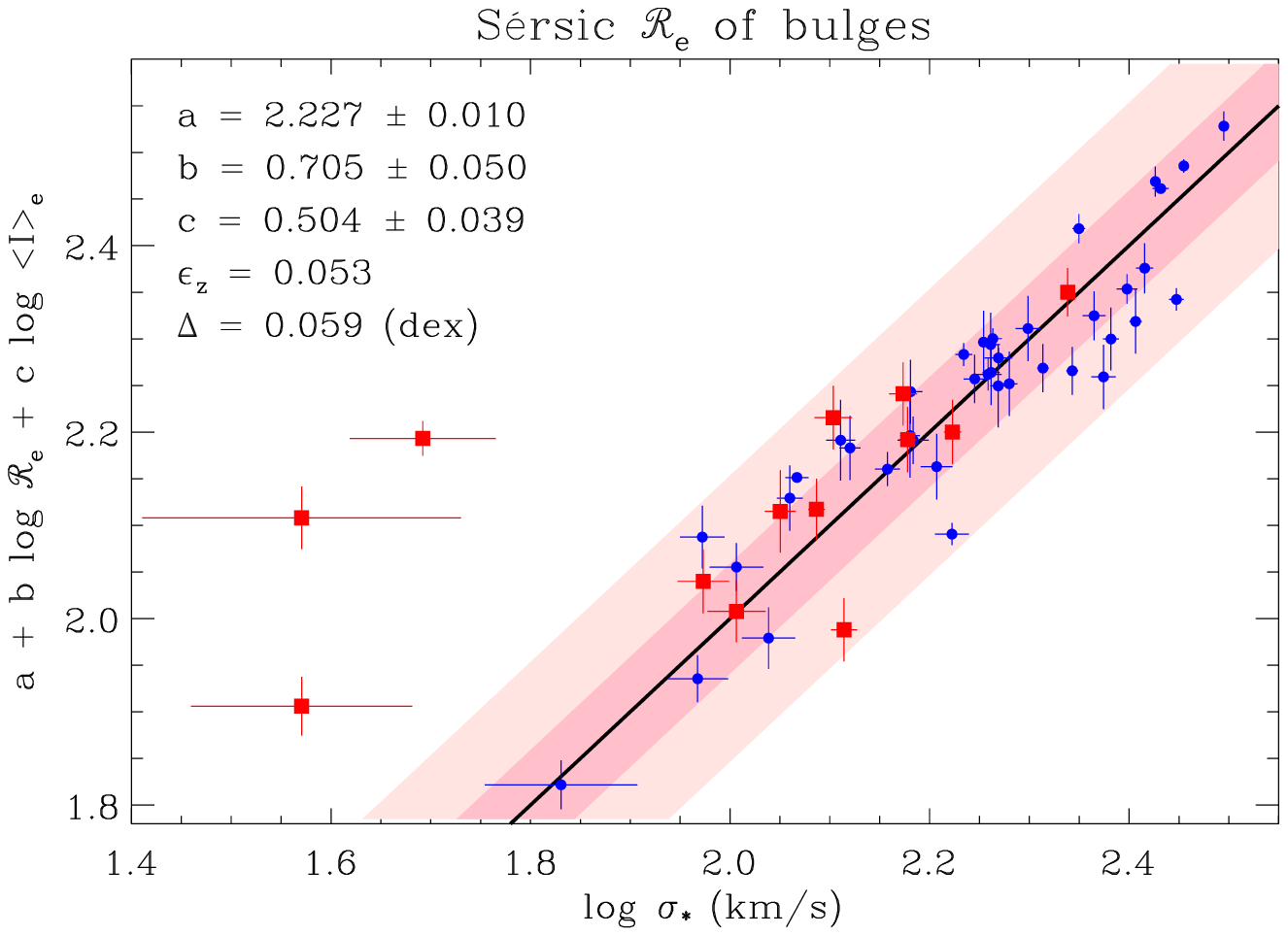}
\caption{Edge-on view of the FP with log $\sigma$ as dependent
  variable and using $\mathcal{R}_{\rm e}$ and $\left<I \right>_e$
  from different photometric models. For a description of panels and
  symbols see Fig.\ref{fig:fp_re}.}
\label{fig:fp_sigma}
\end{figure}

In the FP fits to the three models described above, only
$\sigma_{\star}$ is a fixed parameter common to all three.  We
therefore repeat the fits using $\log\,\sigma_{\star}$ as the
dependent variable, to see whether the minimization process leads to
consistent best-fit planes.  We present the results in Table
\ref{tab:fp_results} and show the results in Fig. \ref{fig:fp_sigma}.
Only the fits obtained by using \sersic\ photometric models and
S\'ersic models of bulges are consistent. We confirm that, with
$\log\,\sigma_{\star}$ as the dependent variable, the tightest FP is
that derived by the S\'ersic bulges. Again, the bulges of LTGs also
lie on the FP, with the exception of the three galaxies with
$\sigma_{\star}< 50$ km s$^{-1}$.

\subsection{Comparison with local FPs}
\label{sec:fp_local}

\subsubsection{Coma cluster}
\label{sec:coma}

We first compare the FP we find for Abell 1689 with that derived for
the Coma cluster by \citet[][hereafter JFK96]{jor96}, which is based
on an orthogonal fit. This is a classic comparison generally found in
literature. For the sake of uniformity, we fit the Coma data with {\tt
  LTS\_PLANEFIT} and use $\log \mathcal{R}_{\rm e}$ as the dependent
variable, as in JFK96.  We take $\sigma_{\star}$ from \citet{jor95b},
and photometric parameters in the Gunn-$r$ from \citet{jor95a} that
were derived from fitting a de Vaucouleur's law to the observed growth
curve. Our best-fit FP is
\begin{eqnarray}
\log\,\mathcal{R}_{\rm e} & = &0.432\,(\pm0.012)+1.263\,(\pm0.073)\,\log
\,\sigma_{\star} \nonumber\\
& & -0.810\,(\pm0.037)\,\log\,\left<I\right>_{\rm e}, 
\end{eqnarray}
which has $a$, $b$, and $c$ values consistent with those of JFK96
  to within 1$\sigma$(we note that the zero-point of the FP in JFK96
  corresponds to ($a-b\,{\rm log}\,\sigma_{\star,0}-c\,\rm log\,
  \left<I \right>_{e,0})$.

The ACS/WFC/F625W image of Abell~1689 at $z=0.183$ corresponds
approximately to the rest-frame $V$-band. We compute an average colour
within the effective radius $(V-R)_{\rm Gunn}=1.22$\,mag from a sample
of fourteen Coma cluster galaxies from \citet{jor95a} and use this
value to derive $\left<\mu\right>_e$ in $V$-band. We
  verified that we could use a common colour within the effective
  radius for E and S0 galaxies, deriving $(V-R)_{\rm Gunn}$ for the
  two classes of galaxies \citep[the morphological type was taken
    from][]{dre80}. We found consistent values. As a further test to
  increase the sample, we derived the average colour within the
  effective radius $(B_{\rm Johnson}-R_{\rm Gunn})=1.15$\,mag for
  thirty-one ETGs \citep[from][]{jor95a} and again colours for Es and
  S0s were in agreement. We then fit the Coma data to obtain the FP
in the $V$-band (hereafter FP$_{\rm Coma}$), that is consistent with
the Gunn-$r$ FP. We present all our derived FP$_{\rm Coma}$ values in
Table \ref{tab:fp_results_comp}. We compare FP$_{\rm Coma}$ with our
derived FP for Abell~1689 by adopting a \devauc\ model for
$\mathcal{R}_{\rm e}$ and $\left<I \right>_{\rm e}$ and adopting
$\log\mathcal{R}_{\rm e}$ as the dependent variable, for the sake of
consistency.  We find that from the local Coma cluster to Abell~1689
there is a decrease in the parameter $b$, from $1.279\pm0.012$ for
Coma to $1.092\pm0.084$ for Abell~1689.  The parameter $c$ is
consistent for the two clusters.  We show the edge-on view of FP$_{\rm
  Coma}$ together with the Abell~1689 data in
Fig.\ \ref{fig:comp_fp}. We also plot the values of parameters $b$ and
$c$ for the two clusters.

\subsubsection{WINGS survey}
We can also compare our FP with that derived from the WIde-field
Nearby Galaxy-cluster Survey \citep[WINGS,][]{fas06,don08}.  We took
spectroscopic and $V$-band photometric data of the ``WINGS/W+S''
sample of 282 galaxies \citep[][private communication\footnote{We note
    that the values of $\sigma_{\star}$ are corrected to the uniform
    aperture $\mathcal{R}_{\rm e}/8$.}]{don08}, which are ETGs
belonging to thirteen nearby clusters in the redshift range
$0.04<z<0.07$. We obtain values for $\mathcal{R}_{\rm e}$ and $\left<I
\right>_e$ by fitting a S\'ersic law to a growth curve.

\begin{figure*}
\includegraphics[scale=.6,angle=0.]{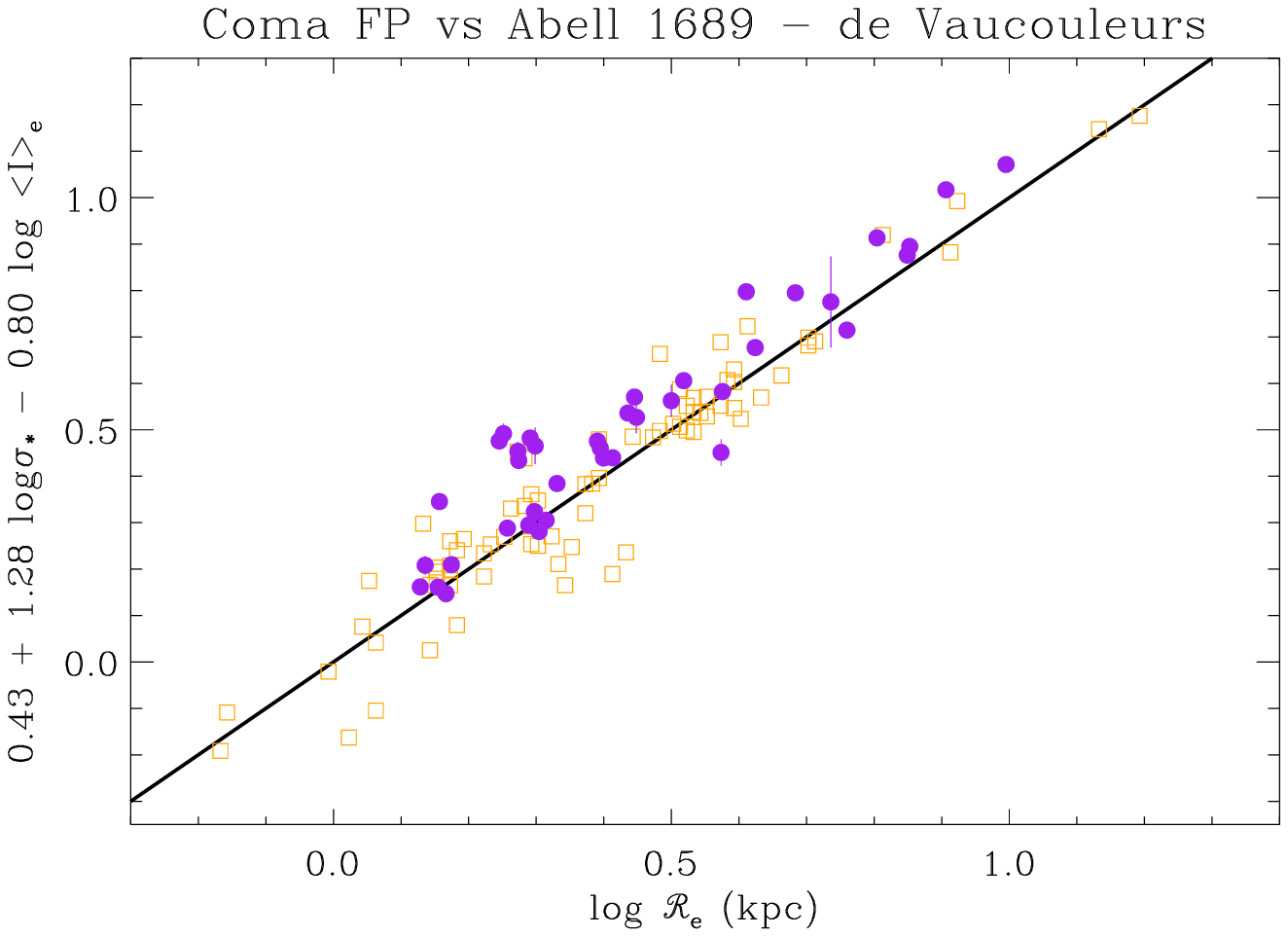}
\includegraphics[scale=.6,angle=0.]{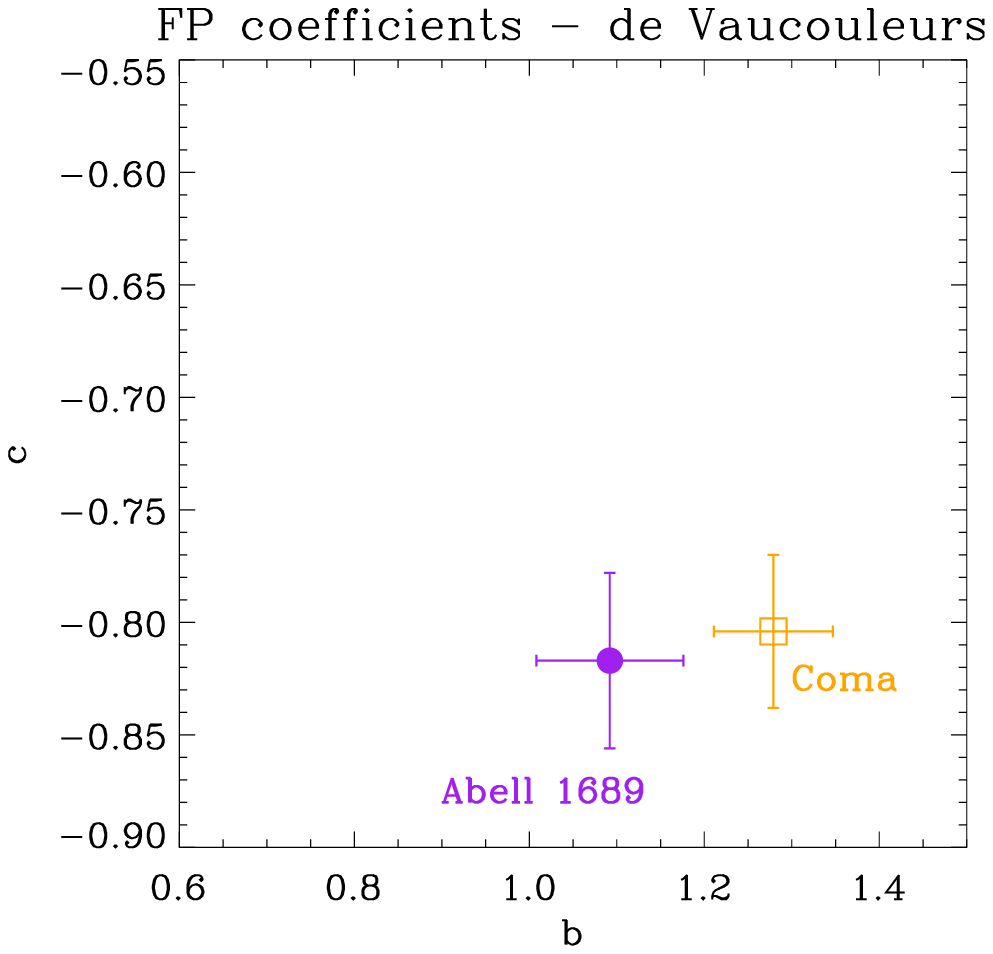}
\includegraphics[scale=.6,angle=0.]{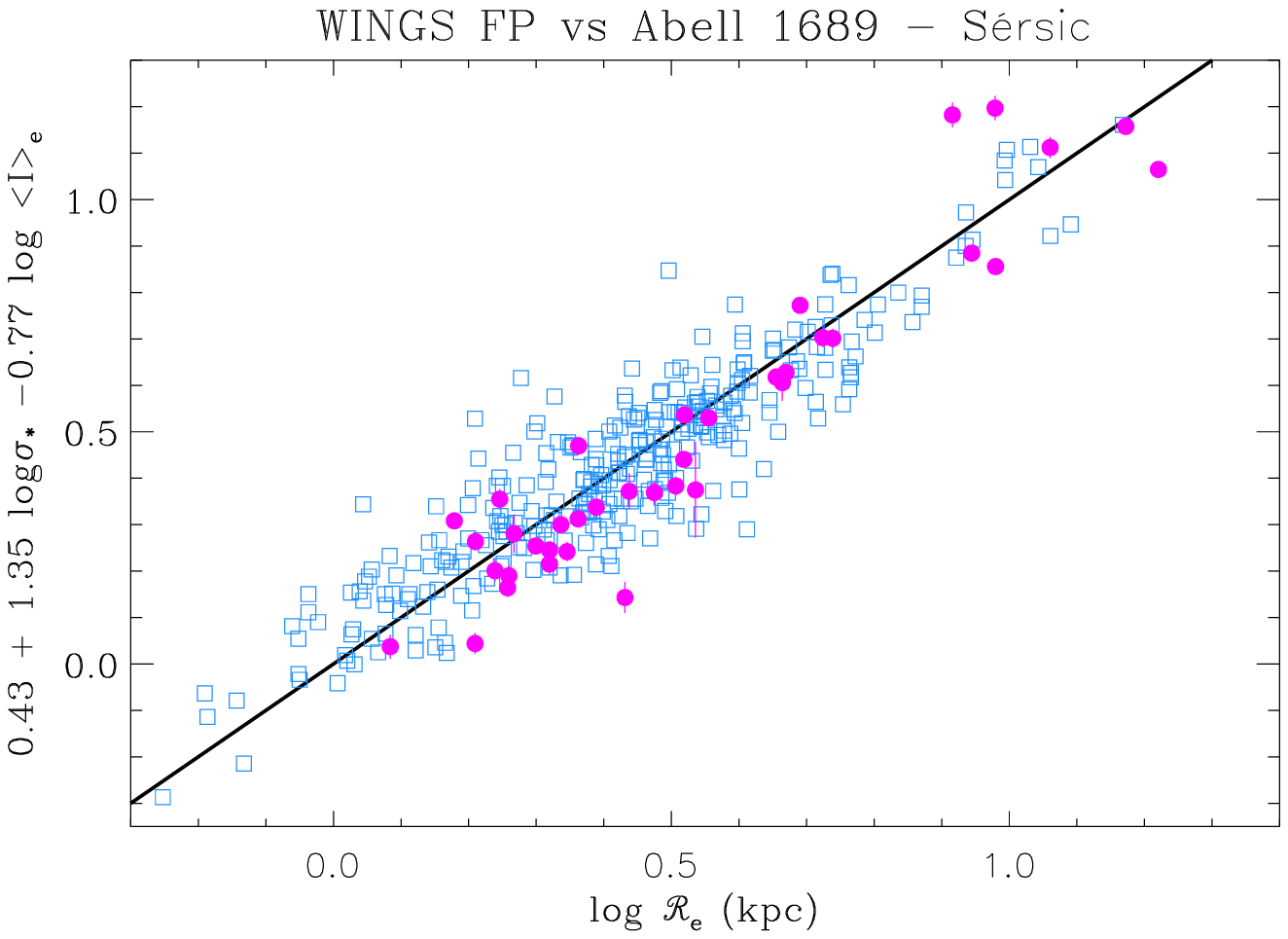}
\includegraphics[scale=.6,angle=0.]{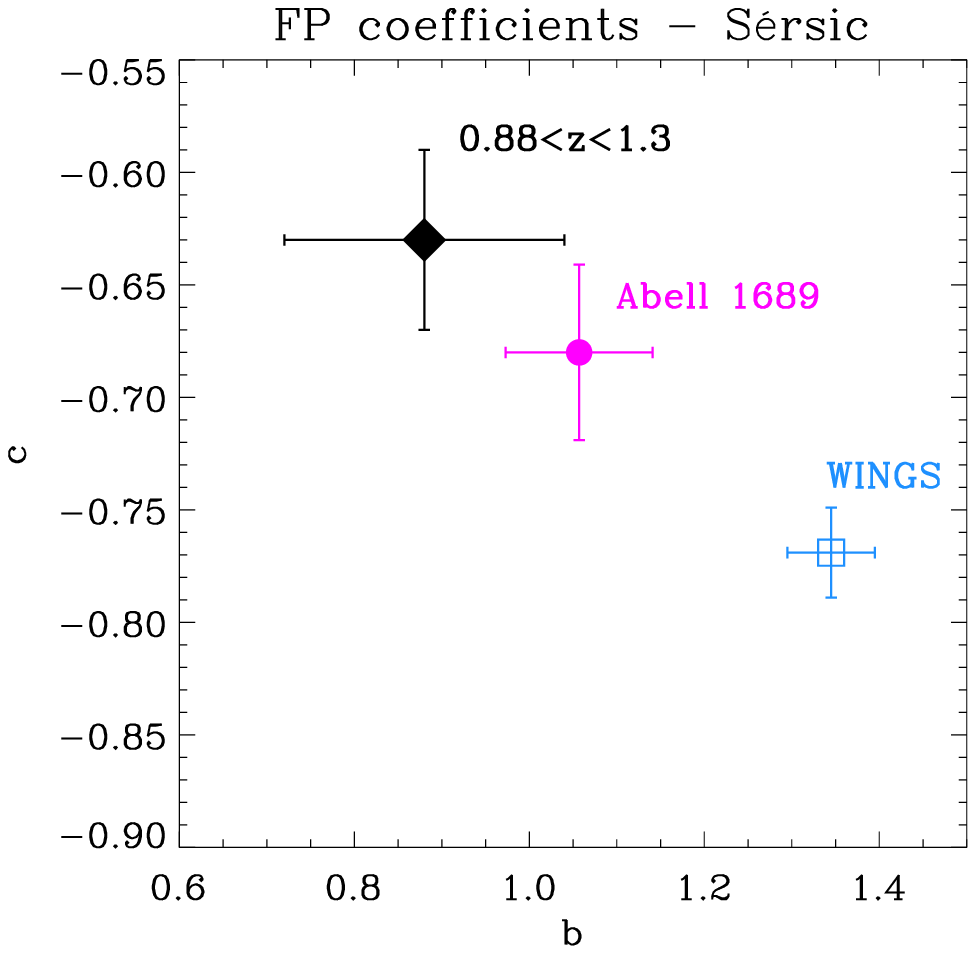}
\caption{Top panels. Left: edge on view of FP$_{\rm Coma}$ (black
  line) obtained for the Coma galaxies (orange squares); the
  Abell~1689 ETGs sample, whose photometric parameters are obtained
  with a \devauc\ model, is shown (purple circles). Right: $b$ and $c$
  FP parameters obtained for the Coma (orange square) and Abell~1689
  ETGs (purple circle) samples, both shown on the left panel.  Bottom
  panels. Left: edge on view of FP$_{\rm WINGS}$ (black line) obtained
  for the WINGS galaxies (light-blue squares); the Abell~1689 ETGs
  sample, whose photometric parameters are obtained with a
  \sersic\ model, is shown (magenta circles). Right: $b$ and $c$ FP
  parameters obtained for the WINGS (light-blue square) and Abell~1689
  ETGs (magenta circle) samples, both shown on the left panel; the FP
  parameters are plotted also for the sample of \citet{dis05} (black
  diamond).}
\label{fig:comp_fp}
\end{figure*}

For a more appropriate comparison, we fit the WINGS data with {\tt
  LTS\_PLANEFIT}, and use log$\mathcal{R}_{\rm e}$ as the dependent
variable, as did \citet{don08}. The derived FP coefficients
(hereafter, FP$_{\rm WINGS}$) are presented in Table
\ref{tab:fp_results}. They are in agreement with those of
\cite{don08}, which are based on an orthogonal fit. We compare the
FP$_{\rm WINGS}$ with that derived for Abell~1689 with
$\mathcal{R}_{\rm e}$ as the dependent variable and use a
\sersic\ photometric model. The edge-on view of the WINGS FP and
Abell~1689 data is shown in Fig. \ref{fig:comp_fp}. We see a decrease
in the value of the parameter $b$ and an increase in the parameter $c$
from the local WINGS FP to the that of Abell~1689\footnote{As a second
  caveat, the WINGS sample has values of $\sigma_{\star}< 95$ km
  s$^{-1}$. In our Abell 1689 analysis, only three galaxies do not
  strictly obey this selection criterion, since S07, S35, and S45 have
  $\sigma_{\star}=67.7$, 92.8, and 93.8 km s$^{-1}$,
  respectively. However, if we derive the Abell 1689 FP excluding
  these galaxies, our conclusions do not change.}.  The parameters $b$
and $c$ for the two samples are plotted in Fig.\ \ref{fig:comp_fp}.

\section{Discussion and conclusions}
\label{sec:end}

We perform a careful photometric analysis of 65 galaxies, specifically
50 ETGs and 15 LTGs, in the cluster Abell~1689 using rest-frame
$V$-band ACS images.  A two-dimensional multi-structure photometric
decomposition of each galaxy provides a complete morphological
classification. For our sample, a \sersic\ model of Es provides a
better fit than a \devauc\ model, as on average, S\'ersic indices
$n\neq4$. This is true also for the bulges of S0s, which are also well
fit by a S\'ersic profile, and S0s are well represented by a
\sedisc\ model. For Ss, we present \sedisc\ models, and for SBs and
SB0s we provide \sedibar\ models.

We use a sample of 40 ETGs to derive the FP by adopting
$\mathcal{R}_{\rm e}$ and $\left<I \right>_{\rm e}$ from different
photometric models, i.e., a \devauc\ model, a \sersic\ model, and a
S\'ersic model for galaxy bulges. We find that the corresponding FP
coefficients are not consistent within $1\sigma$ if we choose log
$\mathcal{R}_{\rm e}$ as the dependent variable. This is partially
confirmed if we choose $\log\sigma_{\star}$ as the dependent variable,
in which case only FPs derived from \sersic\ models and S\'ersic
models of bulges are in agreement. In both cases, the bulges of LTGs
follow the FP, with the exception of three galaxies, out of thirteen,
all with $\sigma_{\star}<50$ km s$^{-1}$. The tightest FP is the one
derived by using a S\'ersic model of the galactic bulges, thus the FP
is better defined by the bulges alone rather than the entire galaxies.

Similar studies have already been published, e.g., \citet{kel00a}
compare the photometric parameters derived by fitting their sample, at
$z=0.33$, with a pure de Vaucouleurs law, a S\'ersic law, and a
combination of a de Vaucouleurs bulge plus exponential disc; while
they find large uncertainties on $\mathcal{R}_{\rm e}$, they conclude
that this does not affect the FP analysis \citep[in][]{kel00b},
because the product $\mathcal{R}_{\rm e}\left<I \right>_{\rm e}^{-c}$
which enters the FP, remains stable. This result was confirmed by
\citet{fer11}, who analysed ETGs in the redshift range
$0.2<z<1.2$. Our investigation differs in that we perform a
  S\'ersic bulge plus exponential disc (plus a Ferrers ellipsoid, in
  case a bar is present) decomposition, and discriminate between Es
  and S0s (Sect. \ref{sec:fp_re}).

We compare the FP for Abell~1689 with the FP derived for local
samples.  We first perform the classic comparison with FP$_{\rm
  Coma}$, where $\mathcal{R}_{\rm e}$ and $\left<I \right>_{\rm e}$
are based on a de Vaucouleurs law fitting procedure. We find a hint of
evolution in the $b$ parameter, in the sense of decreasing with
redshift. The evolution is more evident if we make the comparison with
FP$_{\rm WINGS}$, where the photometric parameters were derived with a
S\'ersic model. The FP of Abell~1689 shows both an offset and
rotation, given that $b$ decreases and $c$ increases with
redshift. Interestingly enough, this trend is in agreement with
\citet{dis05}, who studied a sample of galaxies in the range
$0.88<z<1.3$, in the rest-frame $B$-band, and adopting a
two-dimensional S\'ersic model for the surface brightness
distribution. This study is based on field galaxies, but
  \citet{dis06a,dis06b} show that ETGs are the same in the field
  \citep[using the sample of][]{dis05} and in the clusters
  \citep[using two clusters at z=0.8-0.9 from][]{jor06,jor07}.  We use
  the comparison with \citet{dis05} for consistency in adopting a
  S\'ersic model to derive the photometric parameters which enter the
  FP.  We show their result in Fig. \ref{fig:comp_fp} (bottom-right
panel).  In our two comparisons, two things diverge: (i) the
photometric model, and (ii) the local sample.  As for (i), we find in
our analysis that a \devauc\ model is poorer than a \sersic\ model in
reproducing the surface brightness distribution of ETGs; as for (ii)
we think that the WINGS survey, including data for thirteen clusters,
is more representative of the global behaviour of local galaxies than
the Coma cluster alone. For these reasons we conclude that the FP of
Abell~1689 shows an evolution in both the $b$ and $c$ coefficients, in
the sense described above. A comparison with a local sample in which
Es, S0s, and SB0s are fitted with multiple component surface
brightness distributions will be required to confirm this.

For twenty-nine galaxies in our sample, we measure spatially resolved
kinematics from FLAMES data (Paper II).  In a future paper (Paper IV,
in preparation) we will use the two-dimensional kinematic maps,
alongside ACS photometry to fit dynamical models and measure accurate
dynamical masses \citep{cap07}. We will therefore investigate the
systematic variation of the stellar and dynamical mass-to-light
ratios, and compare these measurements to the prediction of the FP.

\begin{table*}
\begin{minipage}{126mm}
\caption\small{{FP coefficients of Abell~1689 for different dependent
    variables and photometric models.}}
\label{tab:fp_results}
\begin{scriptsize}
\begin{center}
\begin{tabular}{ccccccc}
\hline
\noalign{\vskip 0.1cm}
\multicolumn{7}{c}{$z=a+b\,(x-x_0)+c\,(y-y_0)$} \\
\noalign{\vskip 0.1cm}
\hline
\multicolumn{1}{c}{a} &
\multicolumn{1}{c}{b} &
\multicolumn{1}{c}{c} &
\multicolumn{1}{c}{$\epsilon_z$} &
\multicolumn{1}{c}{$\Delta$} &
\multicolumn{1}{c}{$x_0$} &
\multicolumn{1}{c}{$y_0$} \\
\multicolumn{1}{c}{(1)} &
\multicolumn{1}{c}{(2)} &
\multicolumn{1}{c}{(3)} &
\multicolumn{1}{c}{(4)} &
\multicolumn{1}{c}{(5)} &
\multicolumn{1}{c}{(6)} &
\multicolumn{1}{c}{(7)} \\
\hline
\noalign{\vskip 0.1cm}
\multicolumn{5}{c}{$\log\mathcal{R}_{\rm e}=a+b\,({\rm log}\,\sigma_{\star}-{\rm log}\,\sigma_{\star,0})+c\,({\rm log}\,\left<I \right>_e-{\rm log}\,\left<I \right>_{e,0})$} & \multicolumn{1}{c}{log $\sigma_{\star,0}$} & \multicolumn{1}{c}{log $\left<I \right>_{e,0}$}\\
\noalign{\vskip 0.1cm}
\multicolumn{7}{c}{\devauc\ $\mathcal{R}_{\rm e}$}\\
\noalign{\vskip 0.1cm}
$0.468\pm0.012$ & $1.092\pm0.084$ & $-0.817\pm0.039$ & 0.072 & 0.073 & 2.262 & 2.576 \\
\noalign{\vskip 0.1cm}
\multicolumn{7}{c}{\sersic\ $\mathcal{R}_{\rm e}$}\\
\noalign{\vskip 0.1cm}
$0.532\pm0.015$ & $1.057\pm0.104$ & $-0.680\pm0.033$ & 0.091 & 0.091 & 2.262 & 2.475 \\
\noalign{\vskip 0.1cm}
\multicolumn{7}{c}{\sersic\ $\mathcal{R}_{\rm e}$ of bulges}\\
\noalign{\vskip 0.1cm}
$0.065\pm0.013$ & $1.239\pm0.089$ & $-0.714\pm0.024$ & 0.070 & 0.079 & 2.262 & 3.054 \\
\noalign{\vskip 0.1cm}
\hline
\noalign{\vskip 0.1cm}
\multicolumn{5}{c}{${\rm log}\,\sigma_{\star}=a+b\,(\log\mathcal{R}_{\rm e}-\log\mathcal{R}_{\rm e,0})+c\,({\rm log}\,\left<I \right>_e-{\rm log}\,\left<I \right>_{e,0})$} & \multicolumn{1}{c}{log $\mathcal{R}_{\rm e,0}$} & \multicolumn{1}{c}{log $\left<I \right>_{e,0}$} \\
\noalign{\vskip 0.1cm}
\multicolumn{7}{c}{\devauc\ $\mathcal{R}_{\rm e}$}\\
\noalign{\vskip 0.1cm}
$2.209\pm0.010$ & $0.756\pm0.060$ & $0.635\pm0.054$ & 0.060 & 0.060 & 0.3996 & 2.576 \\
\noalign{\vskip 0.1cm}
\multicolumn{7}{c}{\sersic\ $\mathcal{R}_{\rm e}$}\\
\noalign{\vskip 0.1cm}
$2.243\pm0.012$ & $0.704\pm0.069$ & $0.490\pm0.051$ & 0.074 & 0.076 & 0.5064 & 2.475 \\ 
\noalign{\vskip 0.1cm}
\multicolumn{7}{c}{\sersic\ $\mathcal{R}_{\rm e}$ of bulges}\\
\noalign{\vskip 0.1cm}
$2.227\pm0.010$ & $0.705\pm0.050$ & $0.504\pm0.039$ & 0.053 & 0.059 & 0.01504 & 3.054 \\
\noalign{\vskip 0.1cm}
\hline
\end{tabular}
\end{center}
{\em Note.} Col. (1), col. (2), and col. (3): FP
coefficients. Col. (4): intrinsic scatter. Col. (5): observed scatter
(dex). Col. (6) and col. (7): median of the fitted $x_i$ and $y_i$
values, respectively. Values of $\mathcal{R}_{\rm e}$ used to fit the FP are
in kpc, $\sigma_{\star,0}$ in km/s, and $\left<I \right>_{e,0}$ in
$L_\odot/pc^2$.
\end{scriptsize}
\end{minipage}
\end{table*}

\begin{table*}
\begin{minipage}{126mm}
\caption\small{{Coma and WINGS FP coefficients.}}
\label{tab:fp_results_comp}
\begin{scriptsize}
\begin{center}
\begin{tabular}{ccccccc}
\hline
\noalign{\vskip 0.1cm}
\multicolumn{7}{c}{$\log\mathcal{R}_{\rm e}=a+b\,({\rm log}\,\sigma_{\star}-{\rm log}\,\sigma_{\star,0})+c\,({\rm log}\,\left<I \right>_e-{\rm log}\,\left<I \right>_{e,0})$} \\
\noalign{\vskip 0.1cm}
\hline
\multicolumn{1}{c}{a} &
\multicolumn{1}{c}{b} &
\multicolumn{1}{c}{c} &
\multicolumn{1}{c}{$\epsilon_z$} &
\multicolumn{1}{c}{$\Delta$} &
\multicolumn{1}{c}{log $\sigma_{\star,0}$} &
\multicolumn{1}{c}{log $\left<I \right>_{e,0}$} \\
\multicolumn{1}{c}{(1)} &
\multicolumn{1}{c}{(2)} &
\multicolumn{1}{c}{(3)} &
\multicolumn{1}{c}{(4)} &
\multicolumn{1}{c}{(5)} &
\multicolumn{1}{c}{(6)} &
\multicolumn{1}{c}{(7)} \\
\hline
\noalign{\vskip 0.1cm}
\multicolumn{7}{c}{Coma - $\mathcal{R}_{\rm e}$ from de Vaucouleurs law}\\
\noalign{\vskip 0.1cm}
$0.432\pm0.012$ & $1.279\pm0.068$ & $-0.804\pm0.034$ & 0 & 0.081 & 2.219 & 2.645 \\
\noalign{\vskip 0.1cm}
\multicolumn{7}{c}{WINGS - $\mathcal{R}_{\rm e}$ from S\'ersic law}\\
\noalign{\vskip 0.1cm}
$0.4262\pm0.0056$ & $1.345\pm0.050$ & $-0.769\pm0.020$ & 0.076 & 0.100 & 2.166 & 2.408 \\
\noalign{\vskip 0.1cm}
\hline
\end{tabular}
\end{center}
{\em Note.} Col. (1), col. (2), and col. (3): FP
coefficients. Col. (4): intrinsic scatter. Col. (5): observed scatter
(dex). Col. (6) and col. (7): median of the fitted $x_i$ and $y_i$
values, respectively. Values of $\mathcal{R}_{\rm e}$ used to fit the FP are
in kpc, $\sigma_{\star,0}$ in km/s, and $\left<I \right>_{e,0}$ in
$L_\odot/pc^2$.
\end{scriptsize}
\end{minipage}
\end{table*}

\section*{Acknowledgments}
EDB was supported by grants 60A02-5857/13, 60A02-5833/14,
60A02-4434/15, and CPDA133894 of Padua University. JMA thanks support
from the MINECO through the grant AYA2013-43188-P.  RCWH was supported
by the Science and Technology Facilities Council [STFC grant numbers
  ST/H002456/1, ST/K00106X/1 \& ST/J002216/1]. EDB acknowledges the
Sub-department of Astrophysics, Department of Physics, University of
Oxford and Christ Church College for their hospitality while this
paper was in progress.We thank M. D'Onofrio, G. Fasano, and the
  WINGS team for providing their data. We thank J. Kormendy for
  supplying the values of disc-to-bulge luminosity ratios to define
  the a, b, and c subclasses of S0s and spirals.  We thank
  M. D'Onofrio, S. di Serego Alighieri, B.~M. Peterson, and
  A. Pizzella for the useful discussions.

All of the data presented in this paper were obtained from the
Mikulski Archive for Space Telescopes (MAST). STScI is operated by the
Association of Universities for Research in Astronomy, Inc., under
NASA contract NAS5-26555.





\appendix

\section{Additional figures and tables}
\label{appendix}

\begin{figure*}
\centering
\includegraphics[width=\textwidth, angle=0]{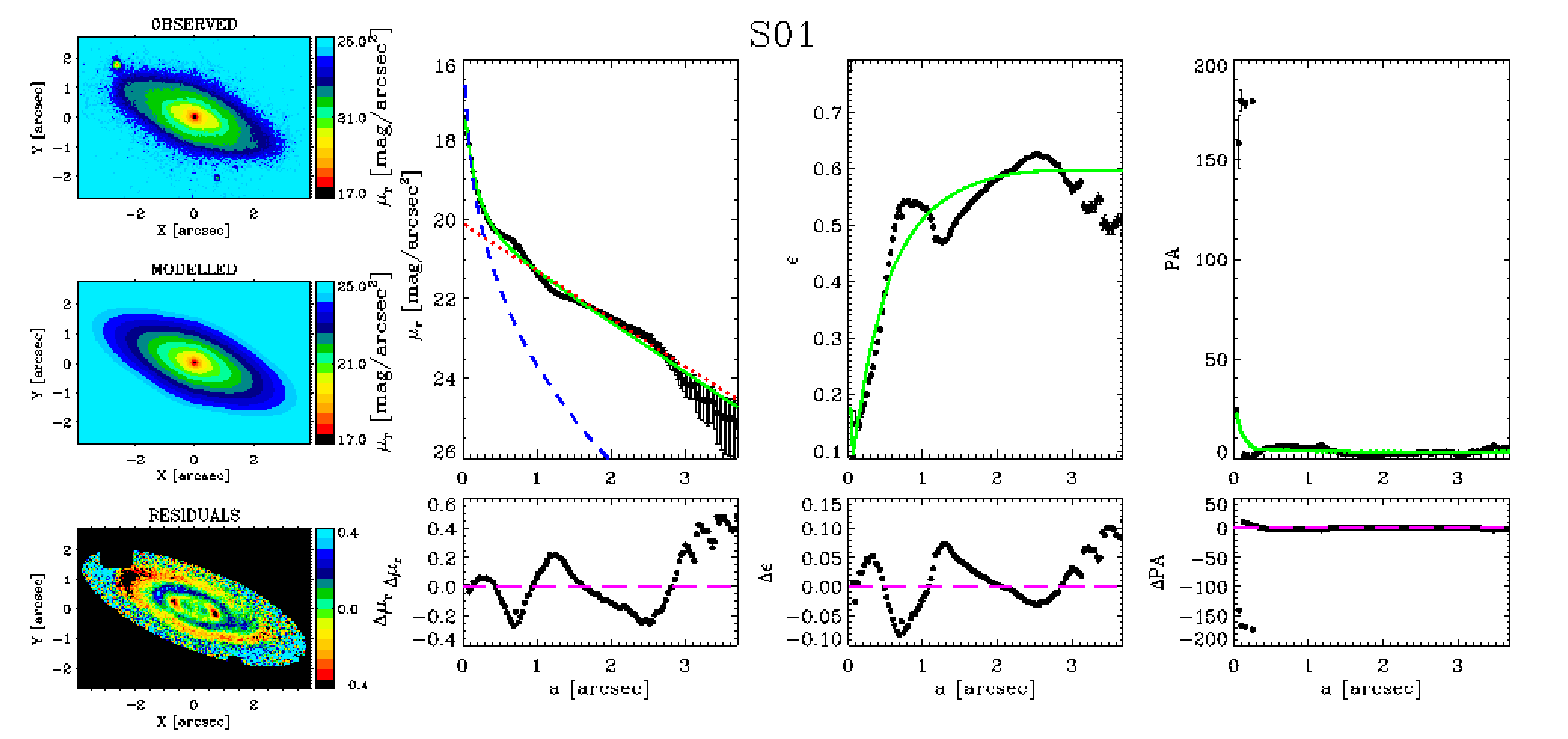}
\caption{Two-dimensional photometric decomposition of the spiral
  galaxy S01 fitted with a \sedisc. Left maps from top to bottom:
  observed, modelled, and residual (observed---modelled)
  surface-brightness distribution of the galaxy. The mask applied to
  the image, containing the pixels rejected in the fit, is highlighted
  in black. Images are oriented as in Fig. \ref{fig:FoV}, i.e., PA of
  Y axis is 115.12$\degr$. Right panels from left to right and top to
  bottom: ellipse-averaged radial profile of surface-brightness,
  ellipticity, and position angle, measured in the observed (black
  dots with error-bars) and modelled image (green solid line). The
  dashed blue and dotted red lines represent the intrinsic
  surface-brightness radial profiles of the bulge and disc,
  respectively, along their semi major axis. The difference between
  the ellipse-averaged radial profiles extracted from the modelled and
  observed images is also shown.}
\label{fig:fit_S01}
\end{figure*}

\begin{figure*}
\centering
\includegraphics[width=\textwidth, angle=0]{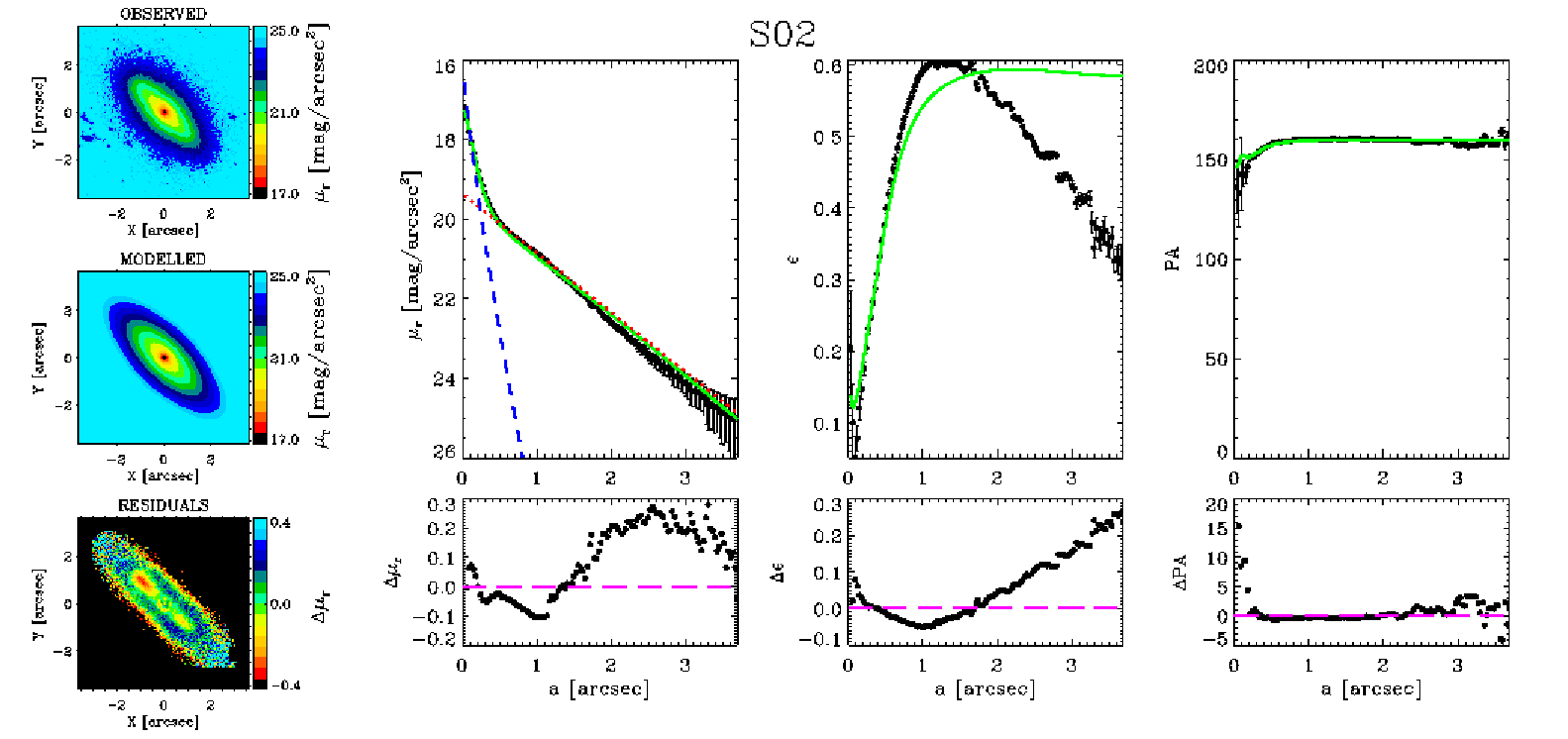}
\caption{As in Fig. \ref{fig:fit_S01} but for galaxy S02 (\sedisc\ model).}
\label{fig:fit_S02}
\end{figure*}

\begin{figure*}
\centering
\includegraphics[width=\textwidth, angle=0]{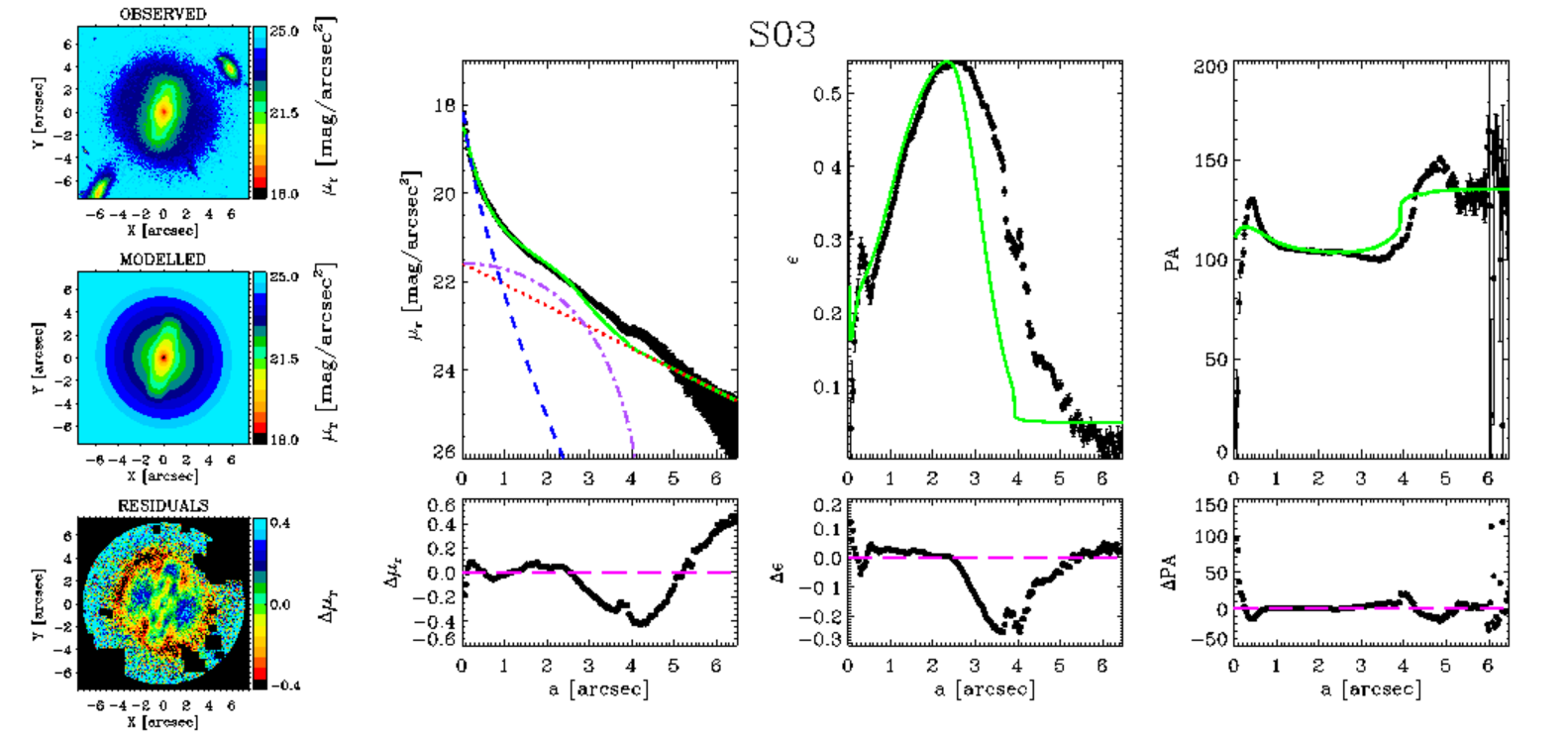}
\caption{As in Fig. \ref{fig:fit_S01} but for galaxy S03 fitted with a
  \sedibar\ model. The dashed-dotted purple line represents the intrinsic
  surface-brightness radial profile of the bar along its semi major
  axis.}
\label{fig:fit_S03}
\end{figure*}

\begin{figure*}
\centering
\includegraphics[width=\textwidth, angle=0]{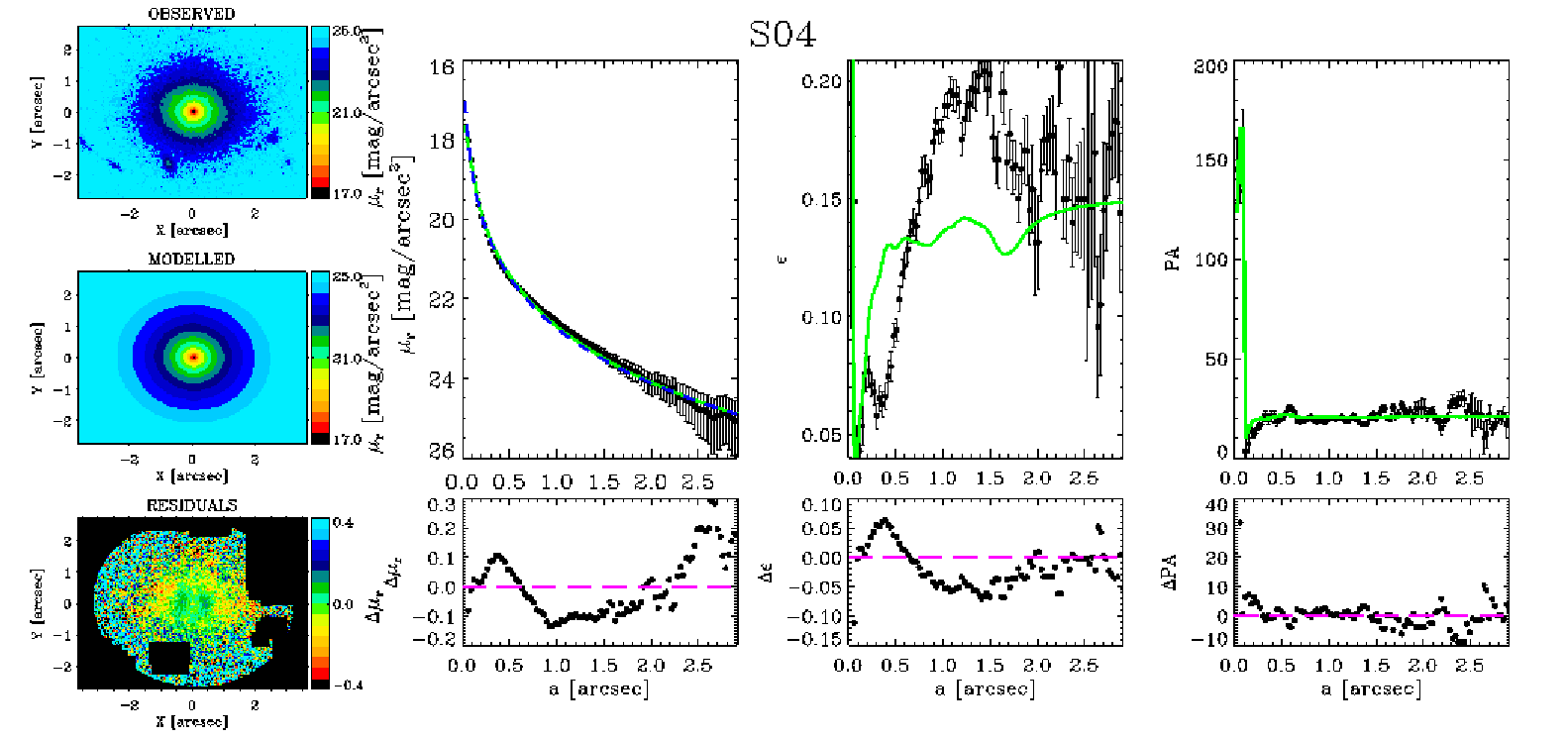}
\caption{As in Fig. \ref{fig:fit_S01} but for galaxy S04 fitted with a
  \sersic\ model.}
\label{fig:fit_S04}
\end{figure*}

\clearpage

\begin{figure*}
\centering
\includegraphics[width=\textwidth, angle=0]{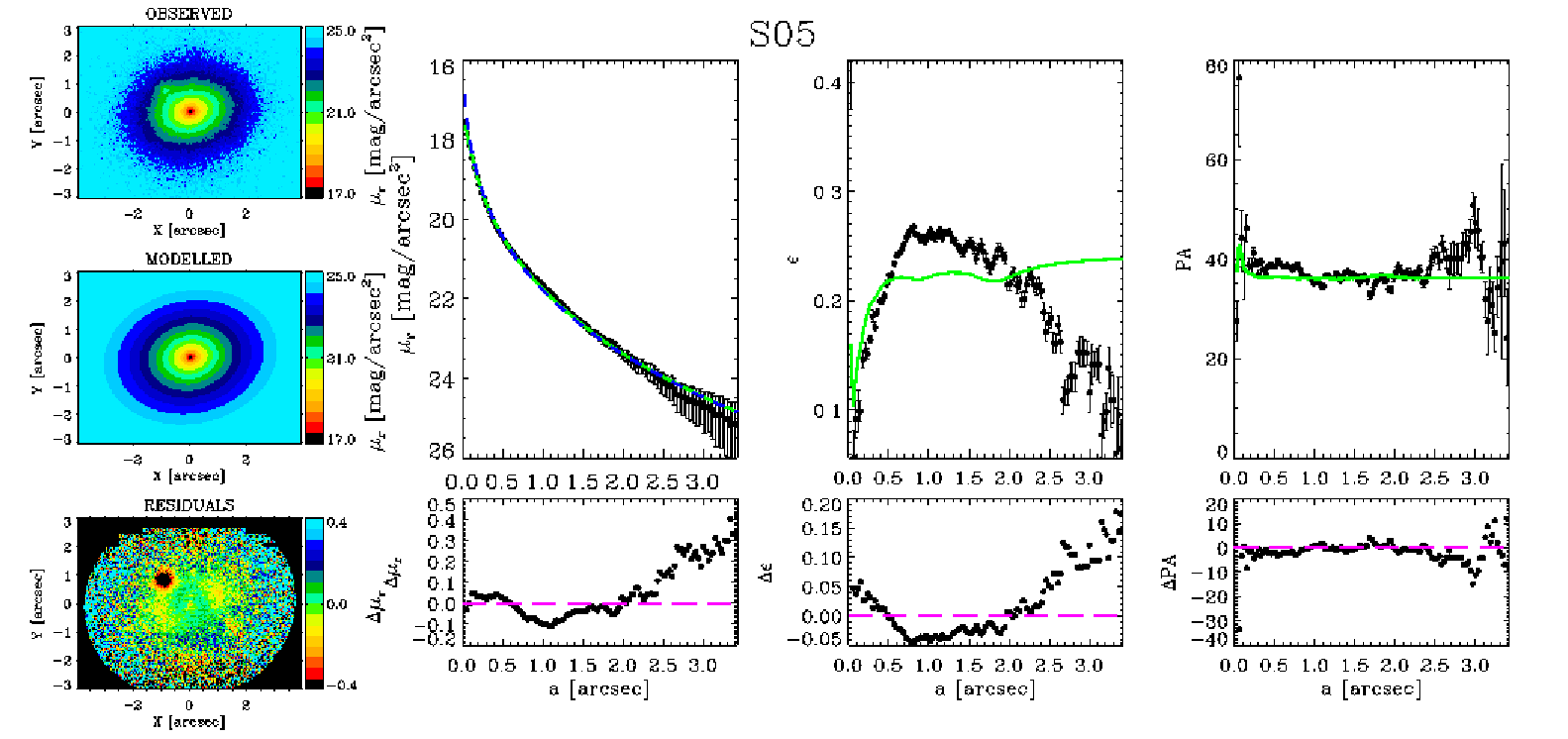}
\caption{As in Fig. \ref{fig:fit_S01} but for galaxy S05 fitted with a
  \sersic\ model.}
\label{fig:fit_S05}
\end{figure*}

\begin{figure*}
\centering
\includegraphics[width=\textwidth, angle=0]{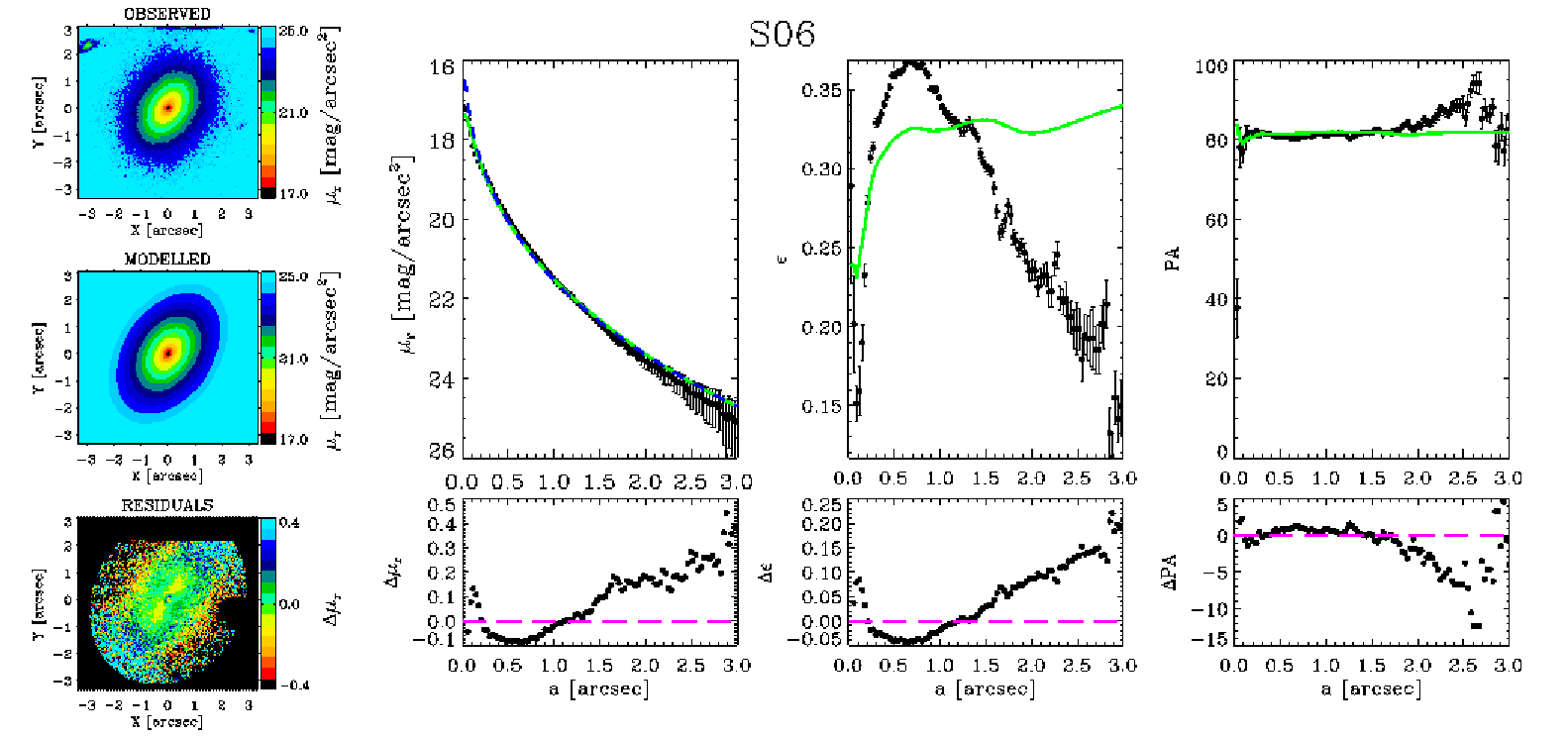}
\caption{As in Fig. \ref{fig:fit_S01} but for galaxy S06 fitted with a
  \sersic\ model.}
\label{fig:fit_S06}
\end{figure*}

\clearpage

\begin{figure*}
\centering
\includegraphics[width=\textwidth, angle=0]{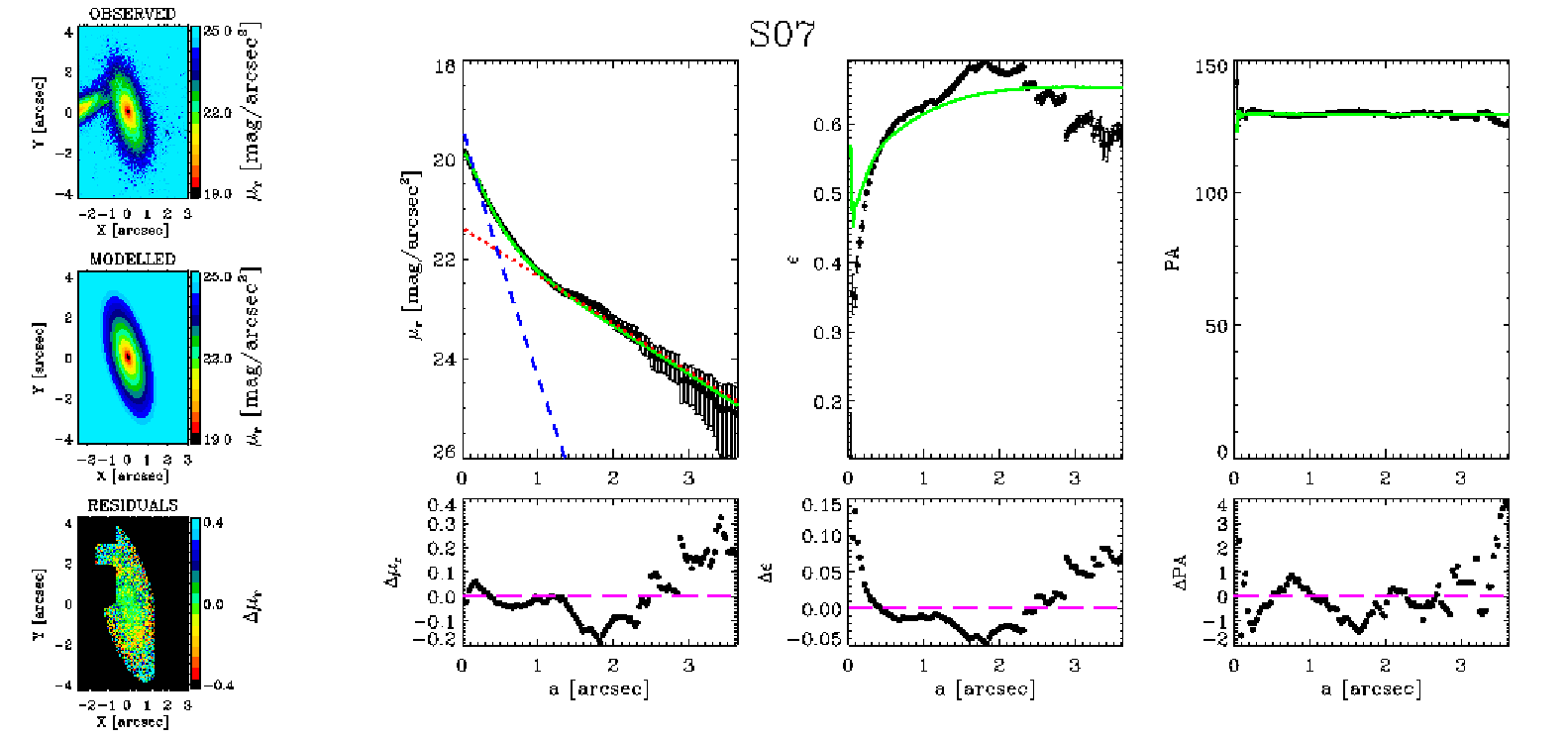}
\caption{As in Fig. \ref{fig:fit_S01} but for galaxy S07 (\sedisc\ model).}
\label{fig:fit_S07}
\end{figure*}

\begin{figure*}
\centering
\includegraphics[width=\textwidth, angle=0]{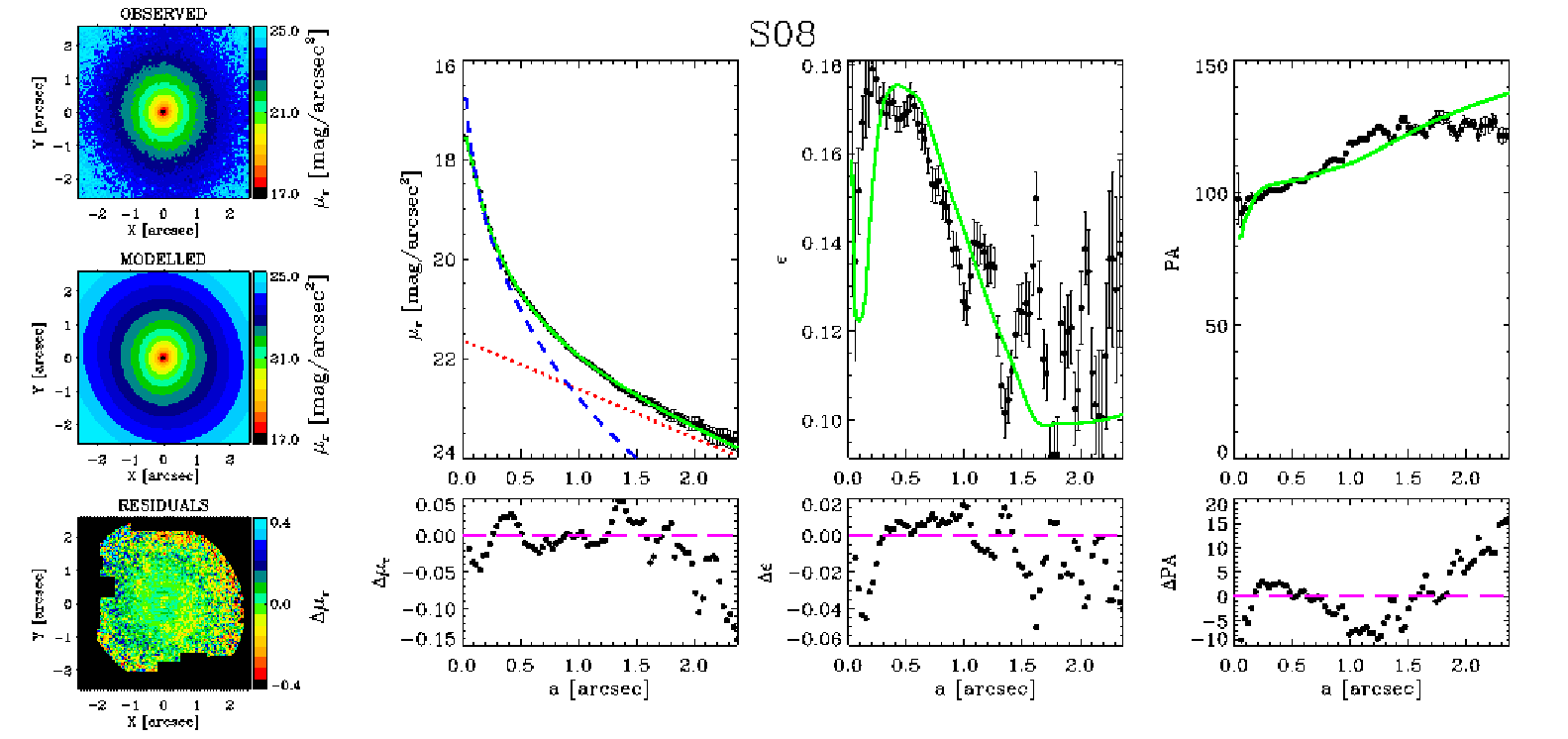}
\caption{As in Fig. \ref{fig:fit_S01} but for galaxy S08 (\sedisc\ model).}
\label{fig:fit_S08}
\end{figure*}

\clearpage

\begin{figure*}
\centering
\includegraphics[width=\textwidth, angle=0]{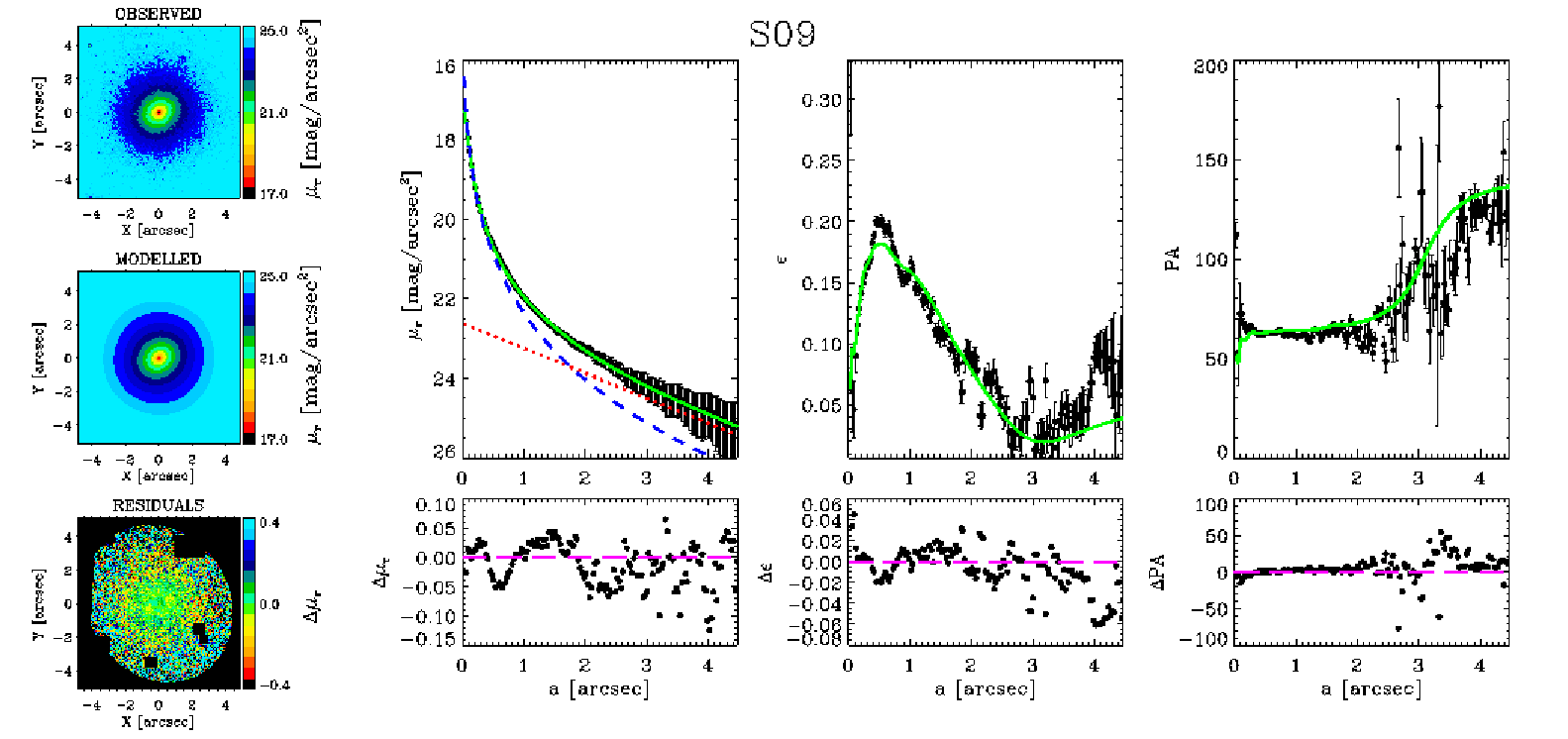}
\caption{As in Fig. \ref{fig:fit_S01} but for galaxy S09 (\sedisc\ model).}
\label{fig:fit_S09}
\end{figure*}

\begin{figure*}
\centering
\includegraphics[width=\textwidth, angle=0]{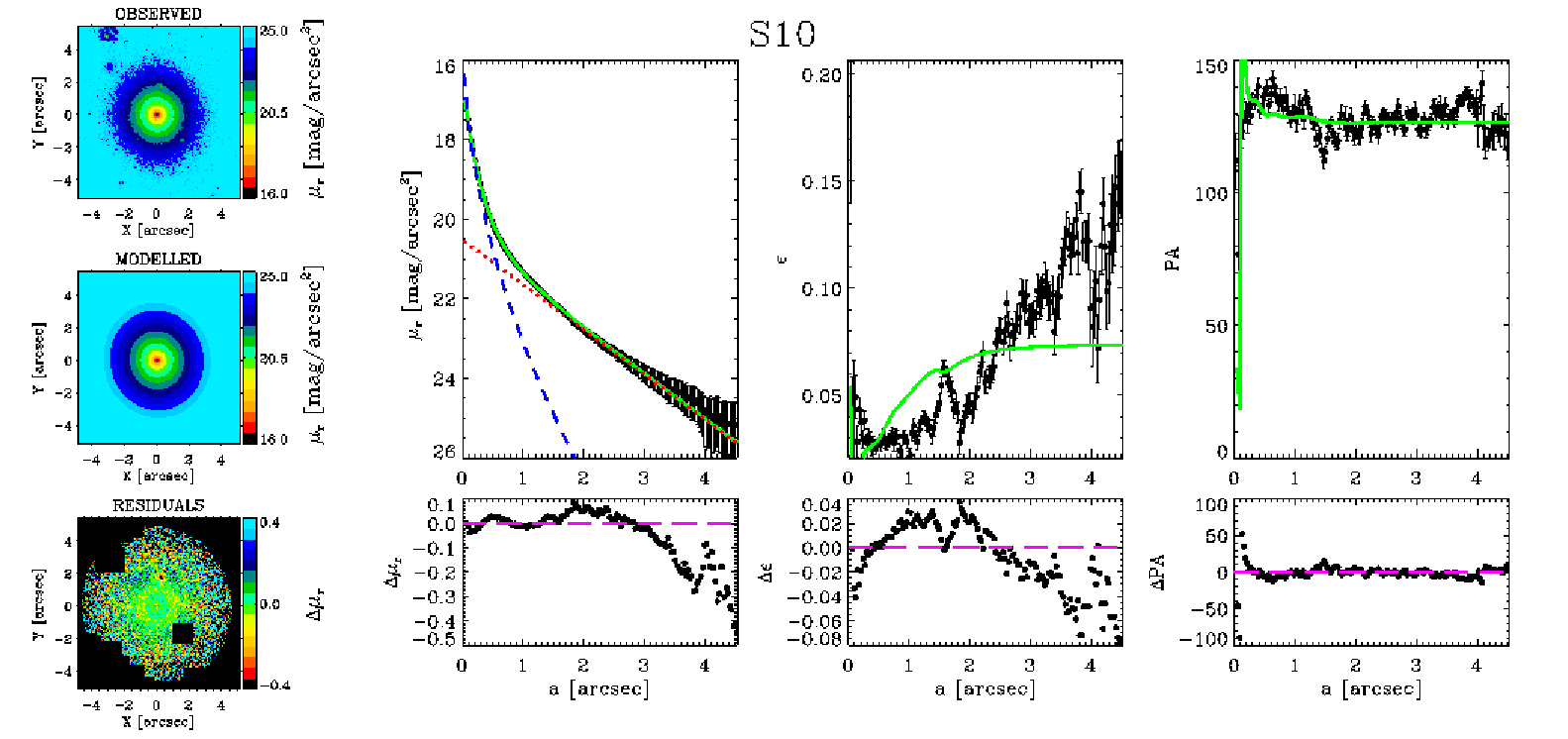}
\caption{As in Fig. \ref{fig:fit_S01} but for galaxy S10 (\sedisc\ model).}
\label{fig:fit_S10}
\end{figure*}

\clearpage

\begin{figure*}
\centering
\includegraphics[width=\textwidth, angle=0]{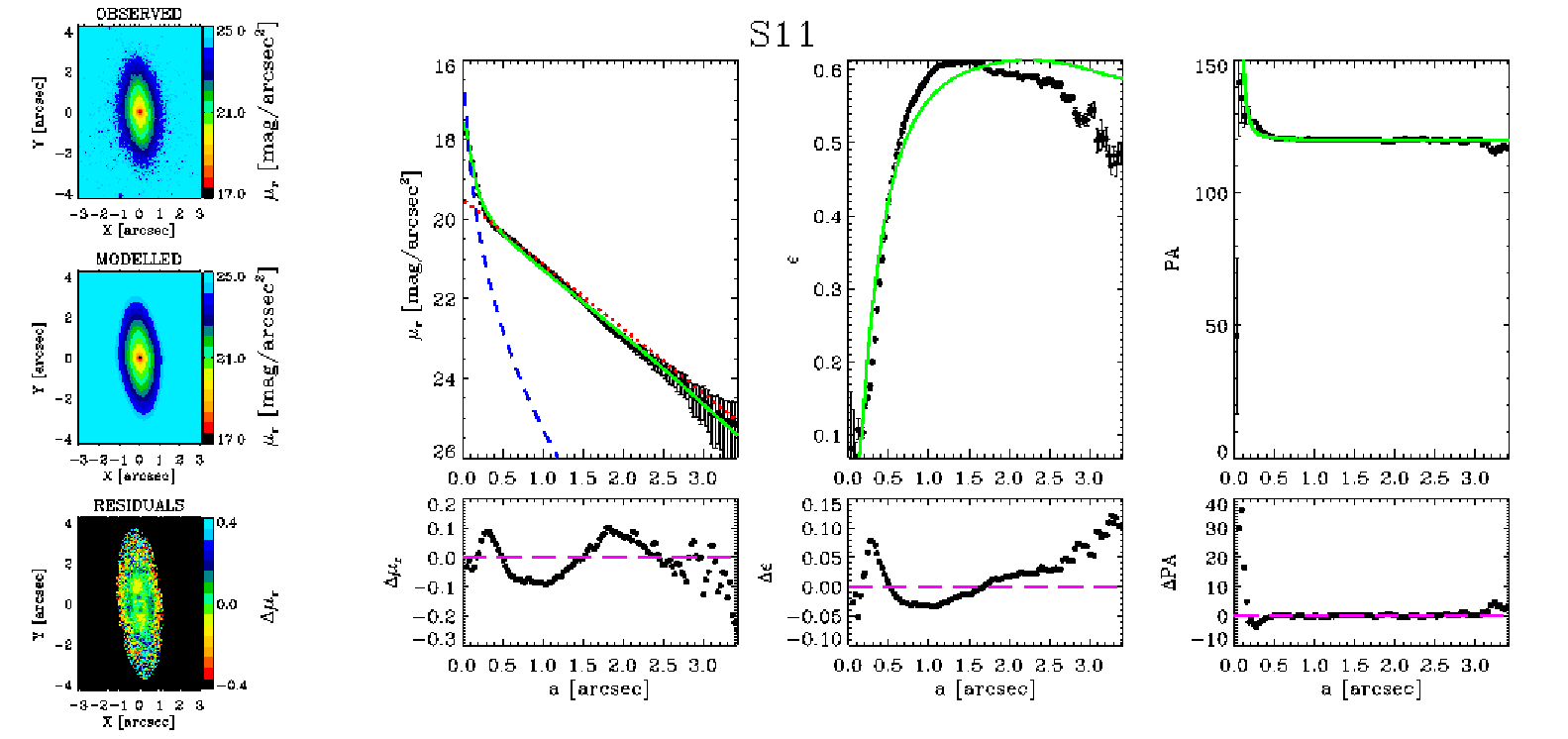}
\caption{As in Fig. \ref{fig:fit_S01} but for galaxy S11 fitted with a
  \sedibar\ model. The dashed-dotted purple line represents the intrinsic
  surface-brightness radial profile of the bar along its semi major
  axis.}
\label{fig:fit_S11}
\end{figure*}

\begin{figure*}
\centering
\includegraphics[width=\textwidth, angle=0]{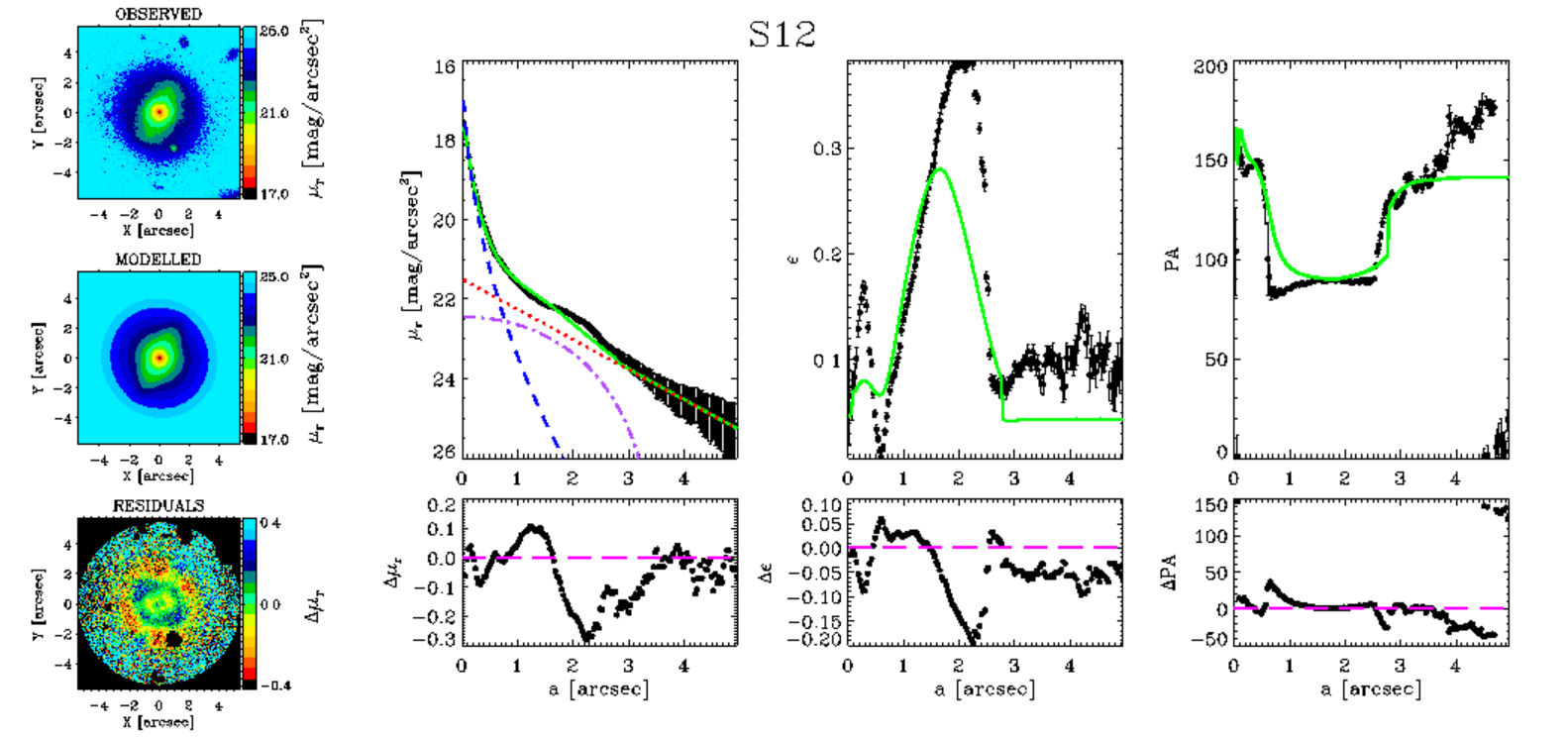}
\caption{As in Fig. \ref{fig:fit_S01} but for galaxy S12 fitted with a
  \sedibar\ model. The dashed-dotted purple line represents the intrinsic
  surface-brightness radial profile of the bar along its semi major
  axis.}
\label{fig:fit_S12}
\end{figure*}

\clearpage

\begin{figure*}
\centering
\includegraphics[width=\textwidth, angle=0]{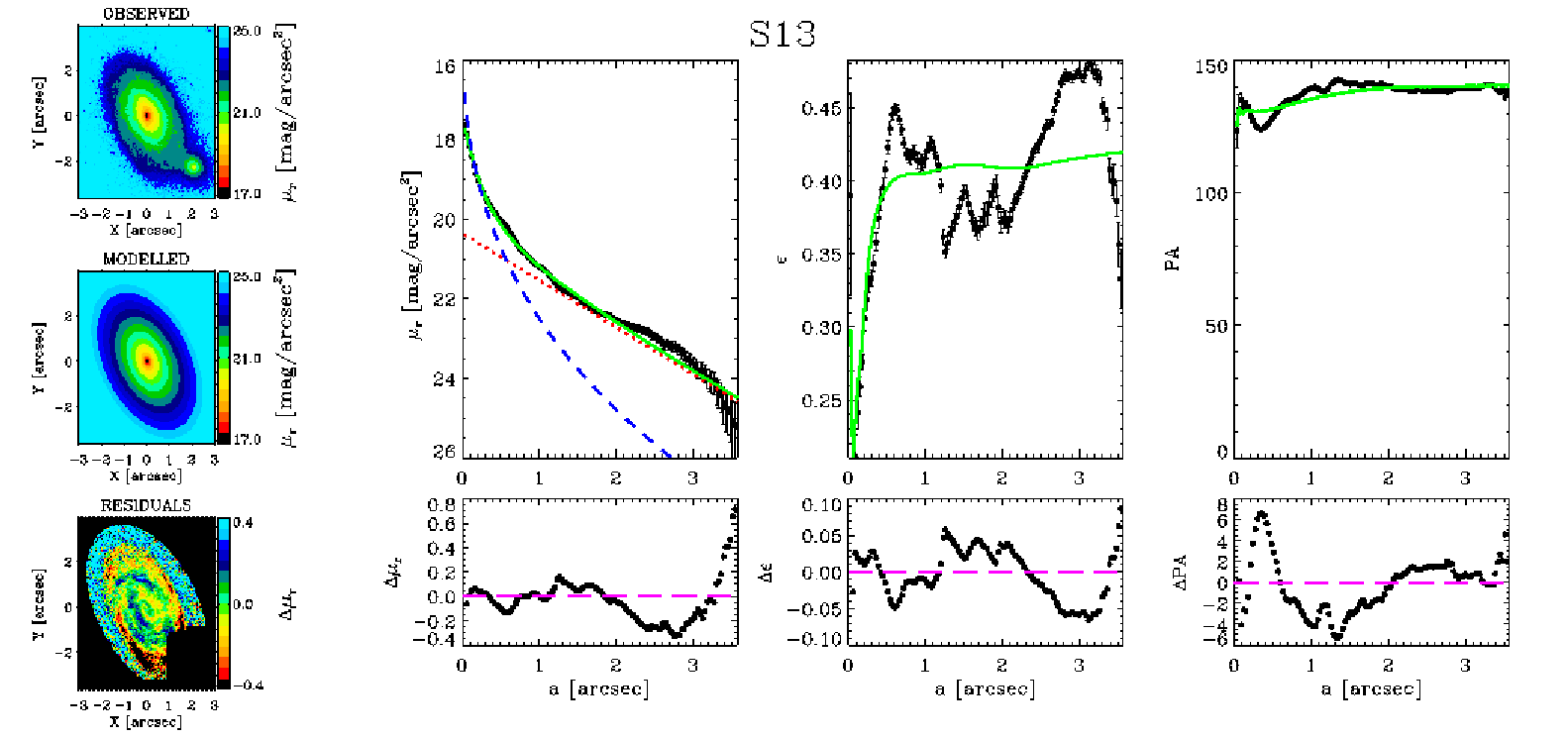}
\caption{As in Fig. \ref{fig:fit_S01} but for galaxy S13 (\sedisc\ model).}
\label{fig:fit_S13}
\end{figure*}

\begin{figure*}
\centering
\includegraphics[width=\textwidth, angle=0]{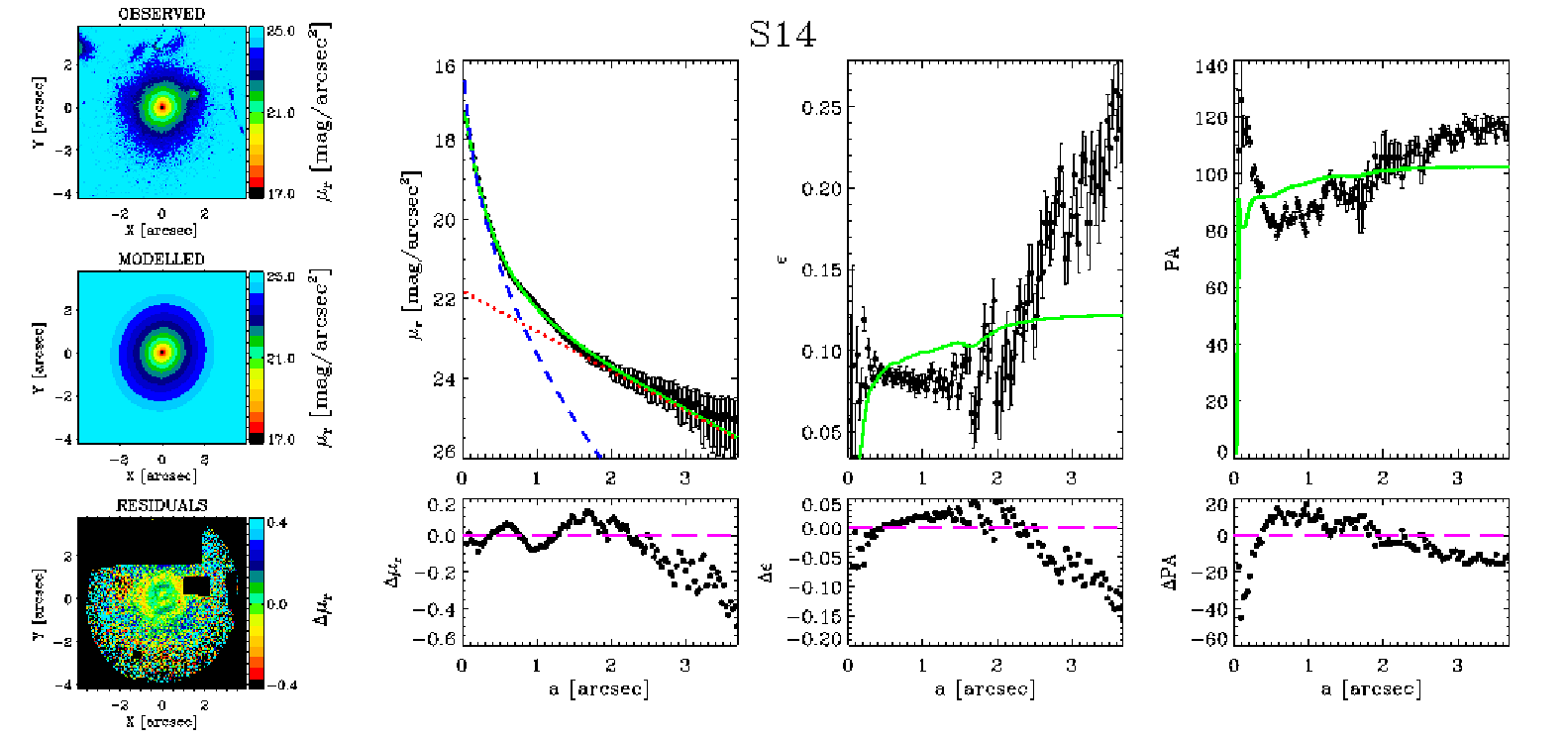}
\caption{As in Fig. \ref{fig:fit_S01} but for galaxy S14 (\sedisc\ model).}
\label{fig:fit_S14}
\end{figure*}

\clearpage

\begin{figure*}
\centering
\includegraphics[width=\textwidth, angle=0]{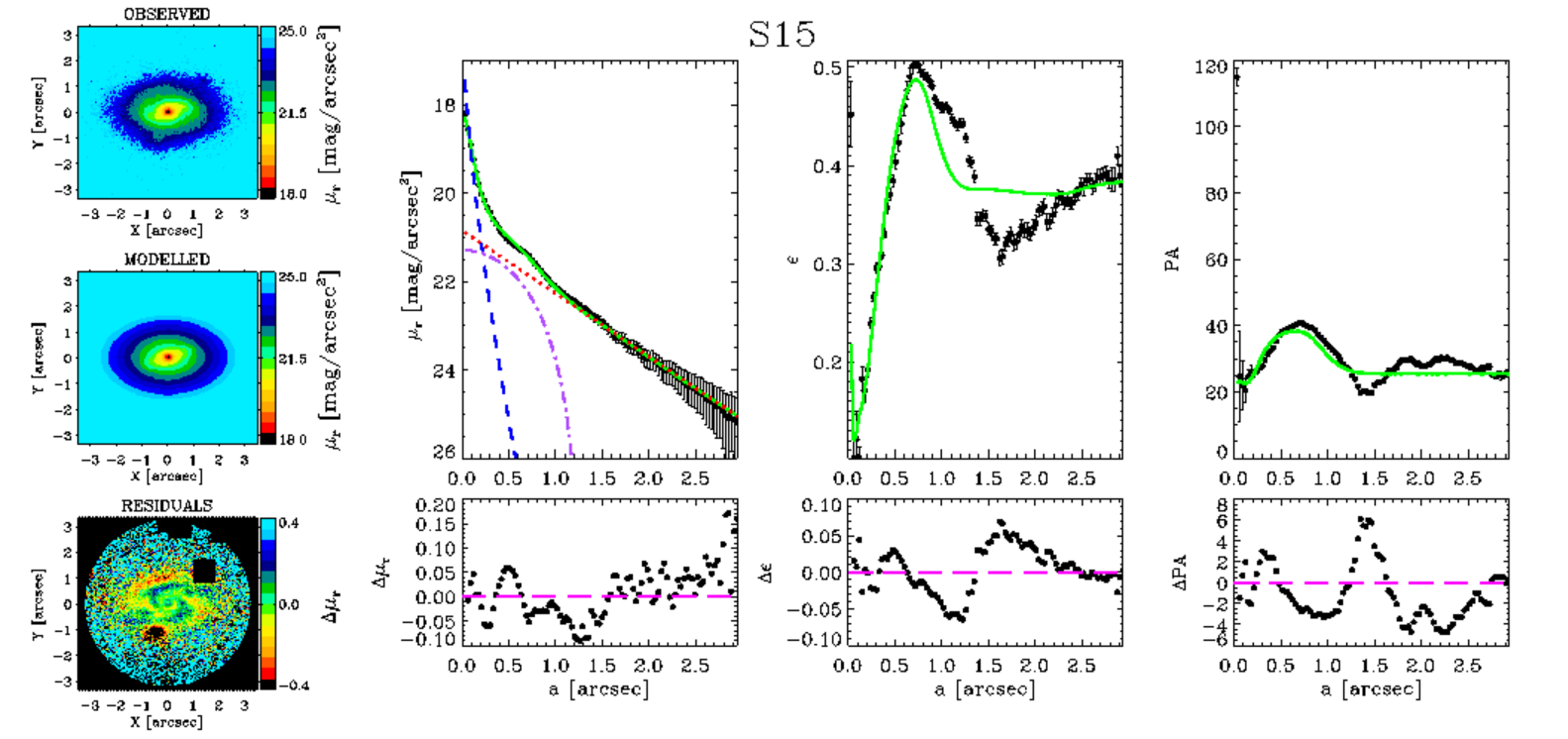}
\caption{As in Fig. \ref{fig:fit_S01} but for galaxy S15 (\sedisc\ model).}
\label{fig:fit_S15}
\end{figure*}

\begin{figure*}
\centering
\includegraphics[width=\textwidth, angle=0]{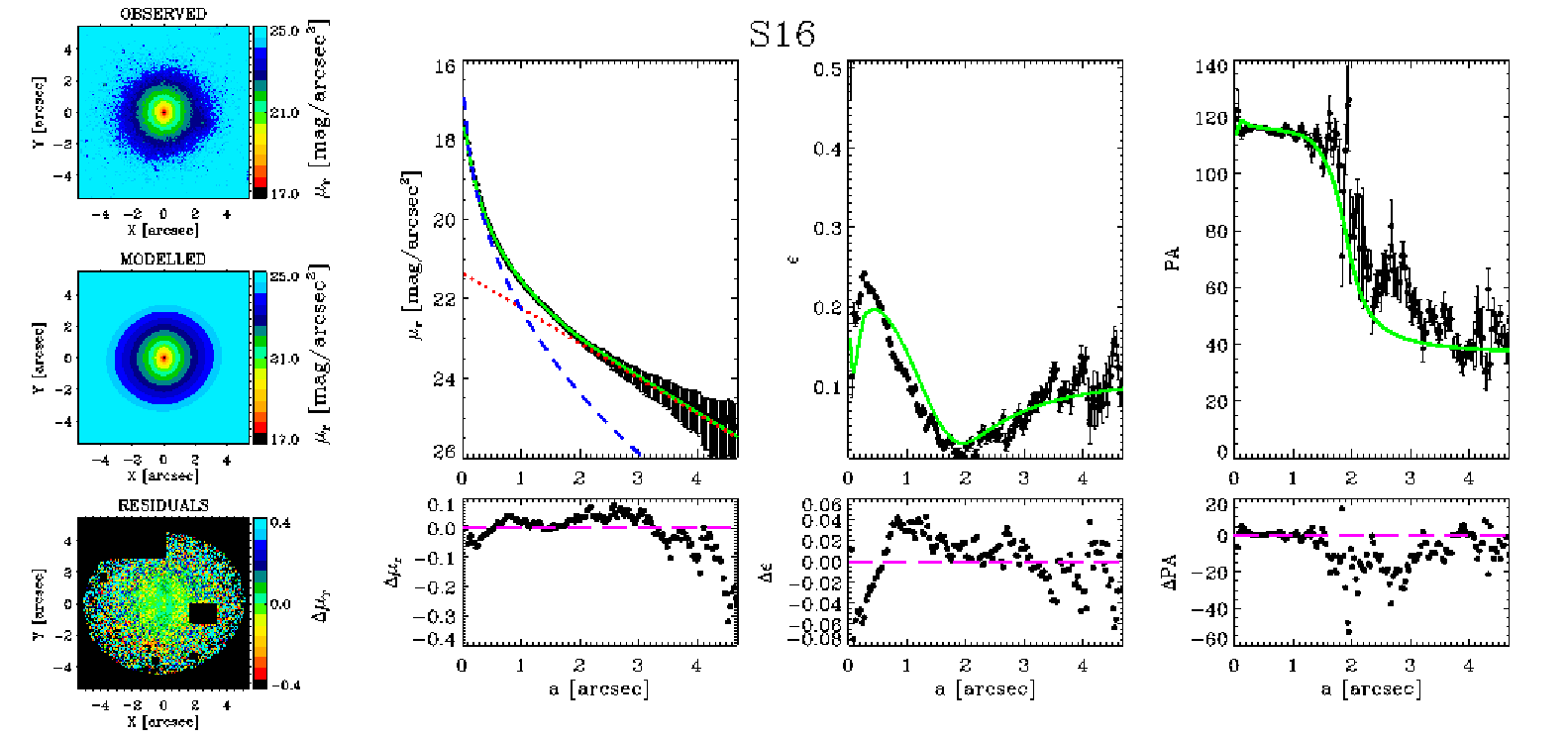}
\caption{As in Fig. \ref{fig:fit_S01} but for galaxy S16 (\sedisc\ model).}
\label{fig:fit_S16}
\end{figure*}

\clearpage

\begin{figure*}
\centering
\includegraphics[width=\textwidth, angle=0]{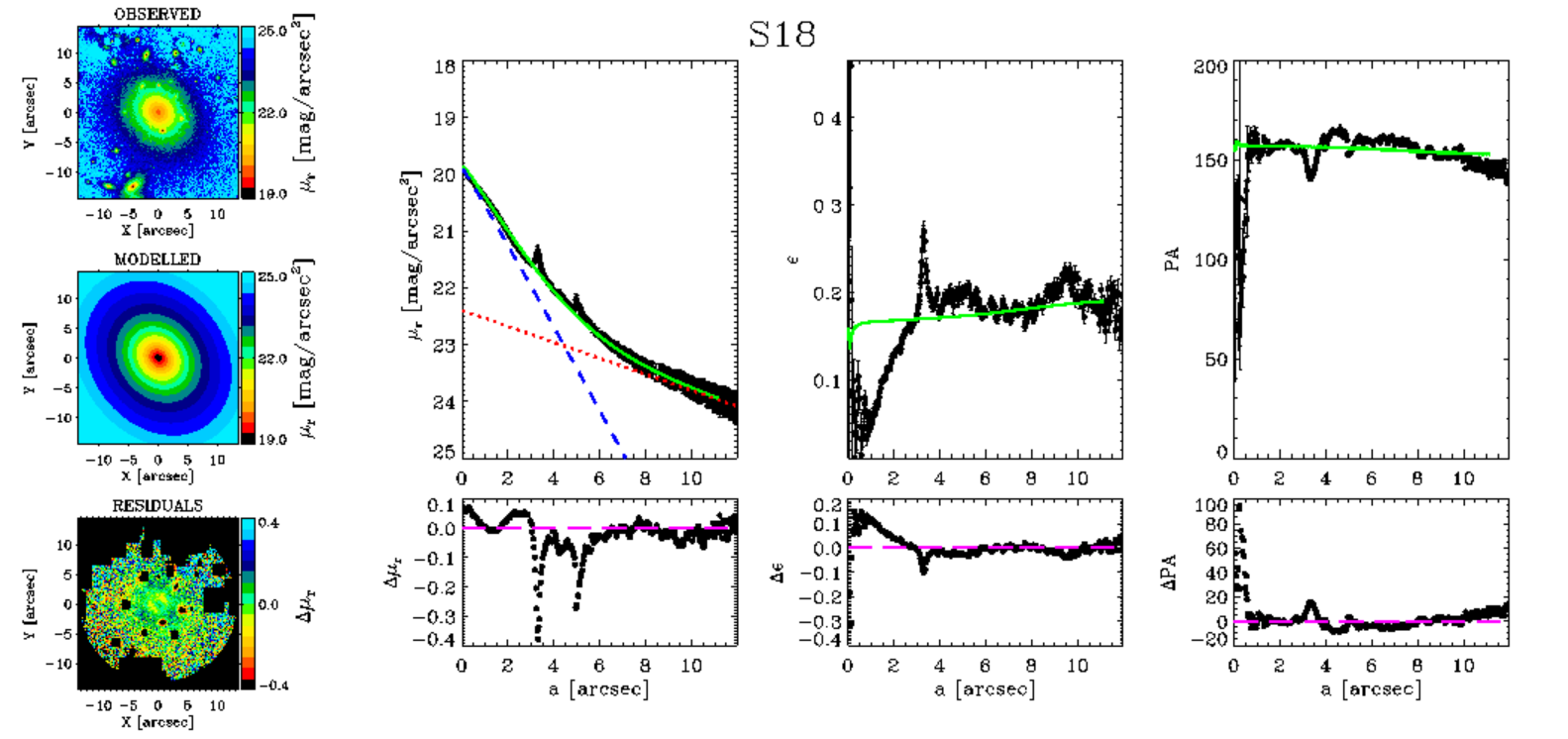}
\caption{As in Fig. \ref{fig:fit_S01} but for galaxy S18 (\sedisc\ model).}
\label{fig:fit_S18}
\end{figure*}

\begin{figure*}
\centering
\includegraphics[width=\textwidth, angle=0]{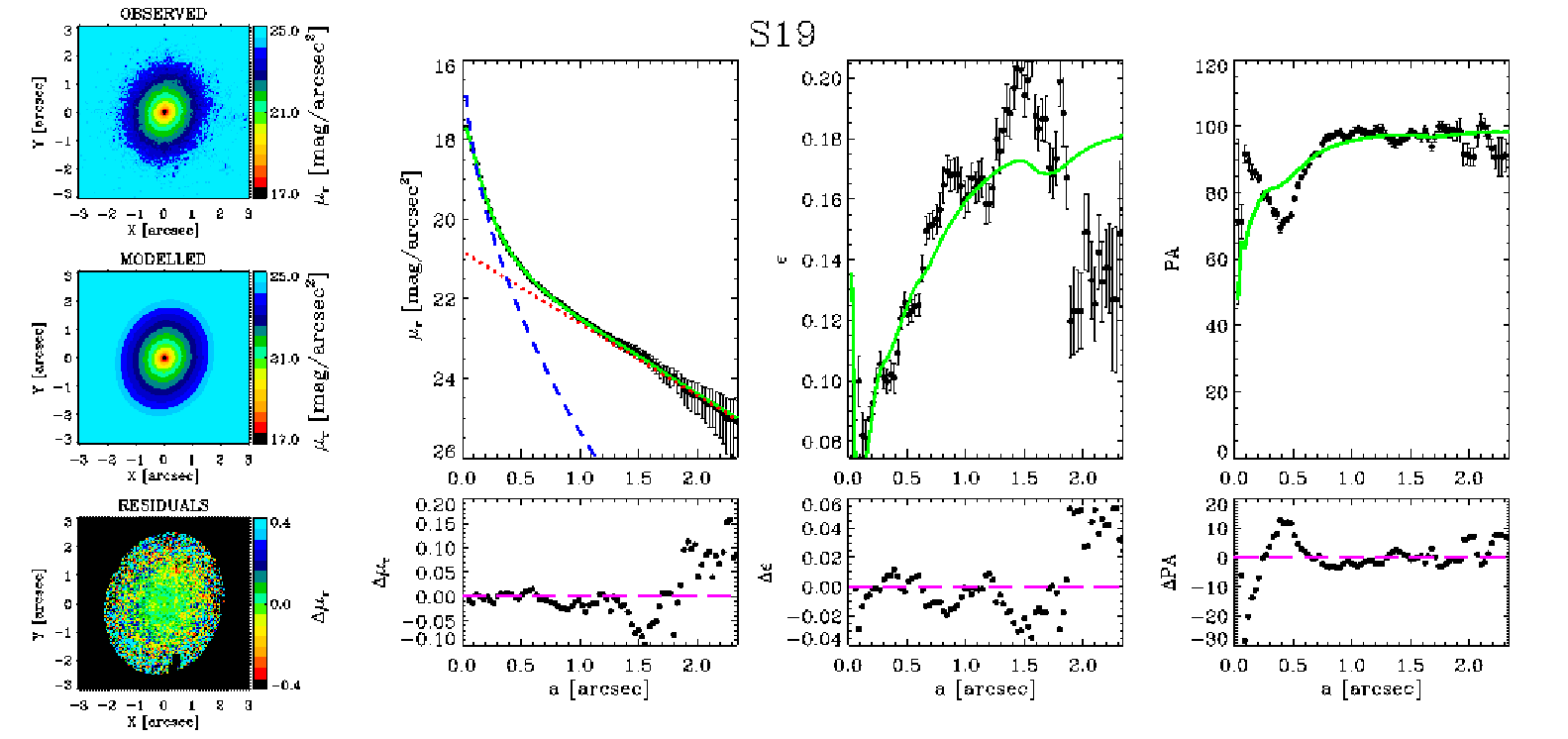}
\caption{As in Fig. \ref{fig:fit_S01} but for galaxy S19 (\sedisc\ model).}
\label{fig:fit_S19}
\end{figure*}

\clearpage

\begin{figure*}
\centering
\includegraphics[width=\textwidth, angle=0]{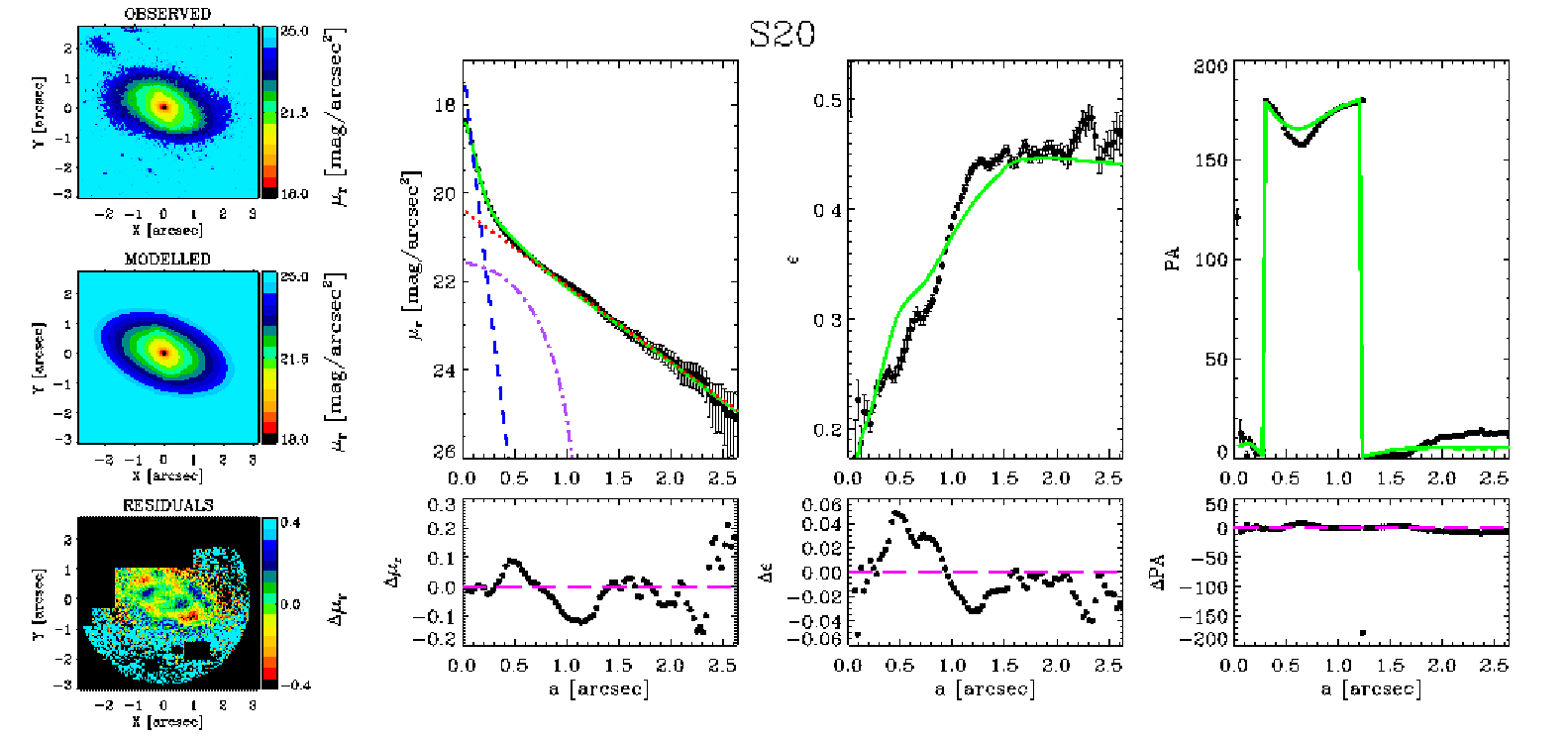}
\caption{As in Fig. \ref{fig:fit_S01} but for galaxy S20 fitted with a
  \sedibar\ model. The dashed-dotted purple line represents the intrinsic
  surface-brightness radial profile of the bar along its semi major
  axis.}
\label{fig:fit_S20}
\end{figure*}

\begin{figure*}
\centering
\includegraphics[width=\textwidth, angle=0]{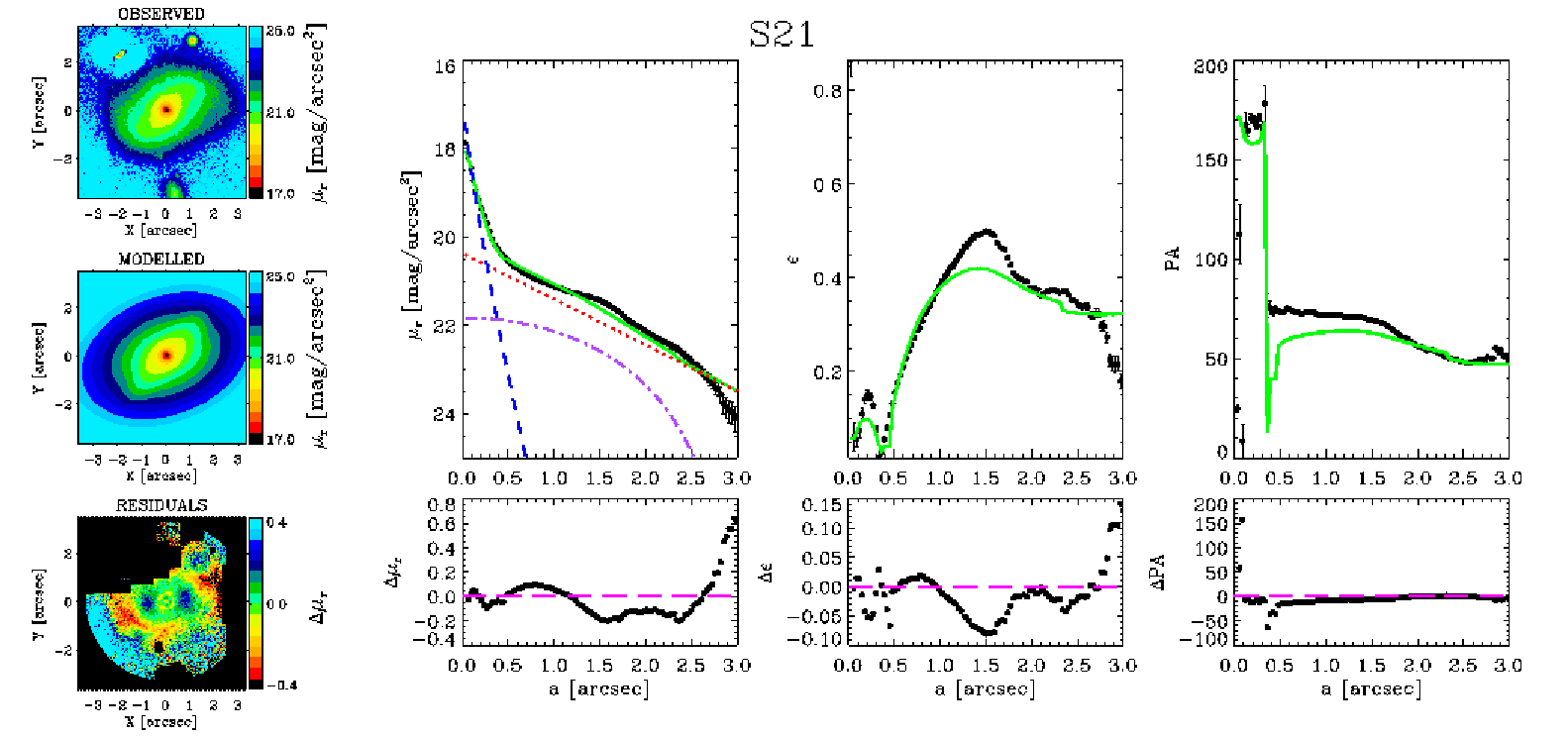}
\caption{As in Fig. \ref{fig:fit_S01} but for galaxy S21 fitted with a
  \sedibar\ model. The dashed-dotted purple line represents the intrinsic
  surface-brightness radial profile of the bar along its semi major
  axis.}
\label{fig:fit_S21}
\end{figure*}

\clearpage

\begin{figure*}
\centering
\includegraphics[width=\textwidth, angle=0]{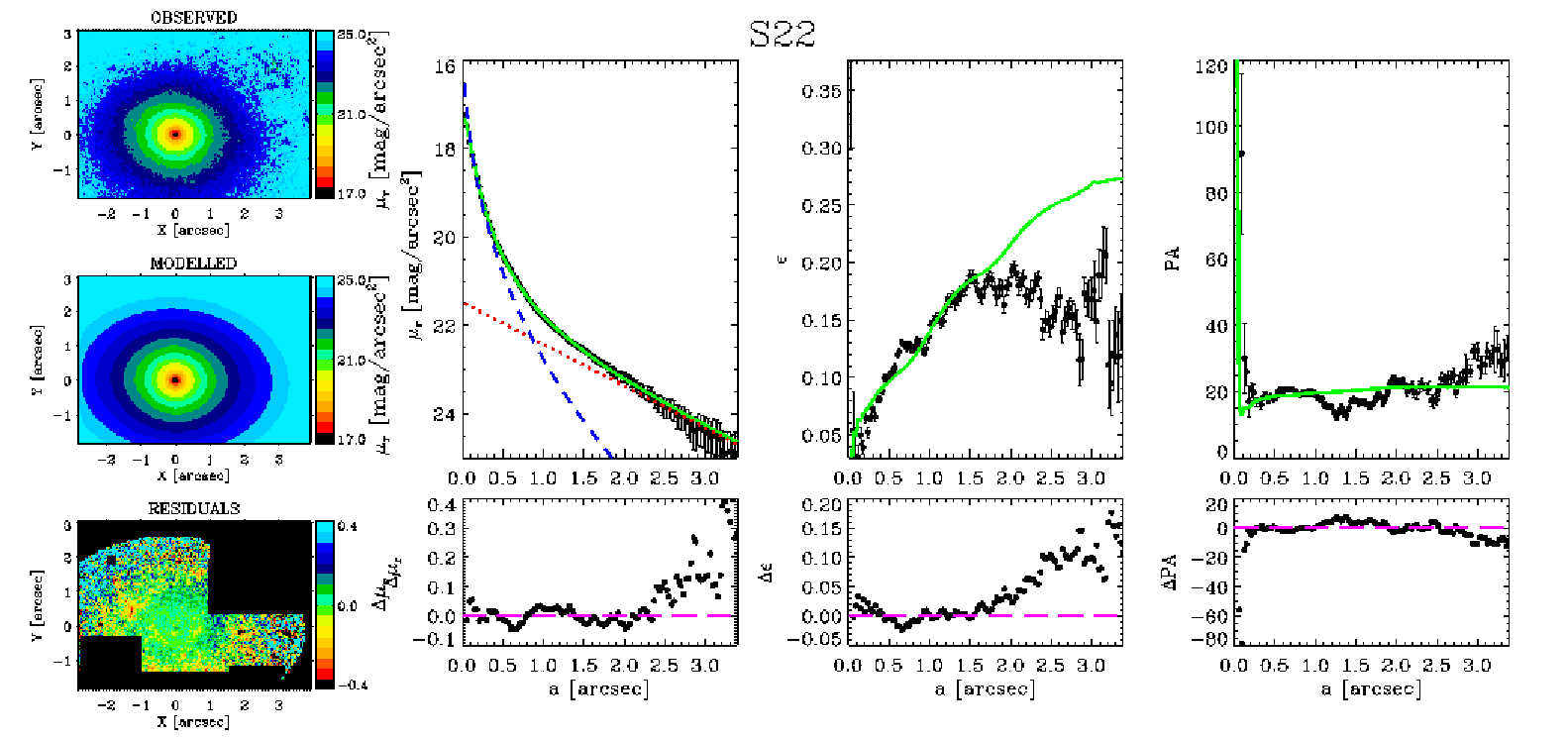}
\caption{As in Fig. \ref{fig:fit_S01} but for galaxy S09 (\sedisc\ model).}
\label{fig:fit_S22}
\end{figure*}

\begin{figure*}
\centering
\includegraphics[width=\textwidth, angle=0]{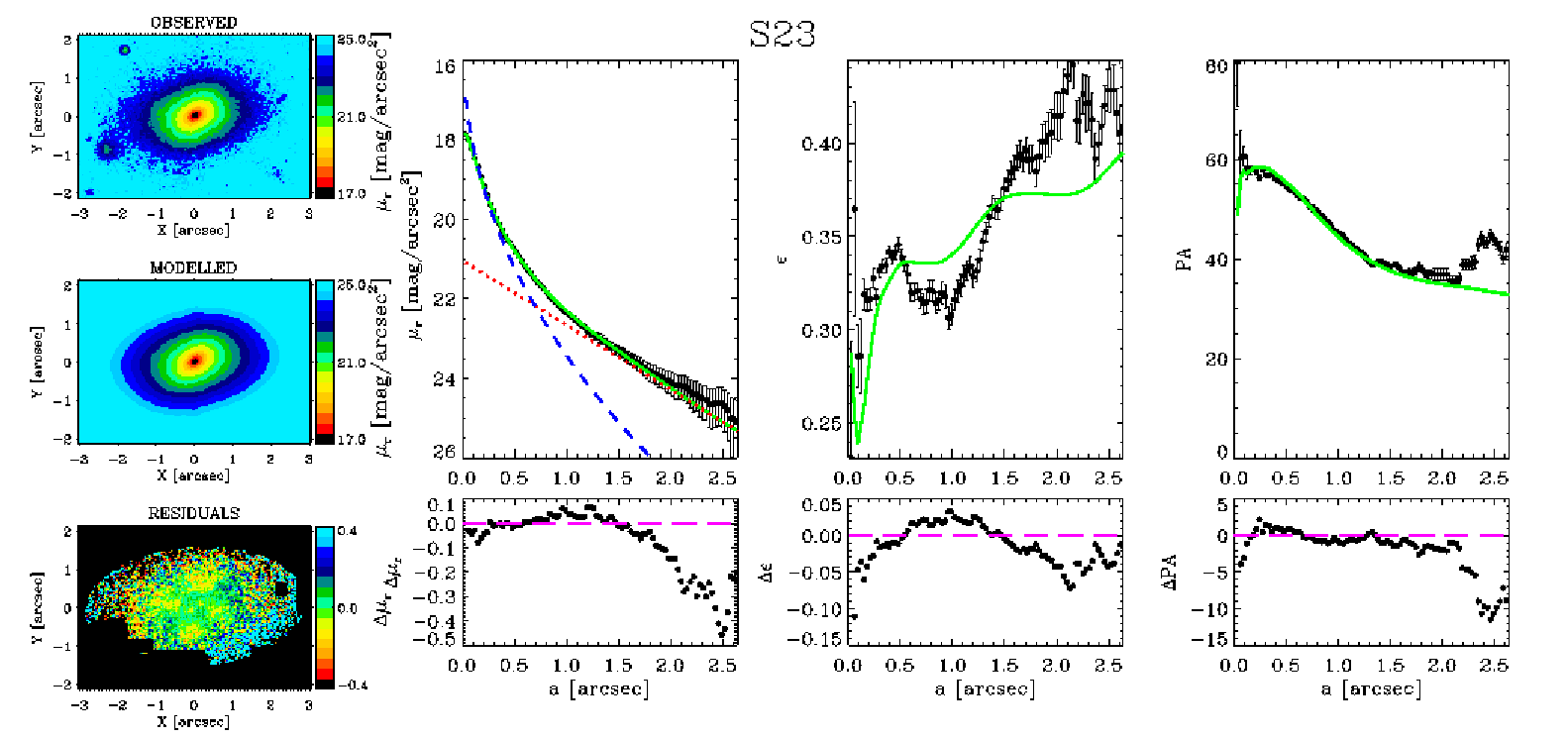}
\caption{As in Fig. \ref{fig:fit_S01} but for galaxy S23 (\sedisc\ model).}
\label{fig:fit_S23}
\end{figure*}

\clearpage

\begin{figure*}
\centering
\includegraphics[width=\textwidth, angle=0]{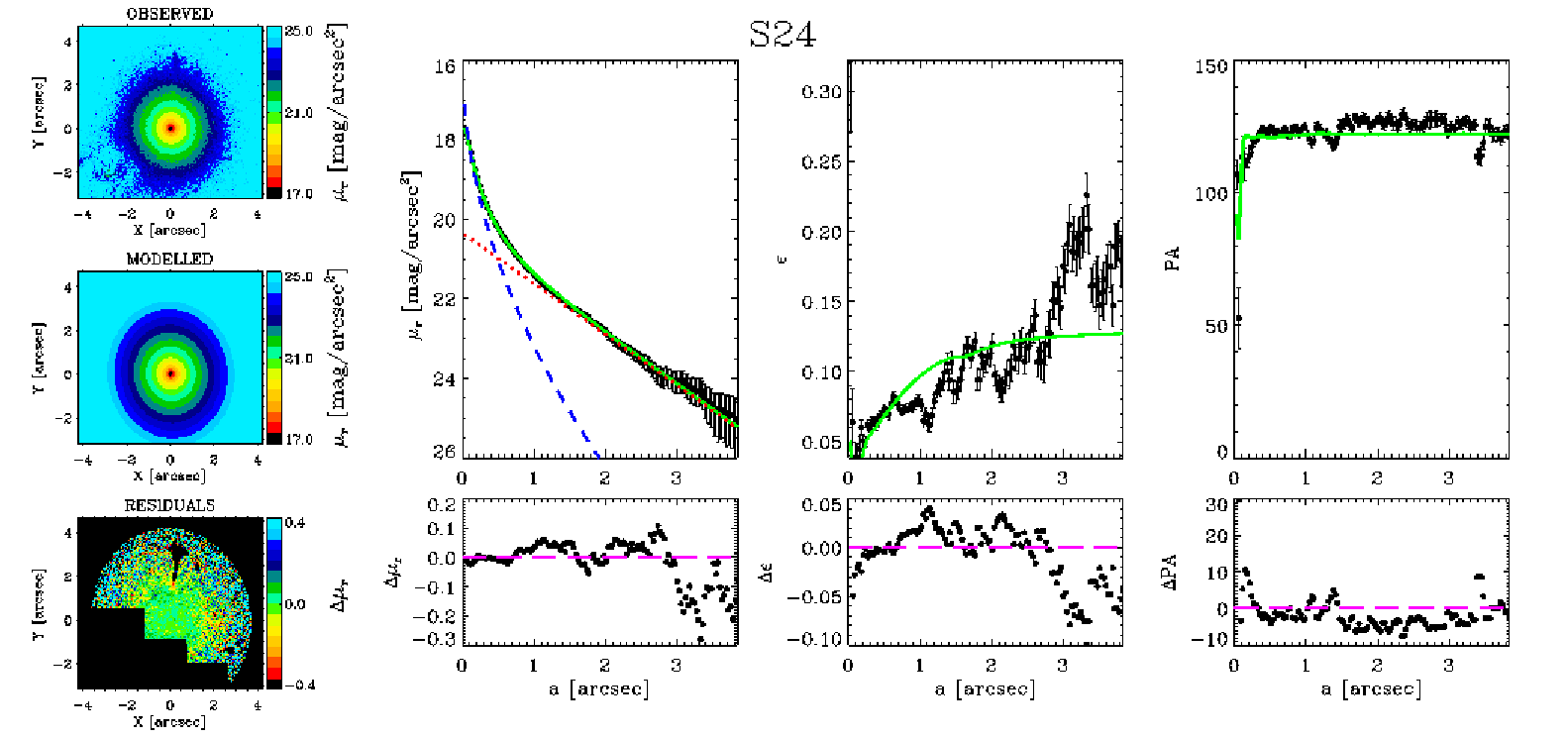}
\caption{As in Fig. \ref{fig:fit_S01} but for galaxy S24 (\sedisc\ model).}
\label{fig:fit_S24}
\end{figure*}

\begin{figure*}
\centering
\includegraphics[width=\textwidth, angle=0]{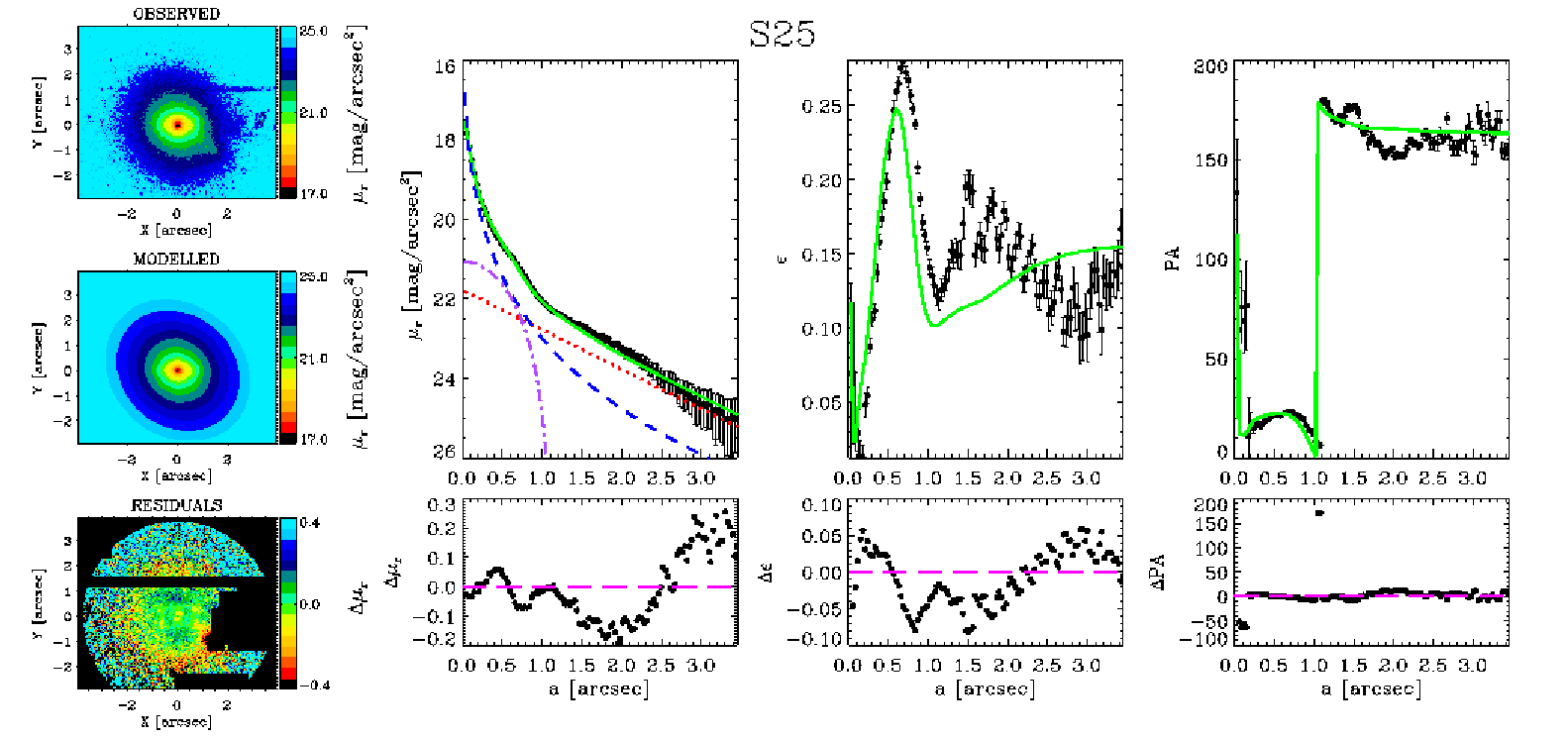}
\caption{As in Fig. \ref{fig:fit_S01} but for galaxy S25 fitted with a
  \sedibar\ model. The dashed-dotted purple line represents the intrinsic
  surface-brightness radial profile of the bar along its semi major
  axis.}
\label{fig:fit_S25}
\end{figure*}

\clearpage

\begin{figure*}
\centering
\includegraphics[width=\textwidth, angle=0]{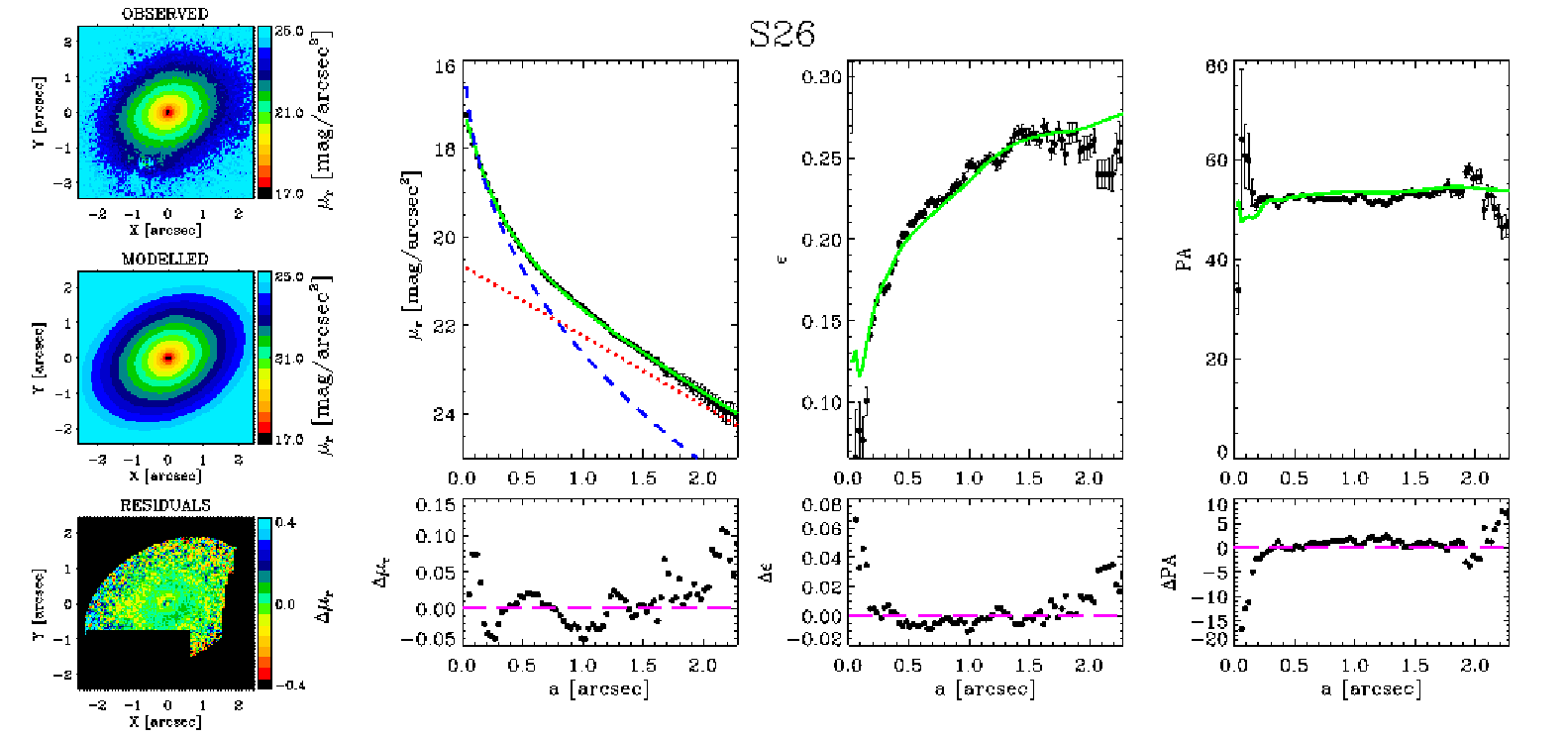}
\caption{As in Fig. \ref{fig:fit_S01} but for galaxy S26 (\sedisc\ model).}
\label{fig:fit_S26}
\end{figure*}

\begin{figure*}
\centering
\includegraphics[width=\textwidth, angle=0]{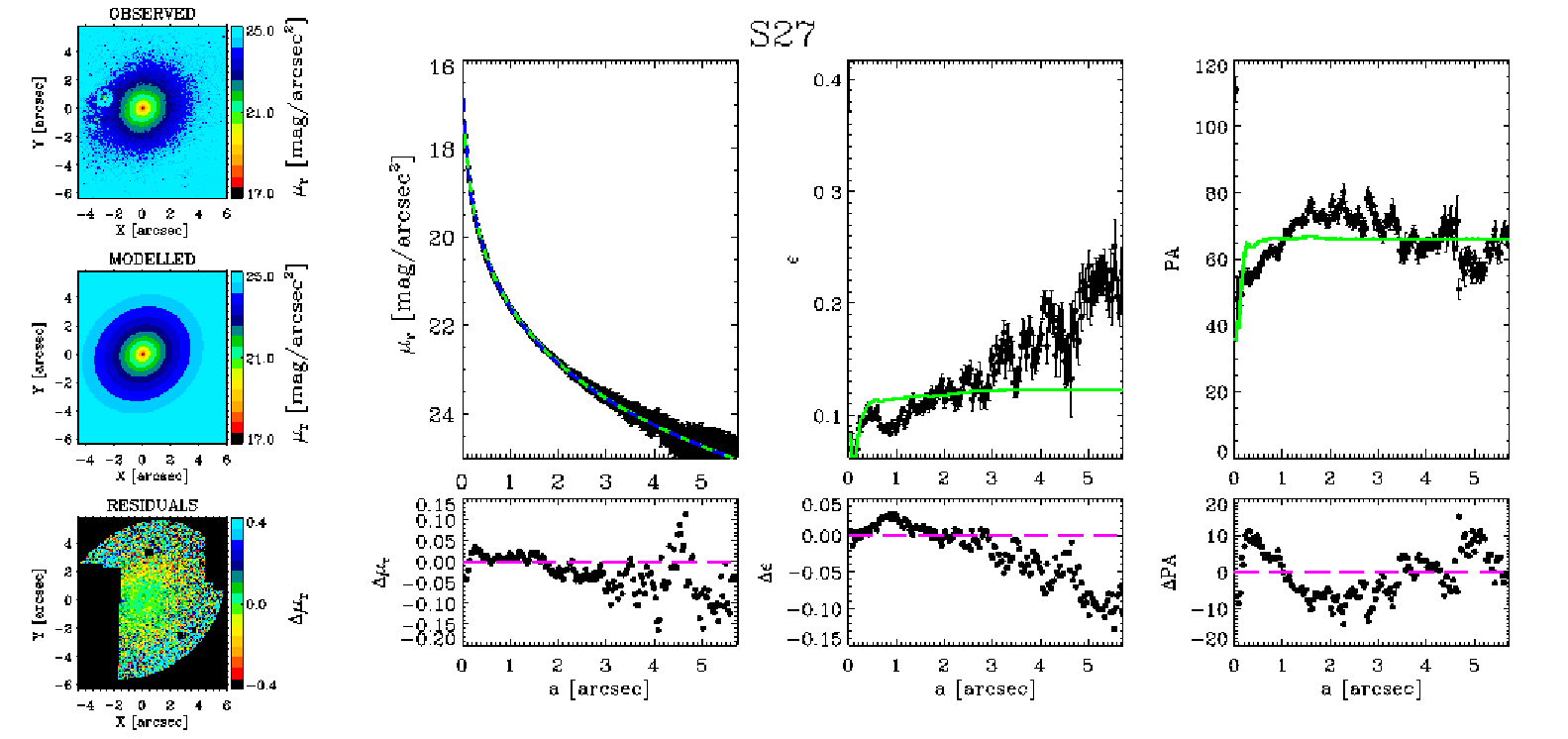}
\caption{As in Fig. \ref{fig:fit_S01} but for galaxy S27 (\sedisc\ model).}
\label{fig:fit_S27}
\end{figure*}

\clearpage

\begin{figure*}
\centering
\includegraphics[width=\textwidth, angle=0]{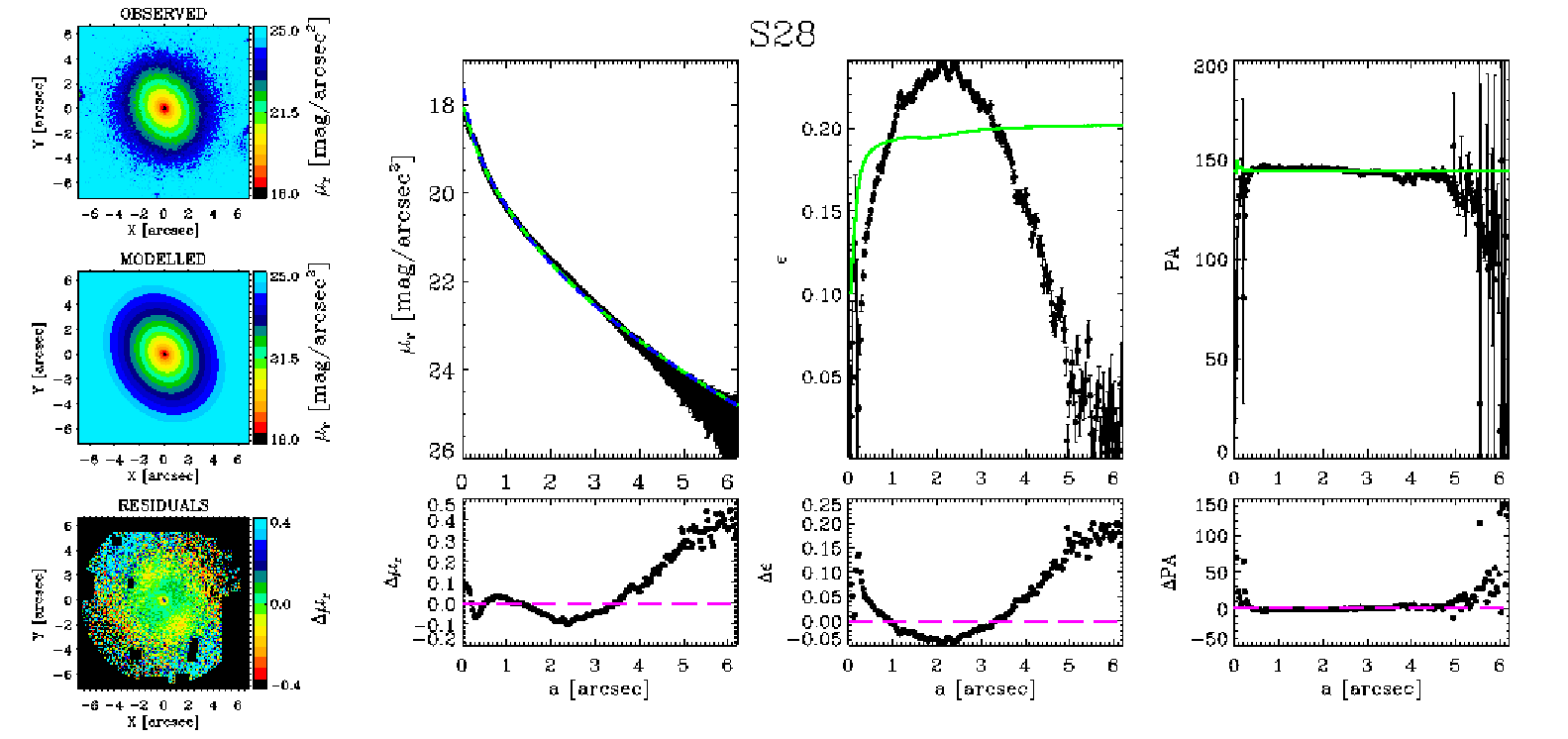}
\caption{As in Fig. \ref{fig:fit_S01} but for galaxy S28 fitted with a \sersic\ model).}
\label{fig:fit_S28}
\end{figure*}

\begin{figure*}
\centering
\includegraphics[width=\textwidth, angle=0]{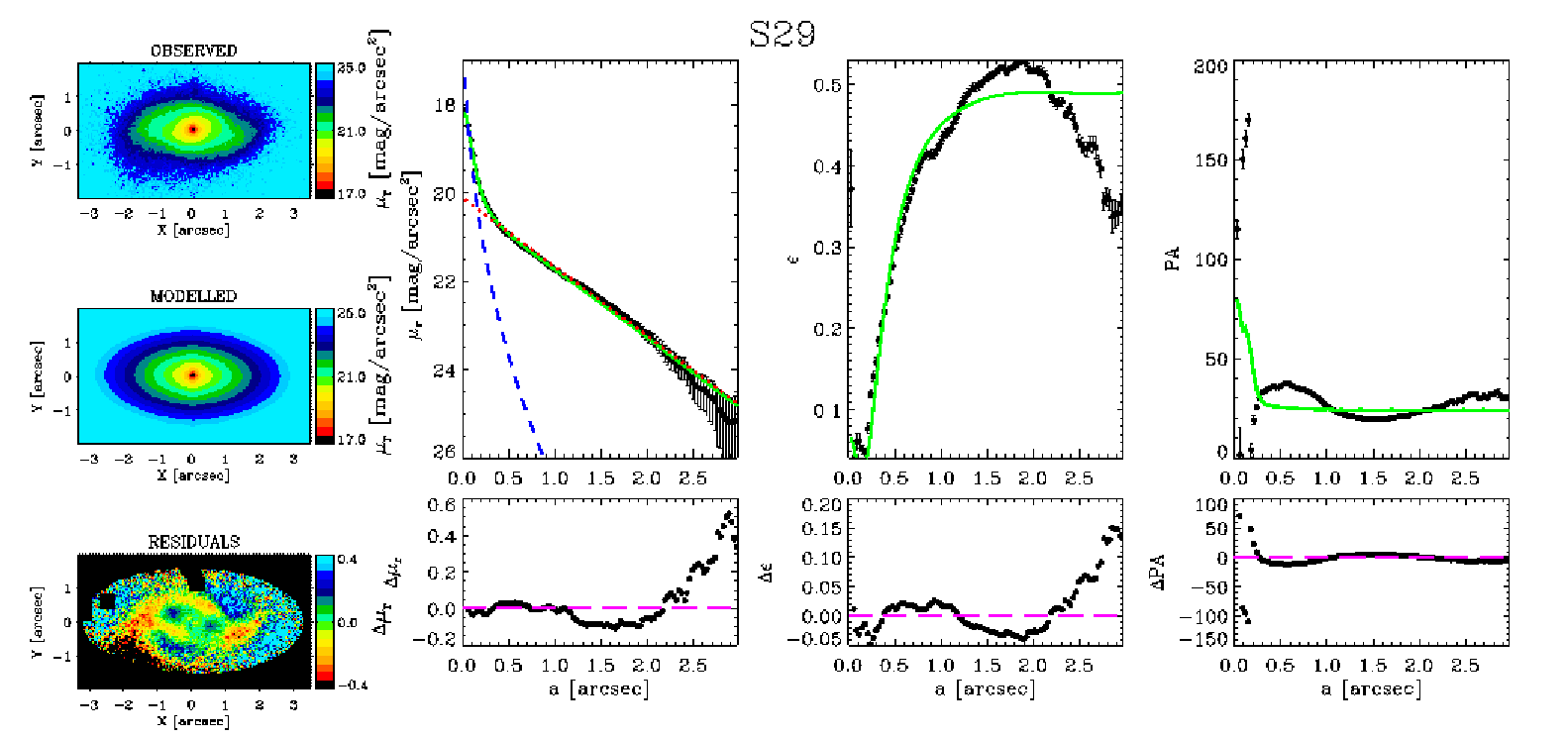}
\caption{As in Fig. \ref{fig:fit_S01} but for galaxy S29 (\sedisc\ model).}
\label{fig:fit_S29}
\end{figure*}

\clearpage

\begin{figure*}
\centering
\includegraphics[width=\textwidth, angle=0]{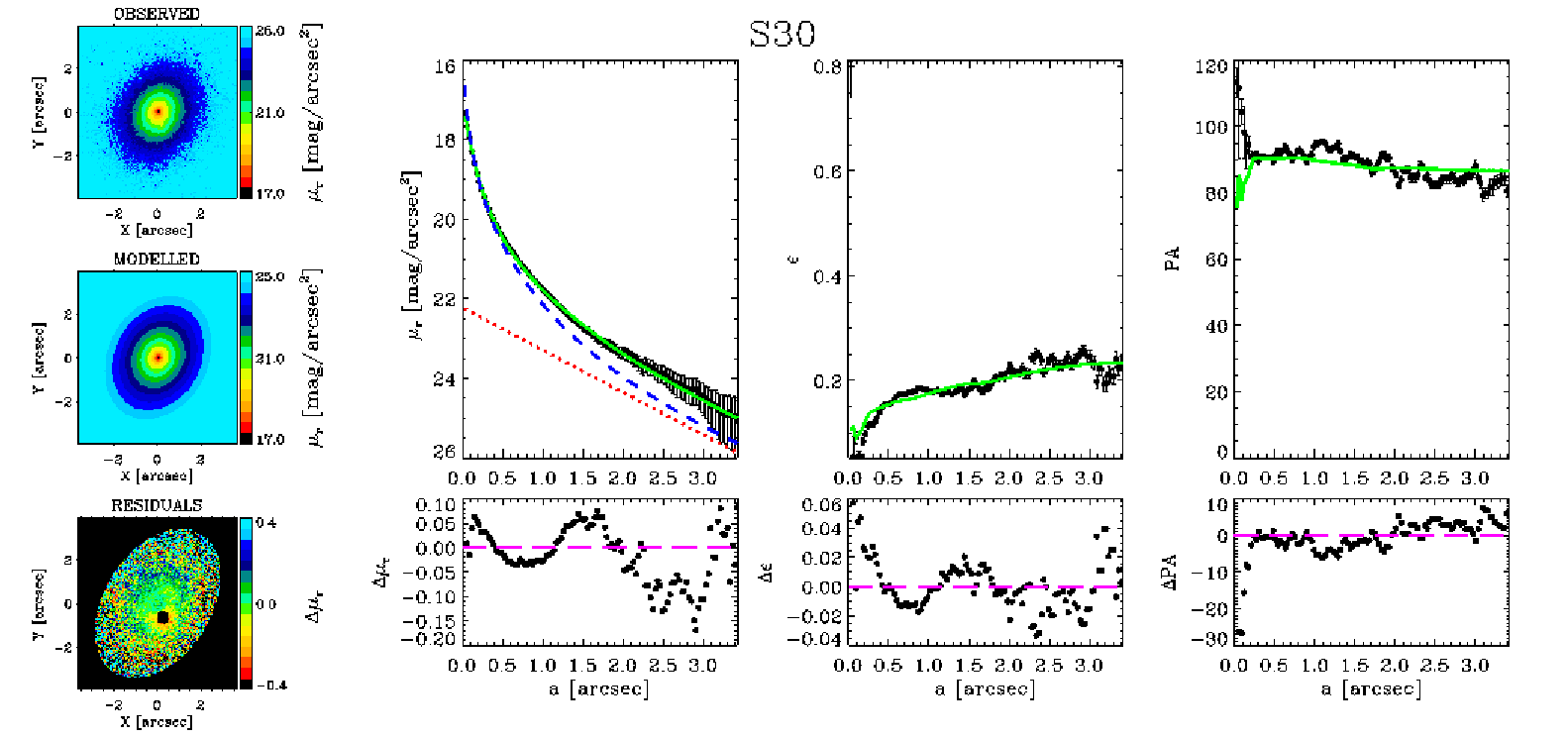}
\caption{As in Fig. \ref{fig:fit_S01} but for galaxy S30 (\sedisc\ model).}
\label{fig:fit_S30}
\end{figure*}

\begin{figure*}
\centering
\includegraphics[width=\textwidth, angle=0]{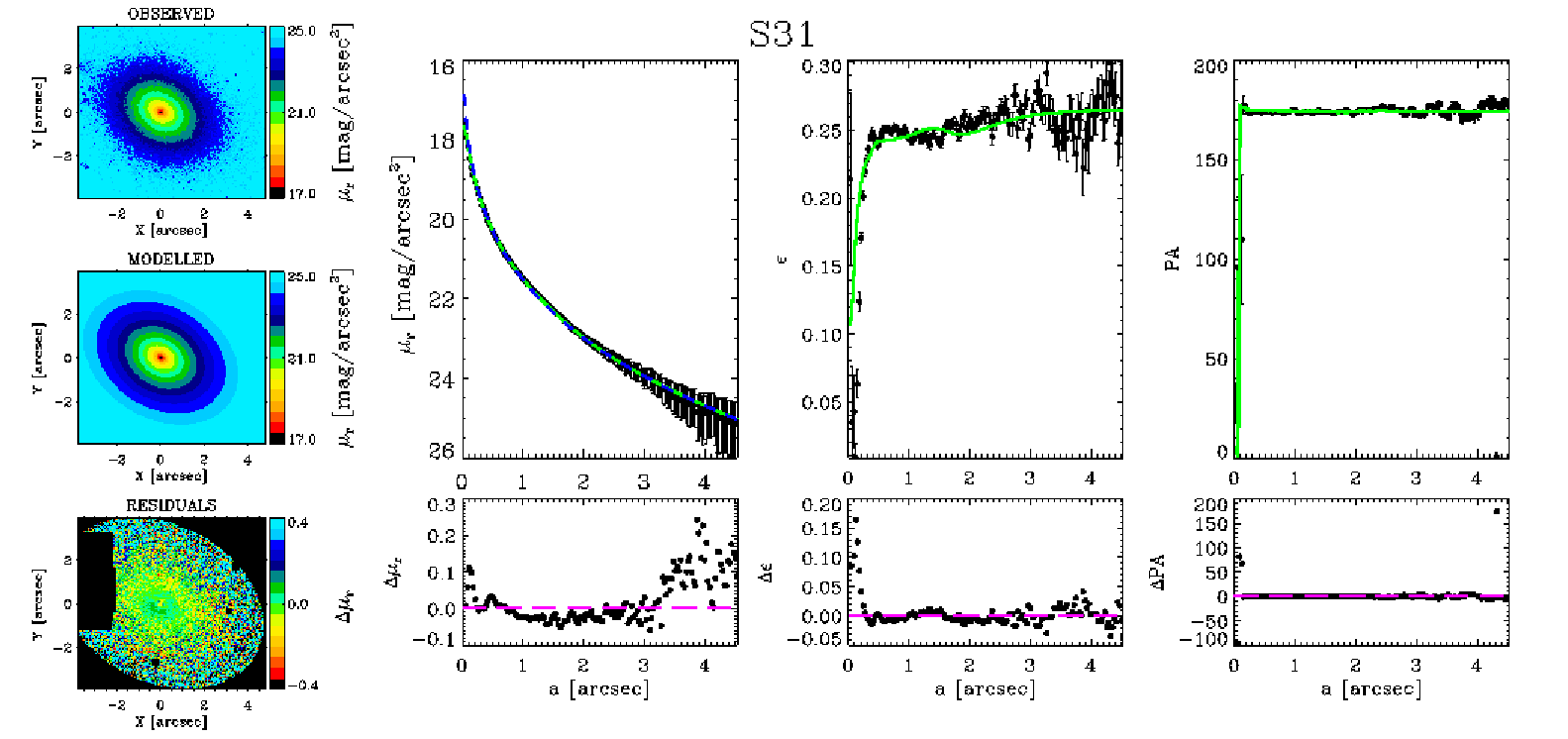}
\caption{As in Fig. \ref{fig:fit_S01} but for galaxy S10 fitted with a \sersic\ model.}
\label{fig:fit_S31}
\end{figure*}

\clearpage

\begin{figure*}
\centering
\includegraphics[width=\textwidth, angle=0]{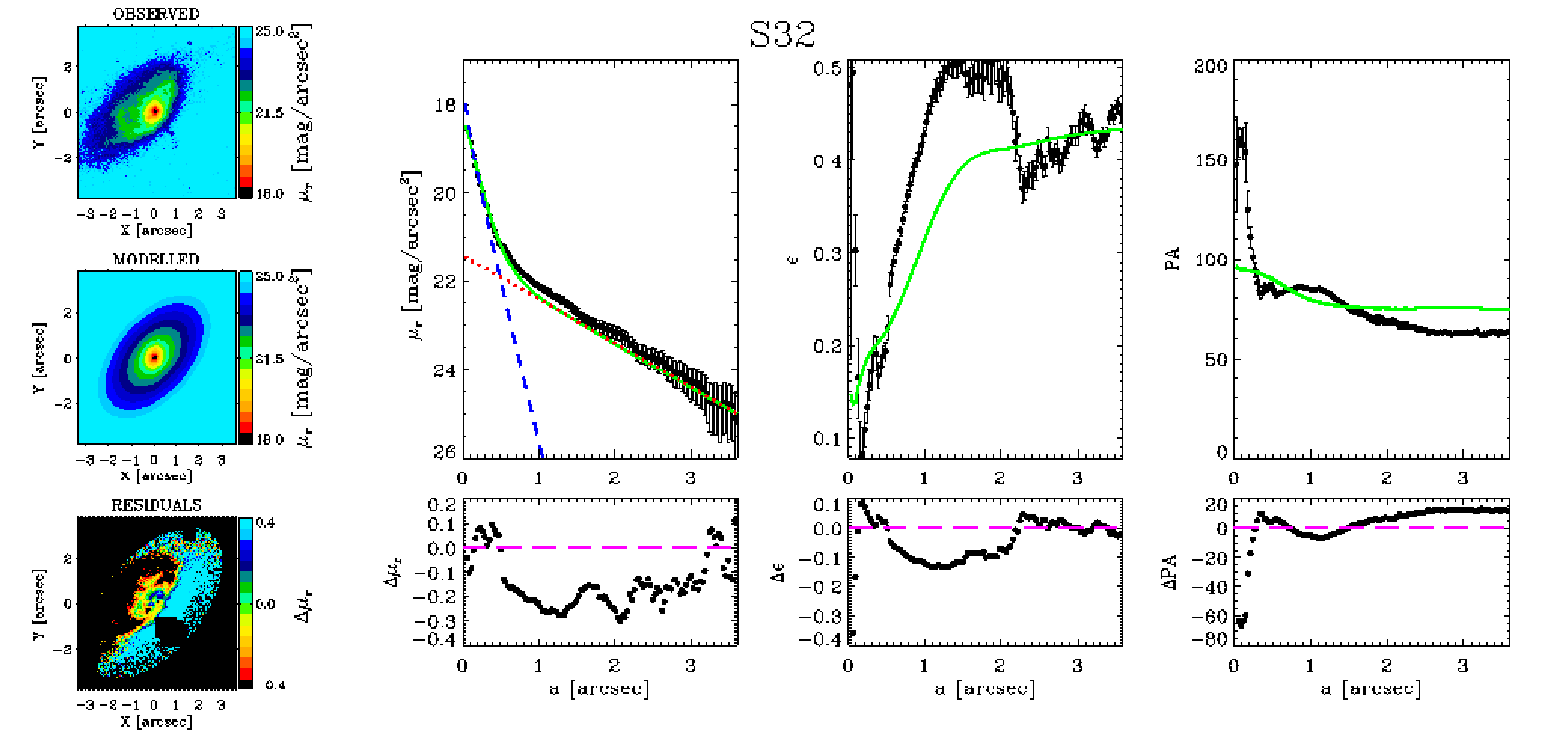}
\caption{As in Fig. \ref{fig:fit_S01} but for galaxy S32 (\sedisc\ model).}
\label{fig:fit_S32}
\end{figure*}

\begin{figure*}
\centering
\includegraphics[width=\textwidth, angle=0]{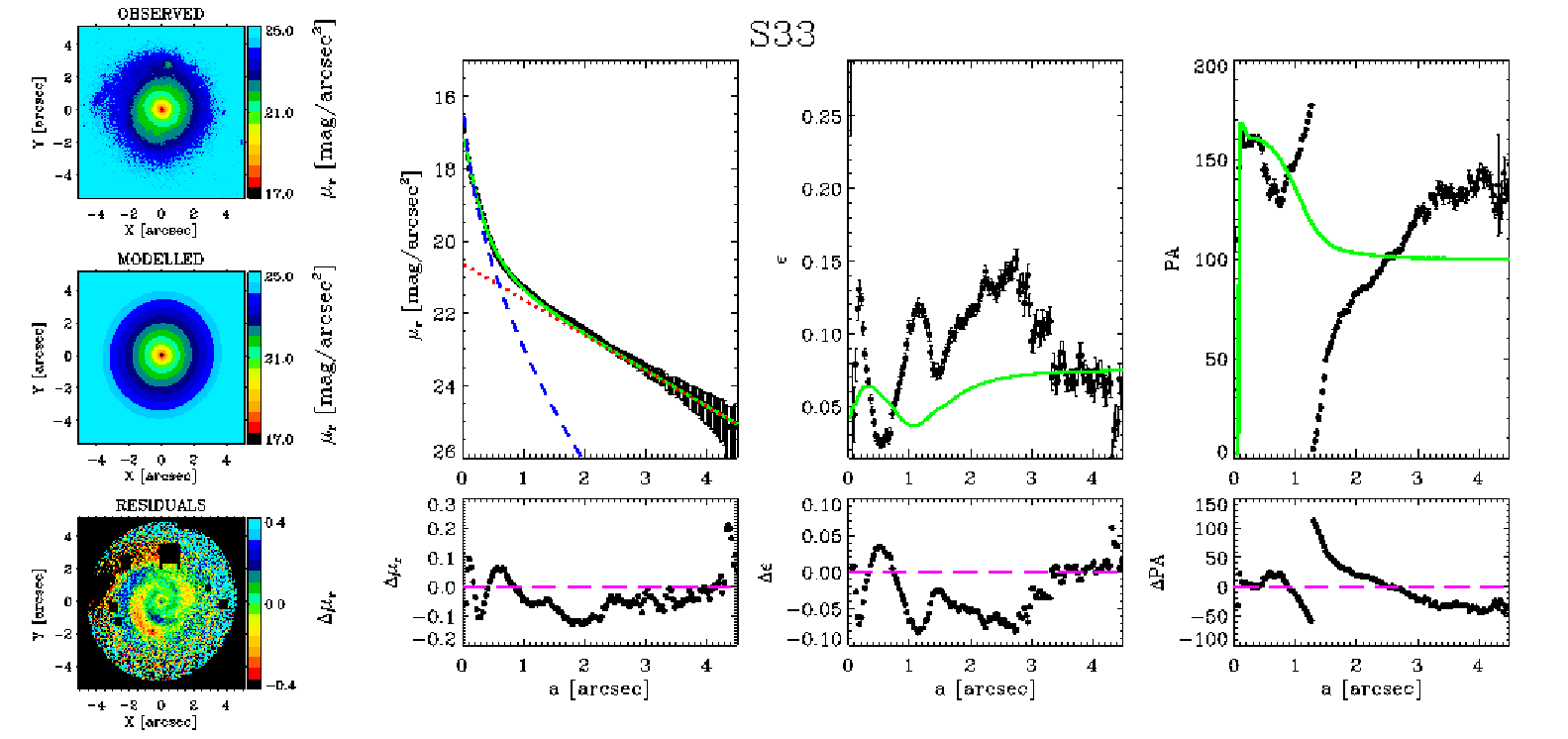}
\caption{As in Fig. \ref{fig:fit_S01} but for galaxy S33 (\sedisc\ model).}
\label{fig:fit_S33}
\end{figure*}

\clearpage

\begin{figure*}
\centering
\includegraphics[width=\textwidth, angle=0]{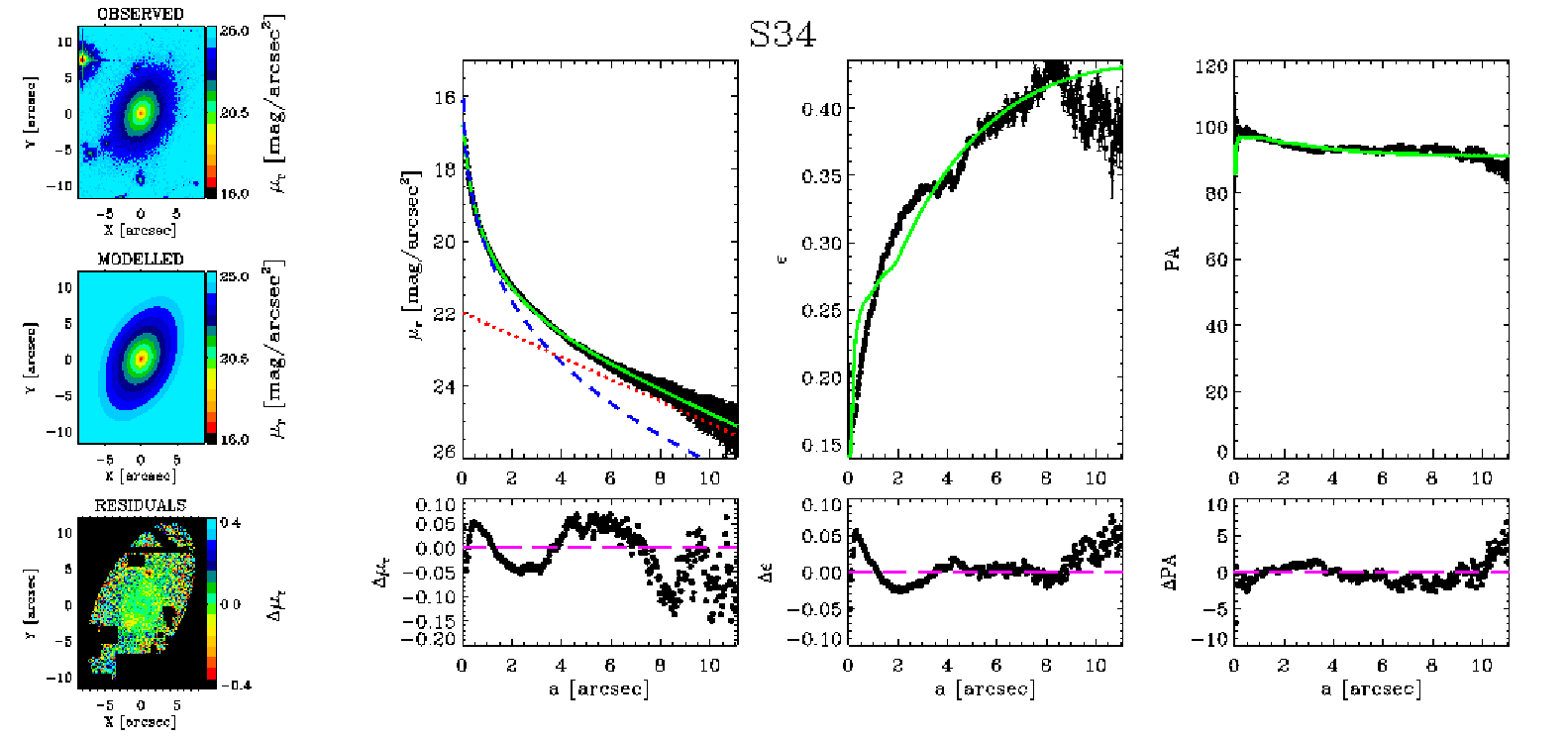}
\caption{As in Fig. \ref{fig:fit_S01} but for galaxy S34 (\sedisc\ model).}
\label{fig:fit_S34}
\end{figure*}

\begin{figure*}
\centering
\includegraphics[width=\textwidth, angle=0]{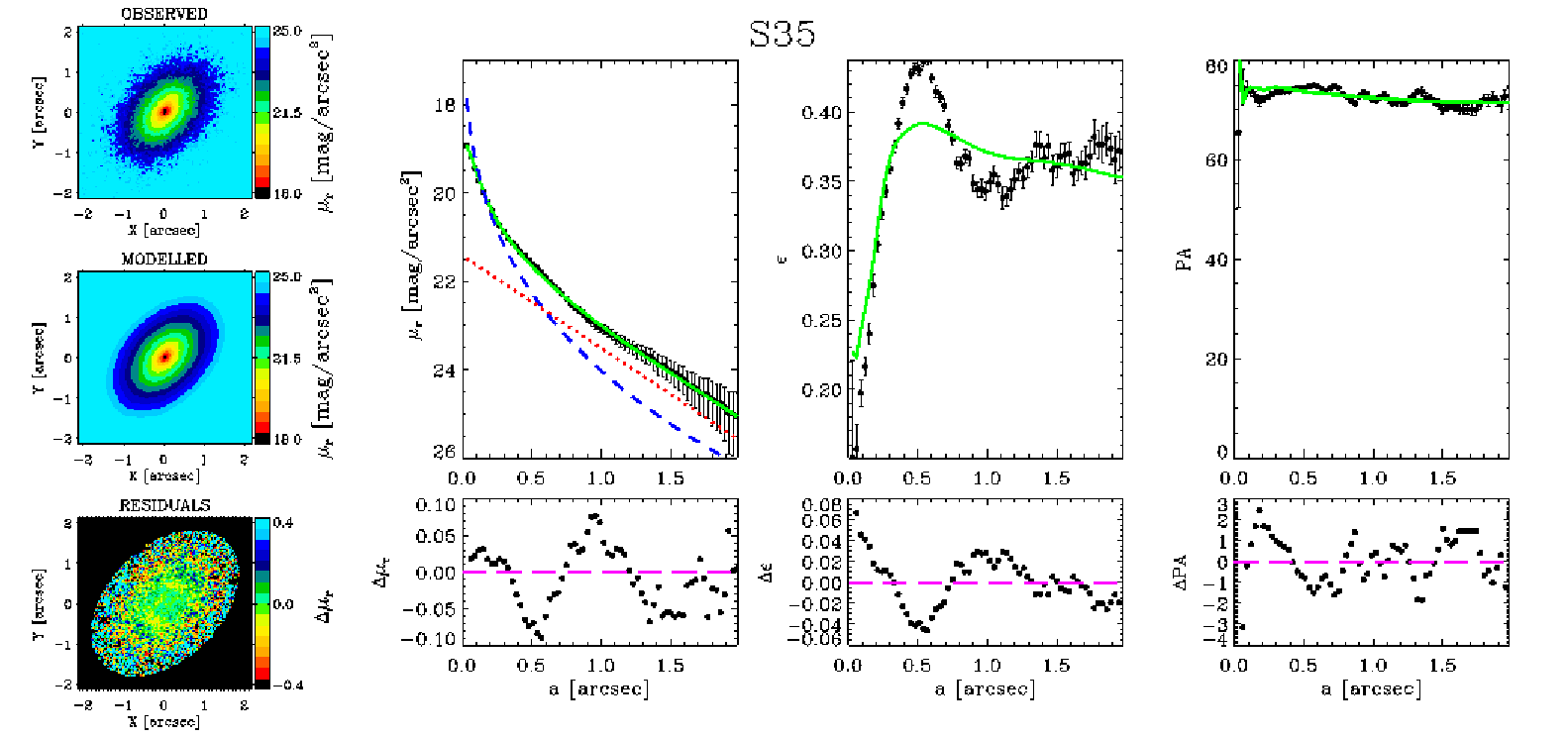}
\caption{As in Fig. \ref{fig:fit_S01} but for galaxy S35 (\sedisc\ model).}
\label{fig:fit_S35}
\end{figure*}

\clearpage

\begin{figure*}
\centering
\includegraphics[width=\textwidth, angle=0]{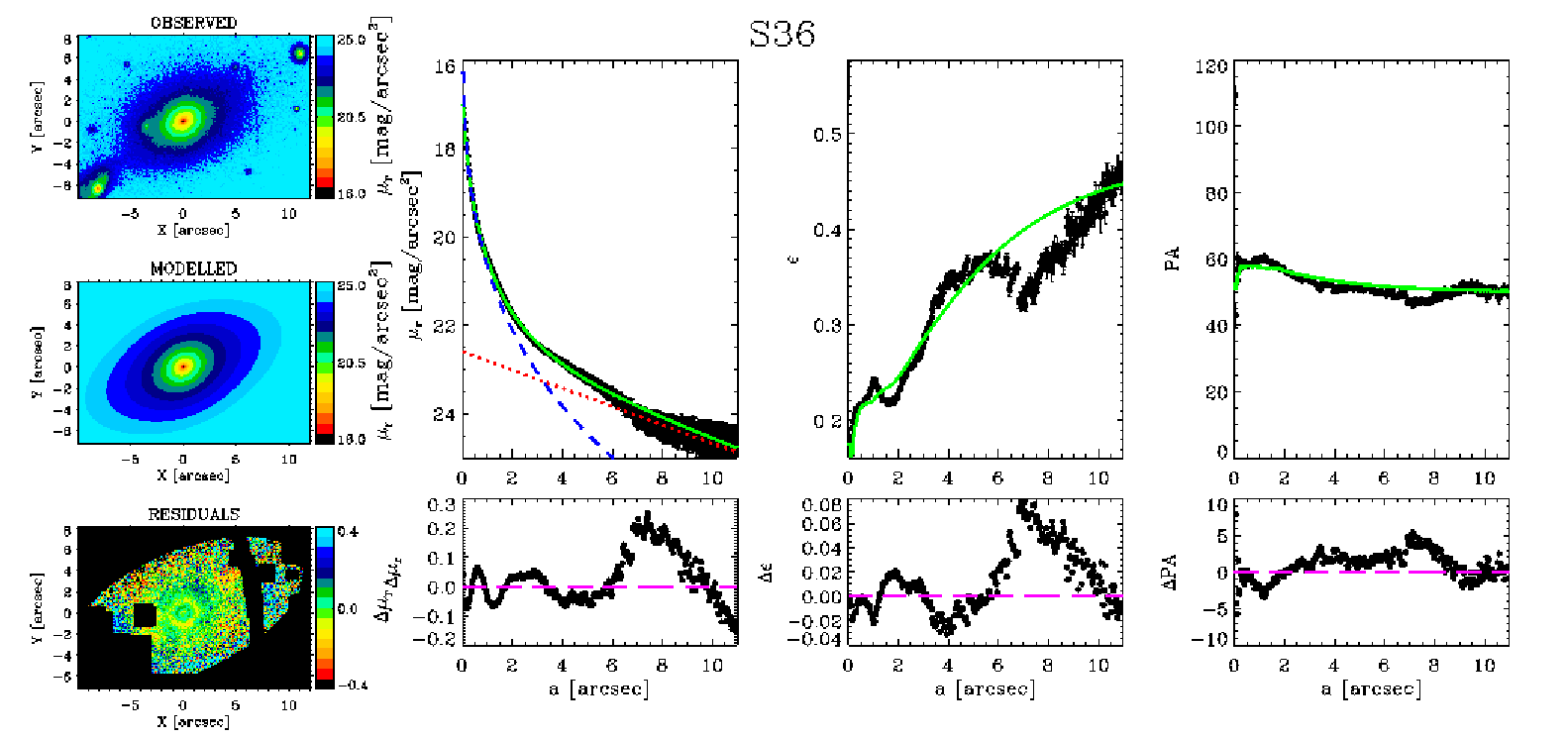}
\caption{As in Fig. \ref{fig:fit_S01} but for galaxy S36 (\sedisc\ model).}
\label{fig:fit_S36}
\end{figure*}

\begin{figure*}
\centering
\includegraphics[width=\textwidth, angle=0]{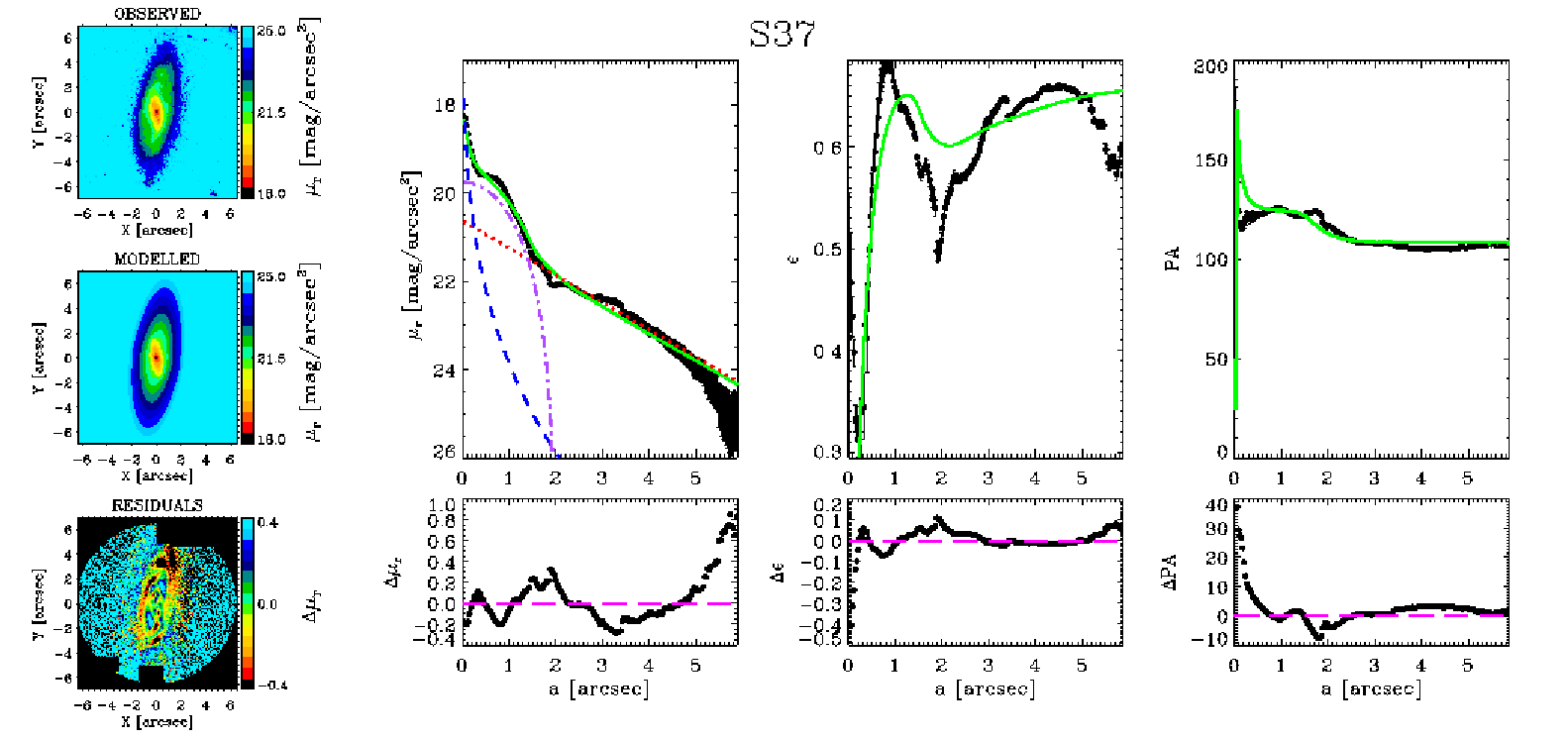}
\caption{As in Fig. \ref{fig:fit_S01} but for galaxy S37 (\sedisc\ model).}
\label{fig:fit_S37}
\end{figure*}

\clearpage

\begin{figure*}
\centering
\includegraphics[width=\textwidth, angle=0]{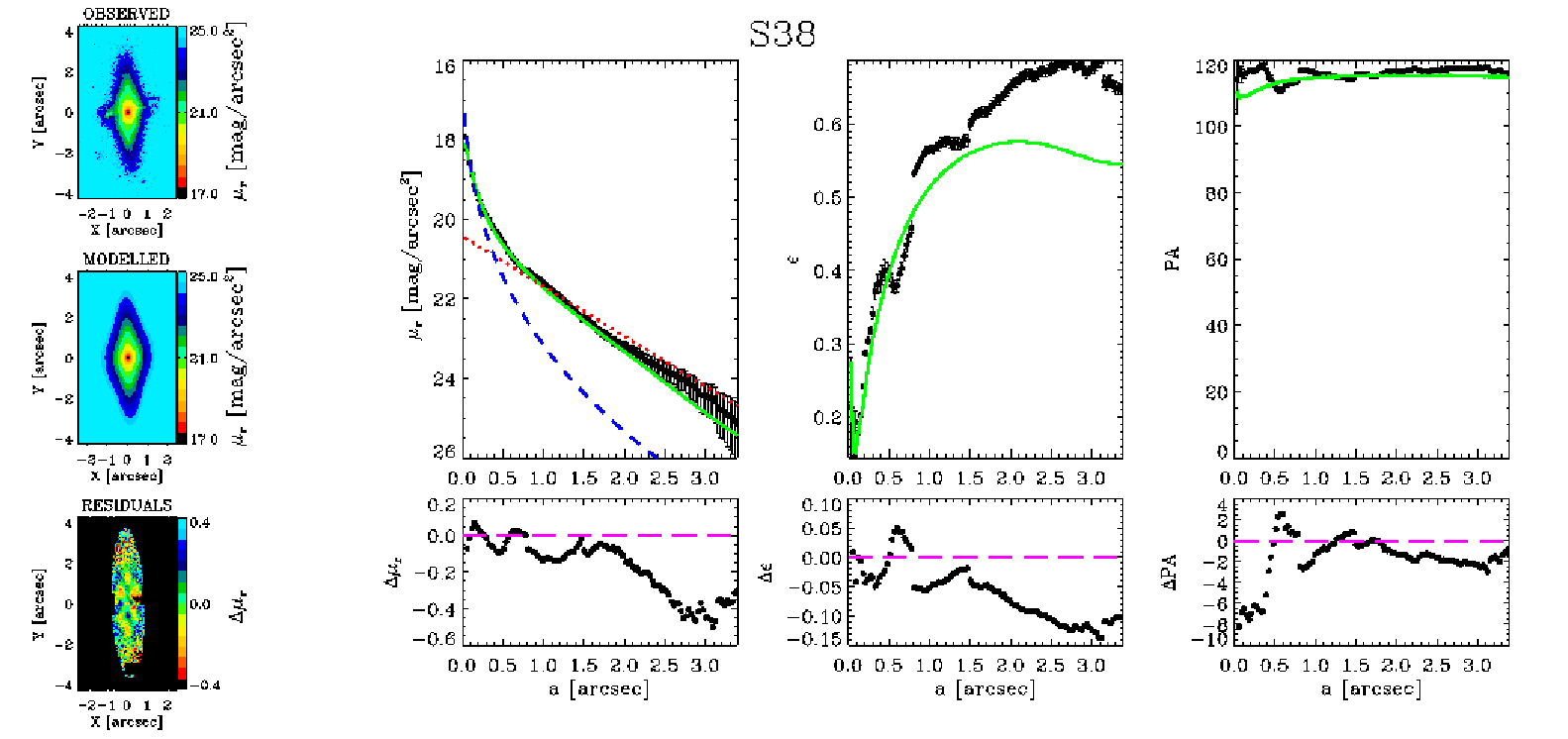}
\caption{As in Fig. \ref{fig:fit_S01} but for galaxy S38 (\sedisc\ model).}
\label{fig:fit_S38}
\end{figure*}

\begin{figure*}
\centering
\includegraphics[width=\textwidth, angle=0]{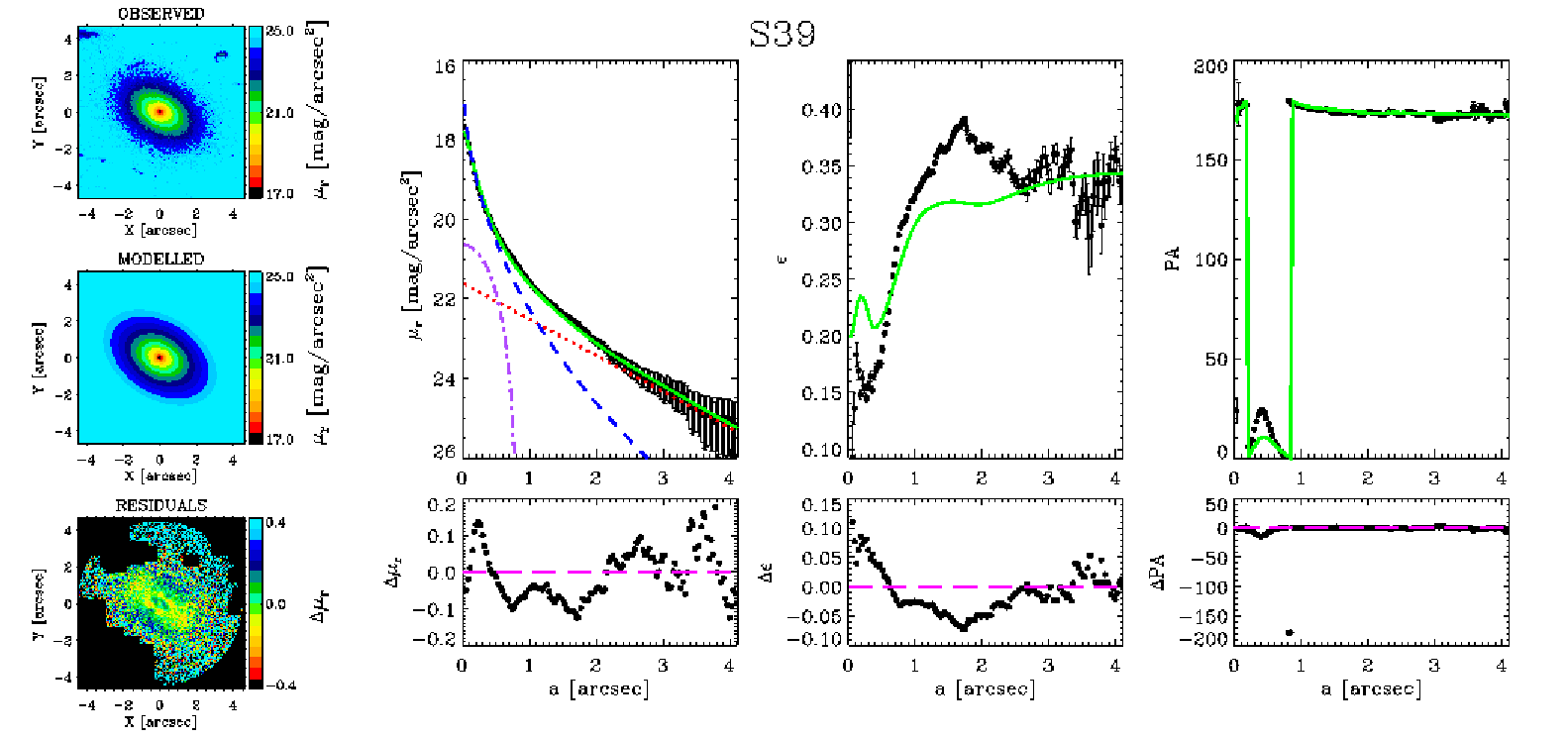}
\caption{As in Fig. \ref{fig:fit_S01} but for galaxy S39 fitted with a
  \sedibar\ model. The dashed-dotted purple line represents the intrinsic
  surface-brightness radial profile of the bar along its semi major
  axis.}
\label{fig:fit_S39}
\end{figure*}

\clearpage

\begin{figure*}
\centering
\includegraphics[width=\textwidth, angle=0]{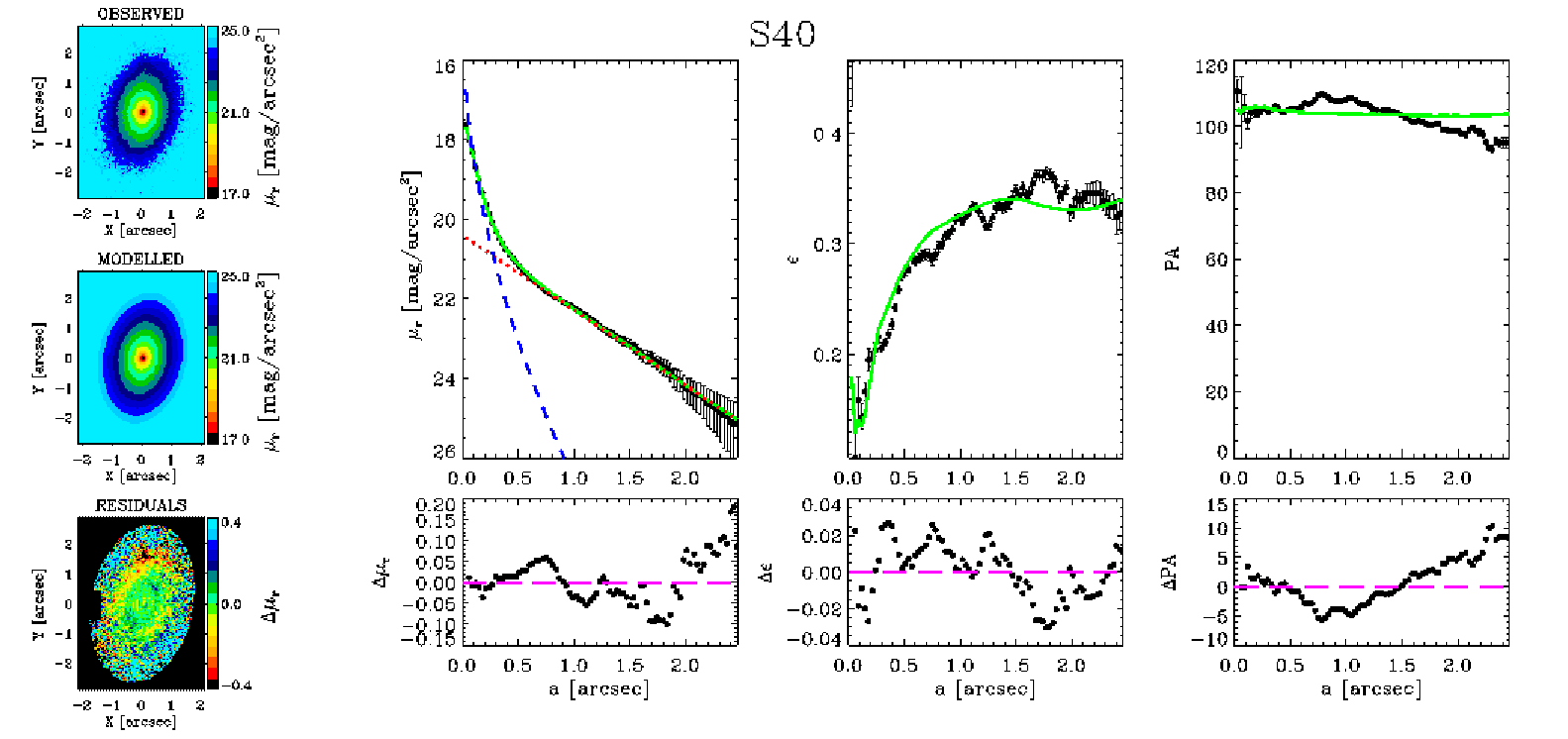}
\caption{As in Fig. \ref{fig:fit_S01} but for galaxy S40 (\sedisc\ model).}
\label{fig:fit_S40}
\end{figure*}

\begin{figure*}
\centering
\includegraphics[width=\textwidth, angle=0]{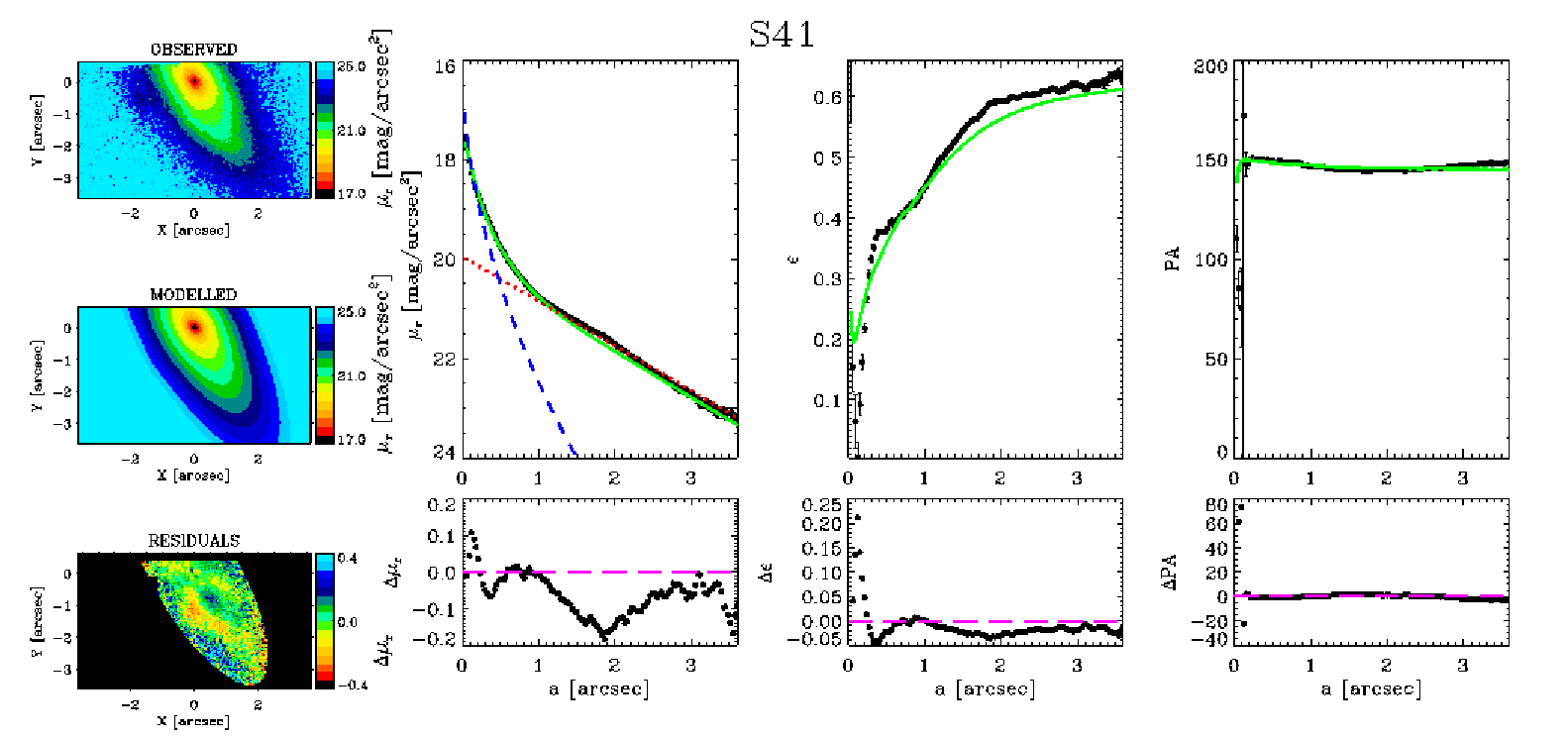}
\caption{As in Fig. \ref{fig:fit_S01} but for galaxy S41 (\sedisc\ model).}
\label{fig:fit_S41}
\end{figure*}

\clearpage

\begin{figure*}
\centering
\includegraphics[width=\textwidth, angle=0]{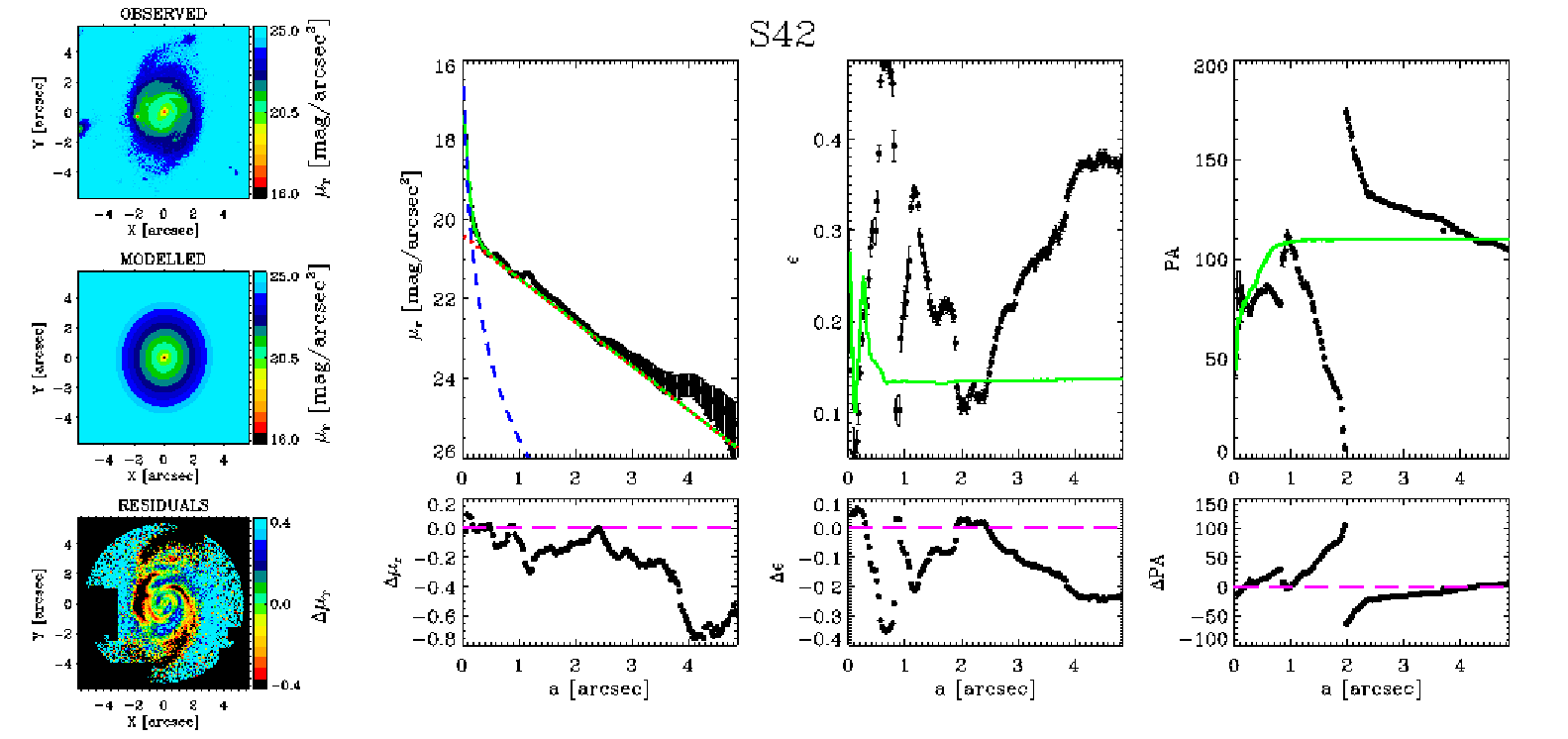}
\caption{As in Fig. \ref{fig:fit_S01} but for galaxy S42 (\sedisc\ model).}
\label{fig:fit_S42}
\end{figure*}

\begin{figure*}
\centering
\includegraphics[width=\textwidth, angle=0]{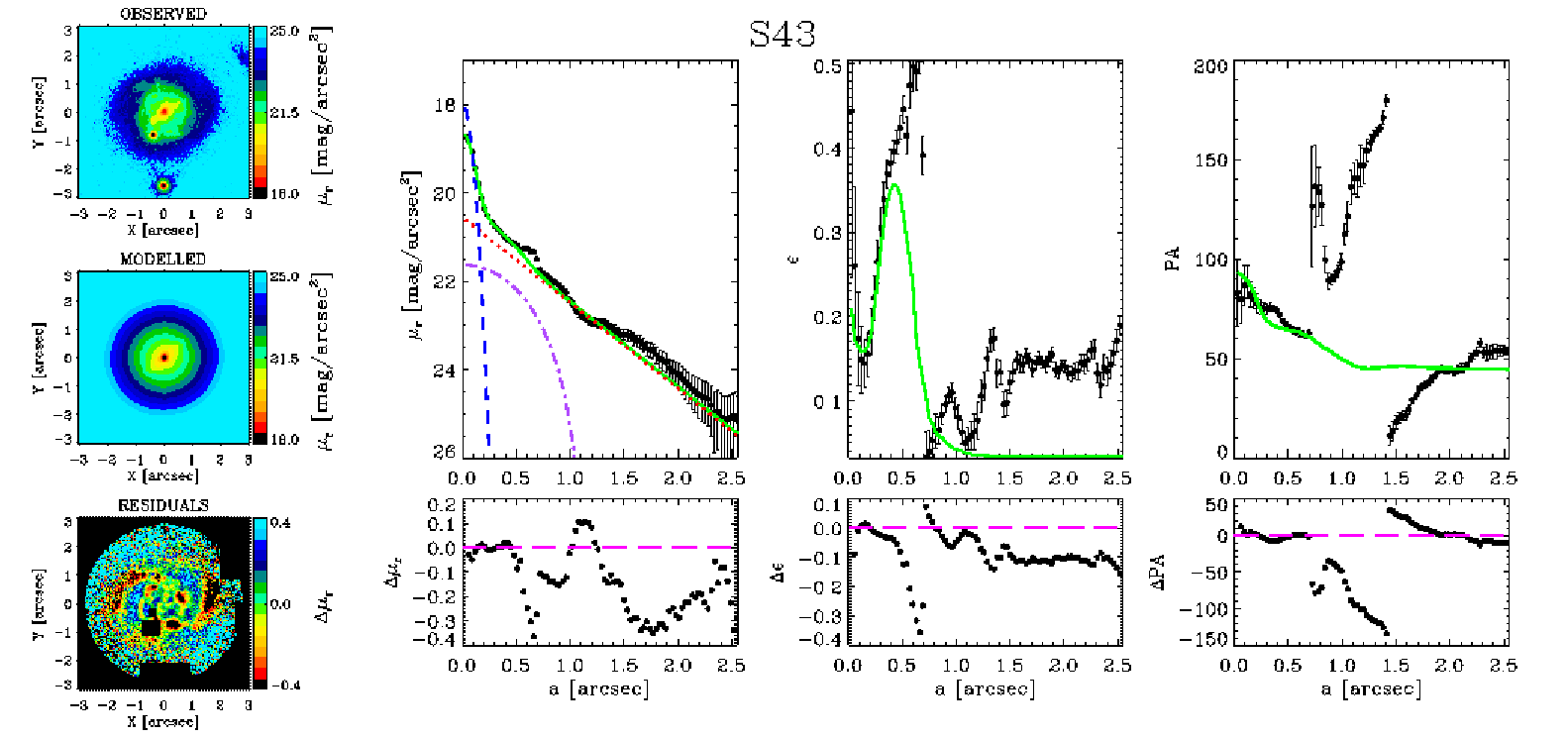}
\caption{As in Fig. \ref{fig:fit_S01} but for galaxy S43 fitted with a
  \sedibar\ model. The dashed-dotted purple line represents the intrinsic
  surface-brightness radial profile of the bar along its semi major
  axis.}
\label{fig:fit_S43}
\end{figure*}

\clearpage

\begin{figure*}
\centering
\includegraphics[width=\textwidth, angle=0]{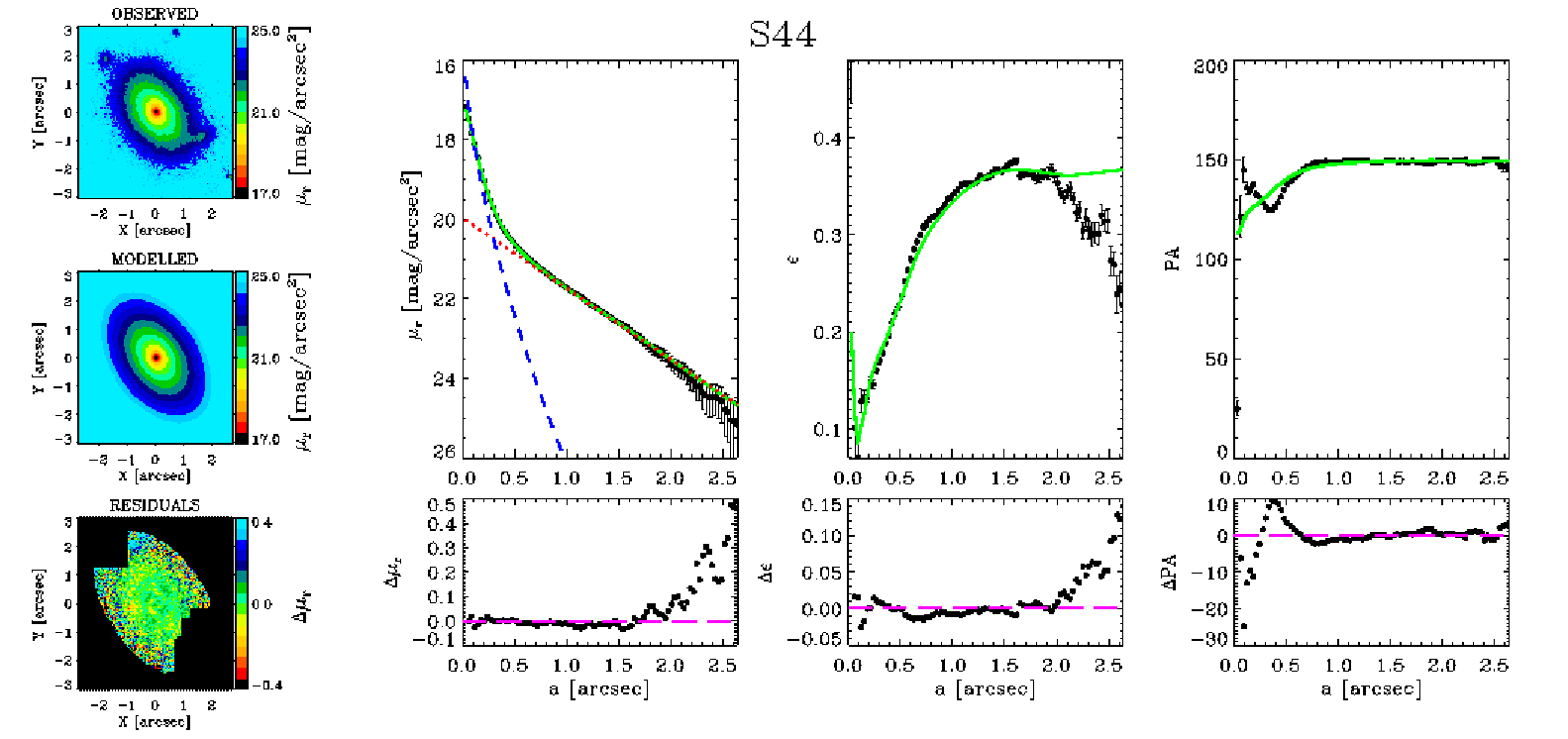}
\caption{As in Fig. \ref{fig:fit_S01} but for galaxy S10 (\sedisc\ model).}
\label{fig:fit_S44}
\end{figure*}

\begin{figure*}
\centering
\includegraphics[width=\textwidth, angle=0]{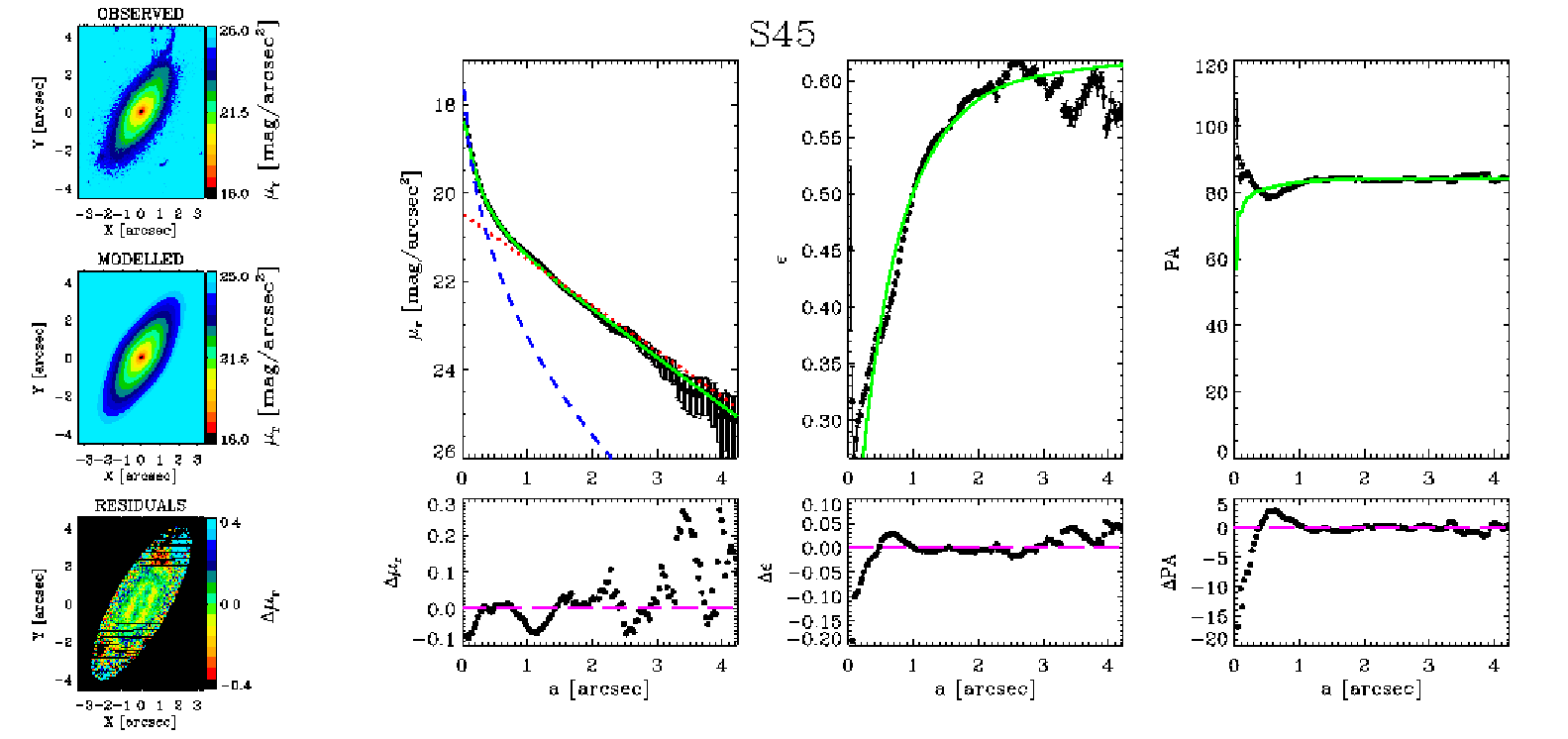}
\caption{As in Fig. \ref{fig:fit_S01} but for galaxy S45 (\sedisc\ model).}
\label{fig:fit_S45}
\end{figure*}

\clearpage

\begin{figure*}
\centering
\includegraphics[width=\textwidth, angle=0]{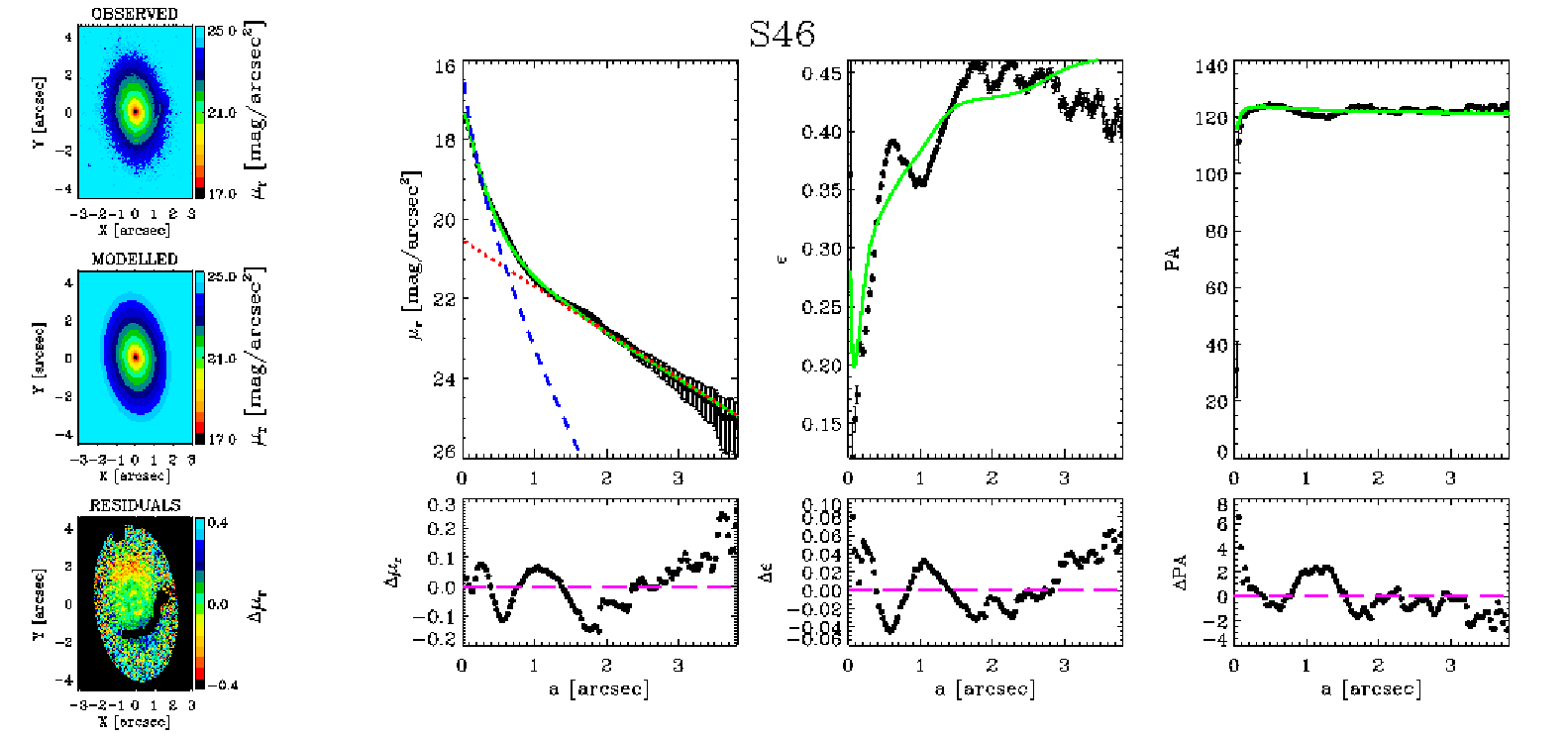}
\caption{As in Fig. \ref{fig:fit_S01} but for galaxy S46 (\sedisc\ model).}
\label{fig:fit_S46}
\end{figure*}

\begin{figure*}
\centering
\includegraphics[width=\textwidth, angle=0]{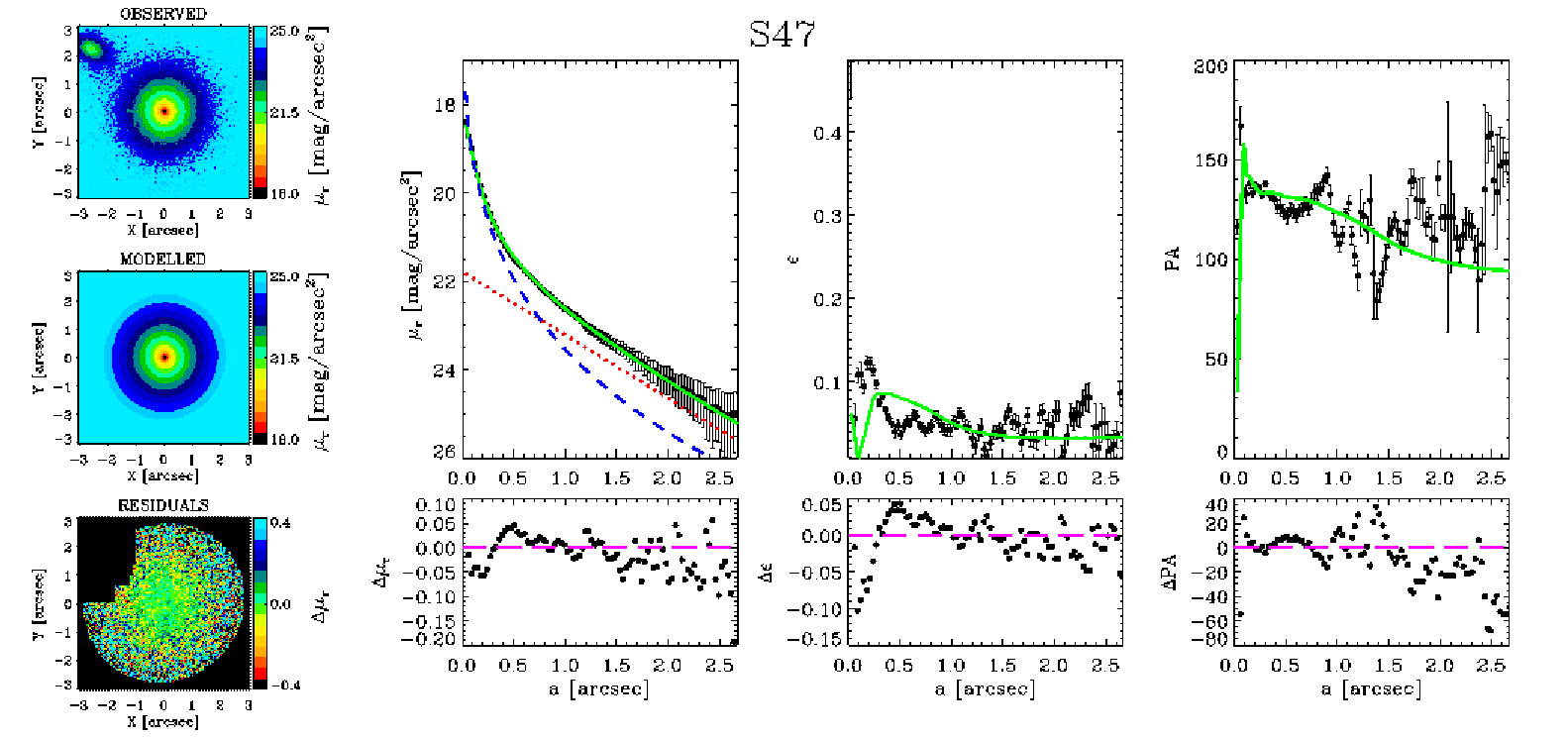}
\caption{As in Fig. \ref{fig:fit_S01} but for galaxy S47 (\sedisc\ model).}
\label{fig:fit_S47}
\end{figure*}

\clearpage

\begin{figure*}
\centering
\includegraphics[width=\textwidth, angle=0]{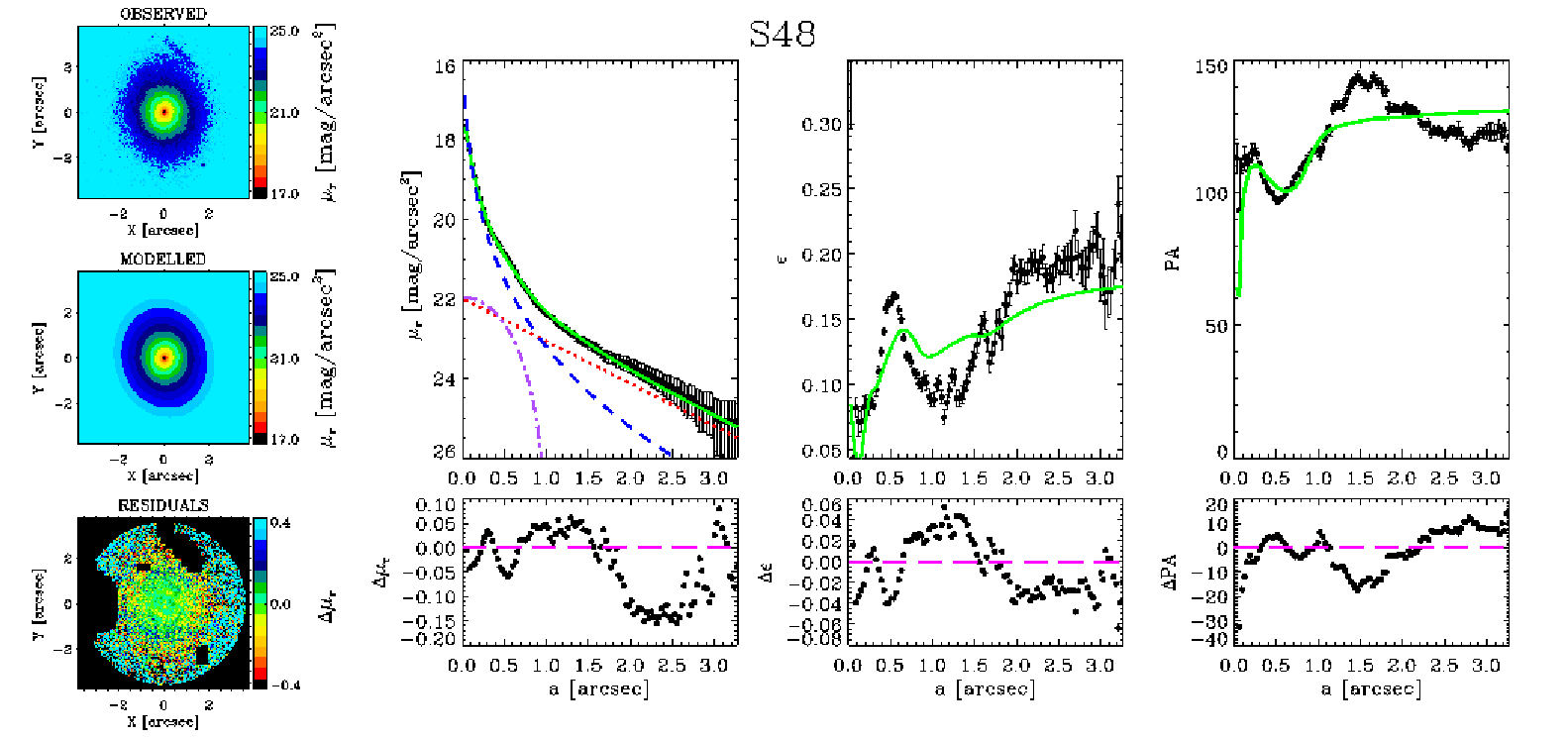}
\caption{As in Fig. \ref{fig:fit_S01} but for galaxy S48 fitted with a
  \sedibar\ model. The dashed-dotted purple line represents the intrinsic
  surface-brightness radial profile of the bar along its semi major
  axis.}
\label{fig:fit_S48}
\end{figure*}

\begin{figure*}
\centering
\includegraphics[width=\textwidth, angle=0]{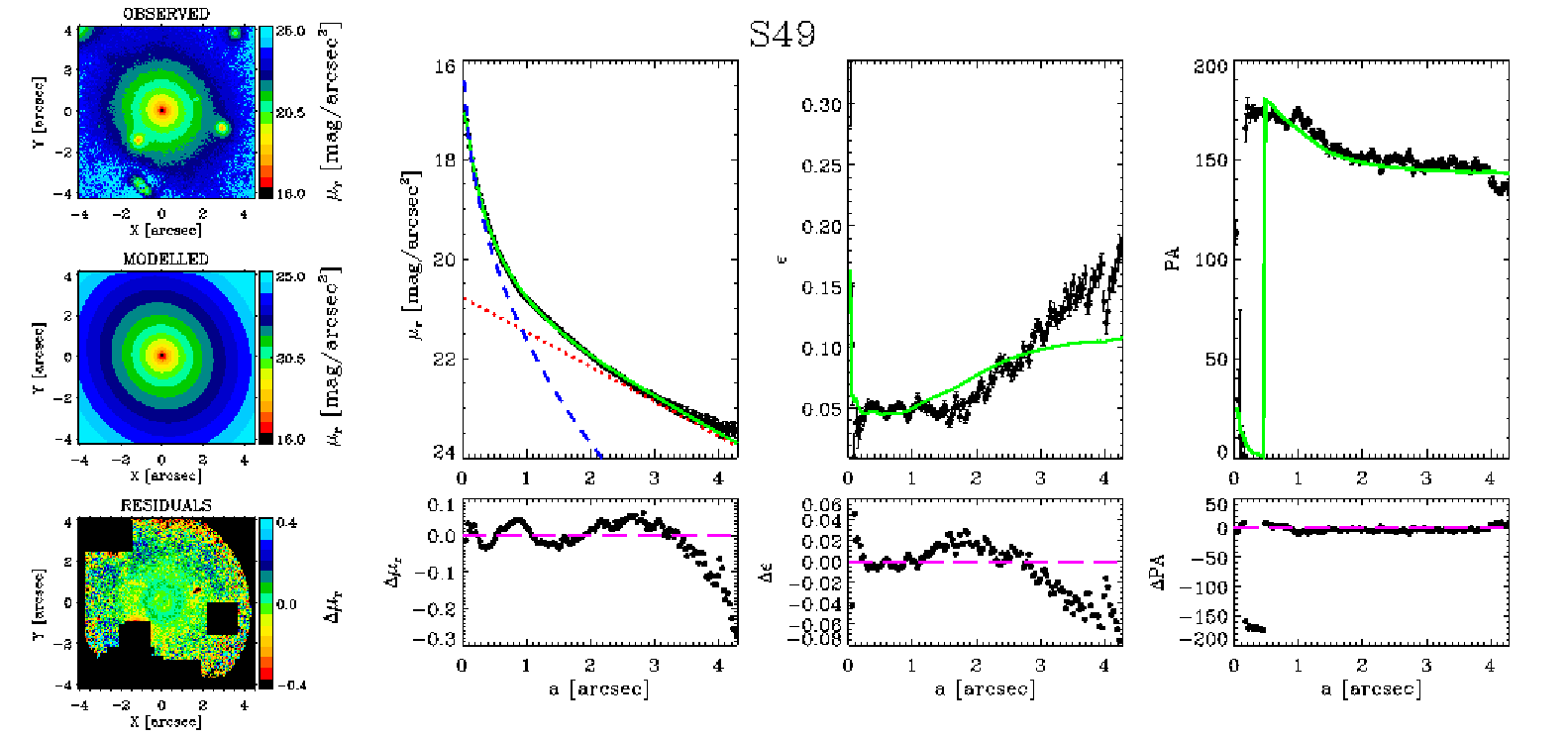}
\caption{As in Fig. \ref{fig:fit_S01} but for galaxy S49 (\sedisc\ model).}
\label{fig:fit_S49}
\end{figure*}

\clearpage

\begin{figure*}
\centering
\includegraphics[width=\textwidth, angle=0]{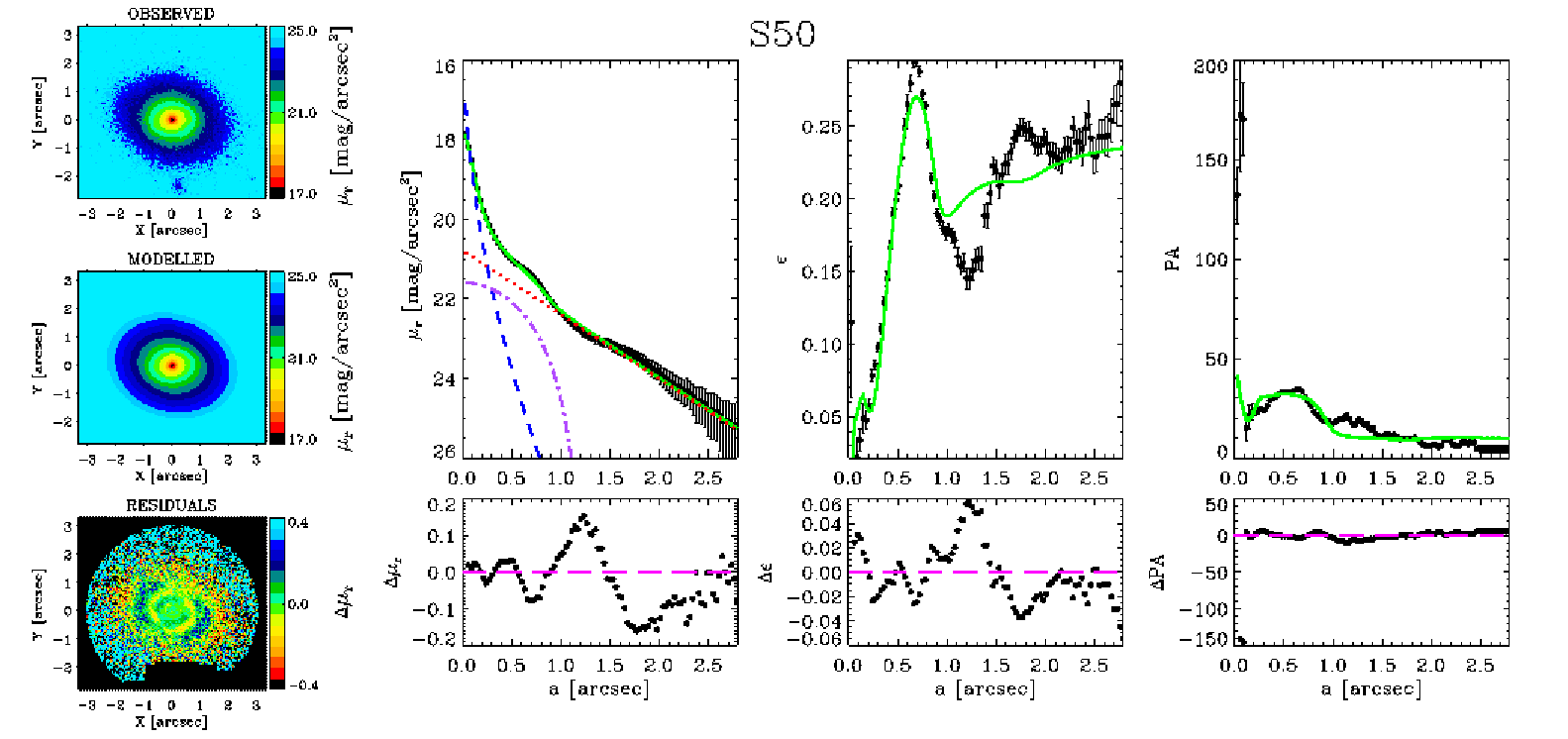}
\caption{As in Fig. \ref{fig:fit_S01} but for galaxy S50 fitted with a
  \sedibar\ model. The dashed-dotted purple line represents the intrinsic
  surface-brightness radial profile of the bar along its semi major
  axis.}
\label{fig:fit_S50}
\end{figure*}

\begin{figure*}
\centering
\includegraphics[width=\textwidth, angle=0]{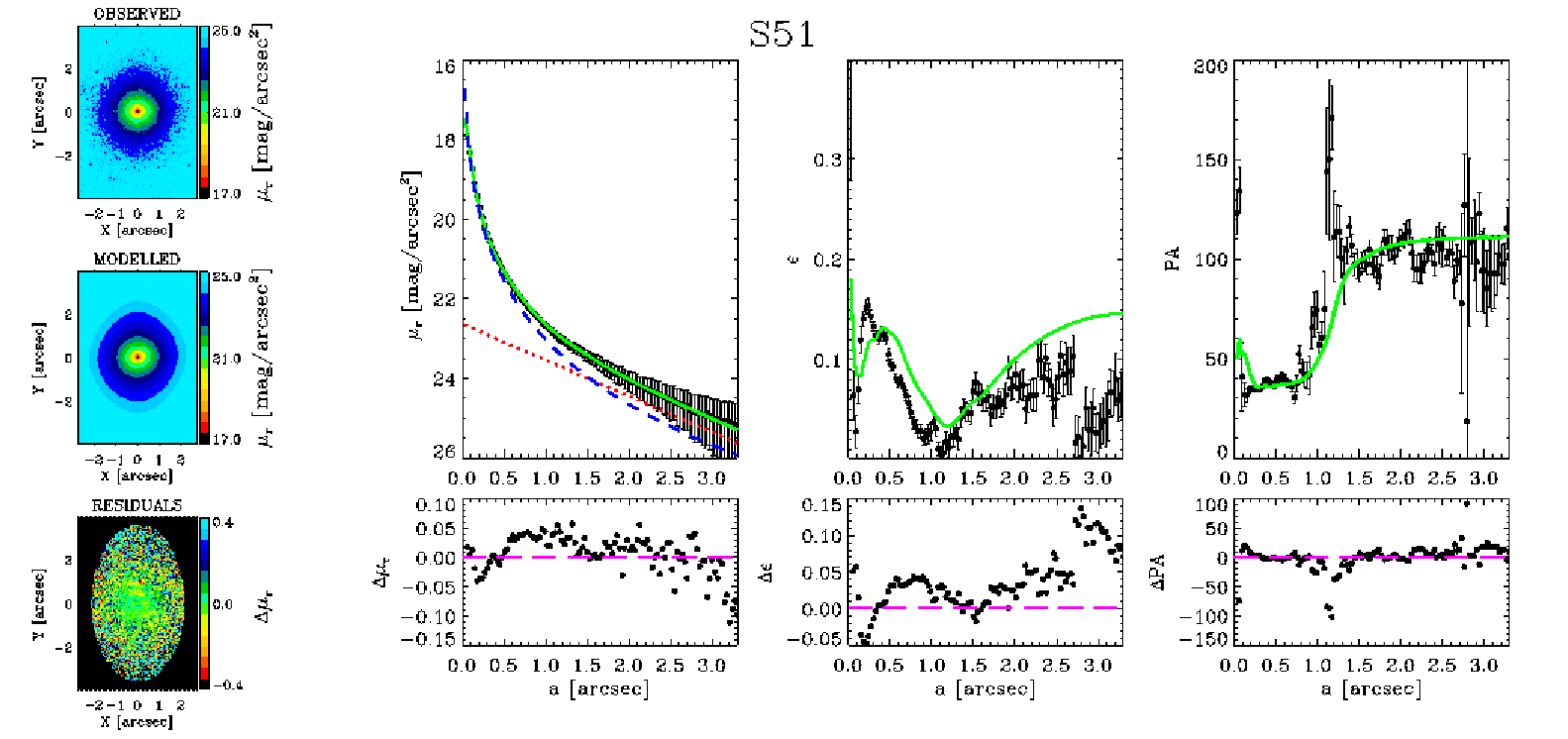}
\caption{As in Fig. \ref{fig:fit_S01} but for galaxy S51 (\sedisc\ model).}
\label{fig:fit_S51}
\end{figure*}

\newpage

\begin{figure*}
\centering
\includegraphics[width=\textwidth, angle=0]{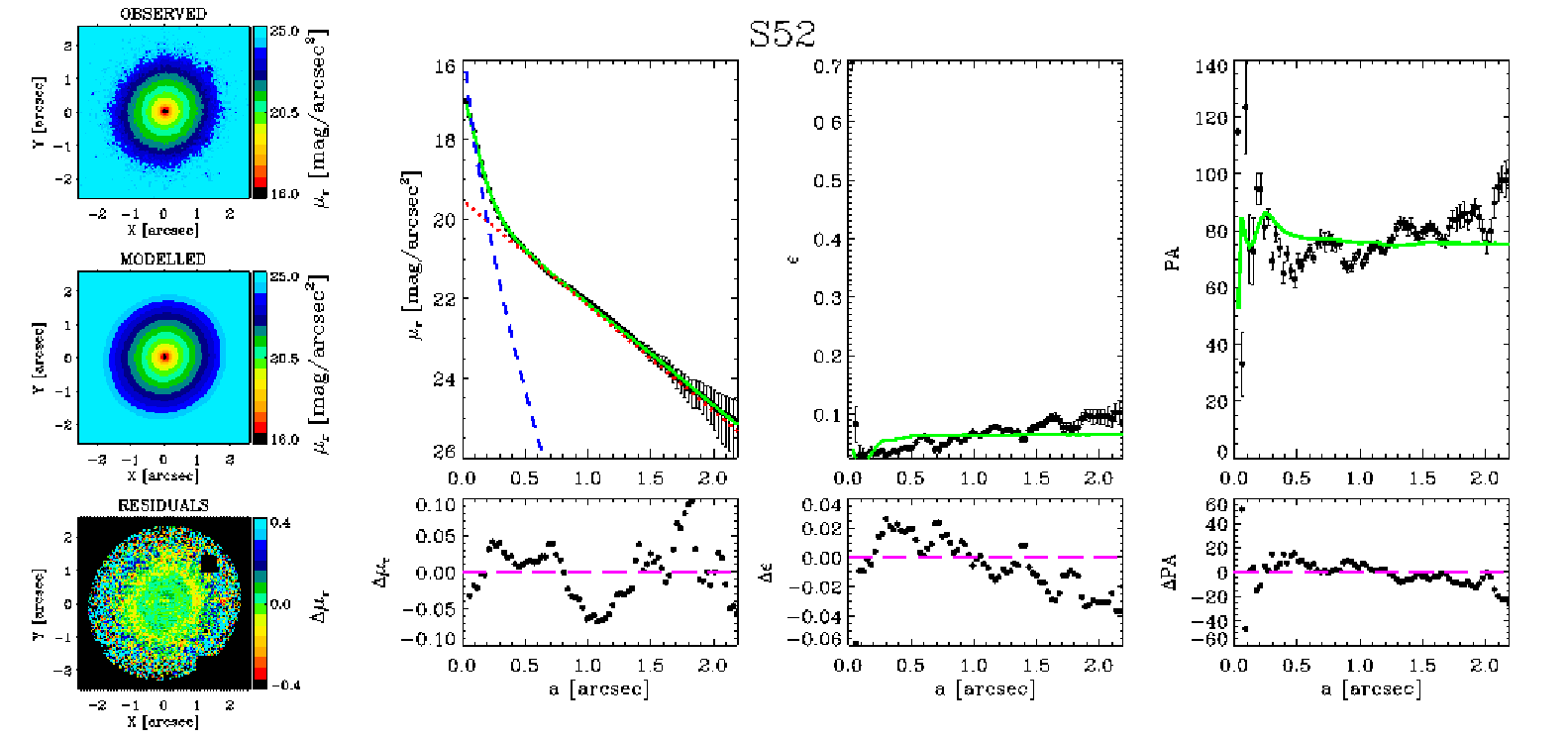}
\caption{As in Fig. \ref{fig:fit_S01} but for galaxy S52 (\sedisc\ model).}
\label{fig:fit_S52}
\end{figure*}

\begin{figure*}
\centering
\includegraphics[width=\textwidth, angle=0]{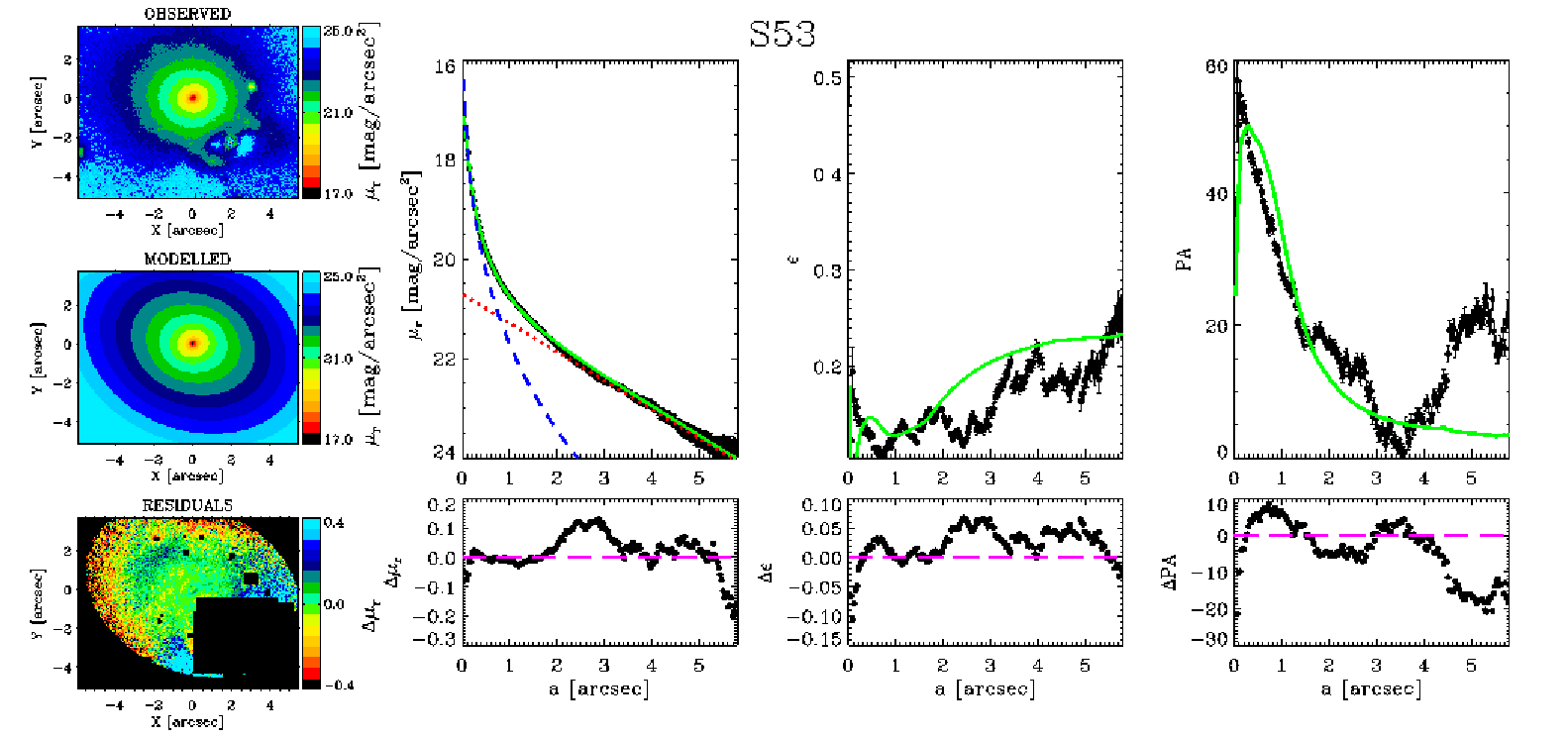}
\caption{As in Fig. \ref{fig:fit_S01} but for galaxy S53 (\sedisc\ model).}
\label{fig:fit_S53}
\end{figure*}.

\clearpage

\begin{figure*}
\centering
\includegraphics[width=\textwidth, angle=0]{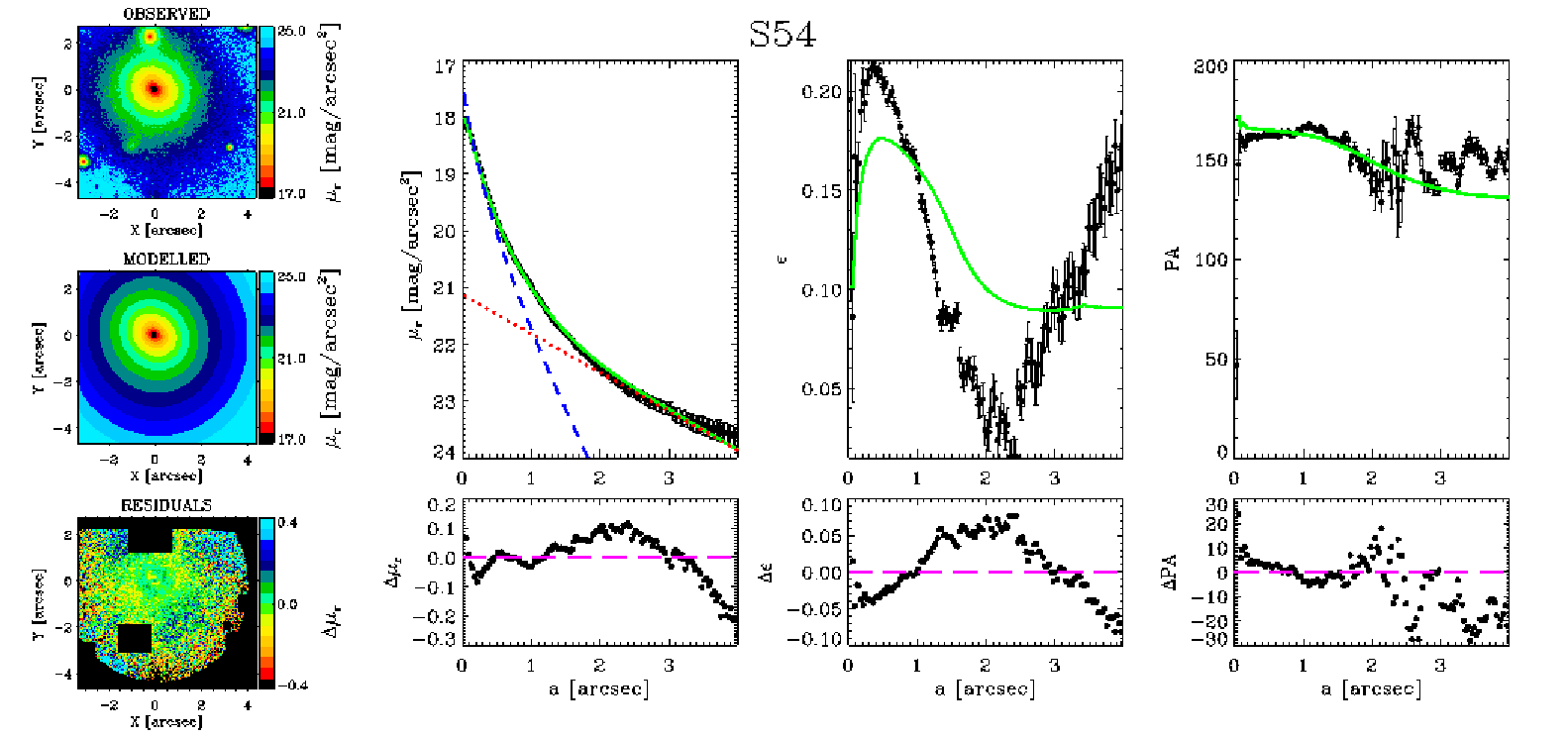}
\caption{As in Fig. \ref{fig:fit_S01} but for galaxy S54 (\sedisc\ model).}
\label{fig:fit_S54}
\end{figure*}

\begin{figure*}
\centering
\includegraphics[width=\textwidth, angle=0]{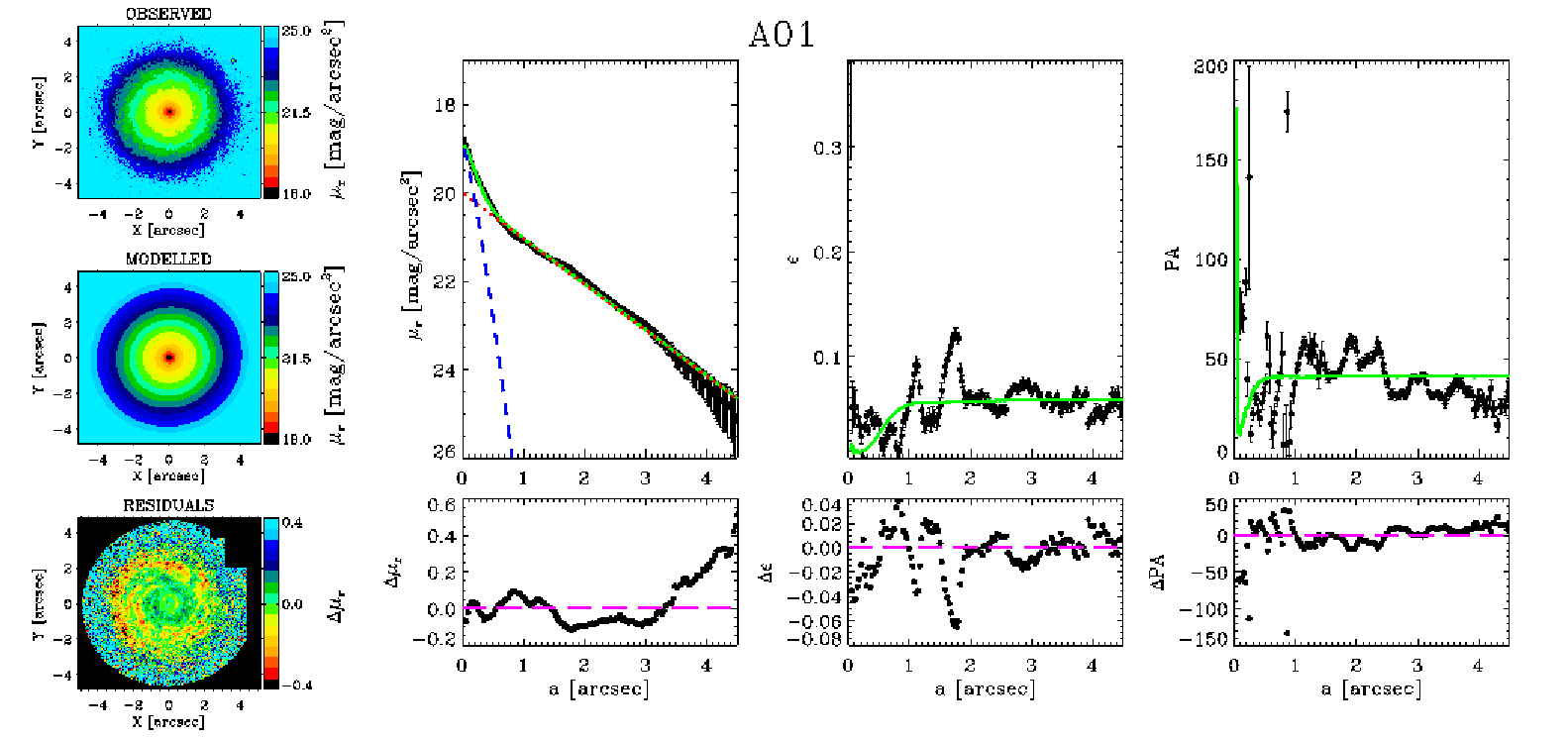}
\caption{As in Fig. \ref{fig:fit_S01} but for galaxy A01 (\sedisc\ model).}
\label{fig:fit_A01}
\end{figure*}

\clearpage

\begin{figure*}
\centering
\includegraphics[width=\textwidth, angle=0]{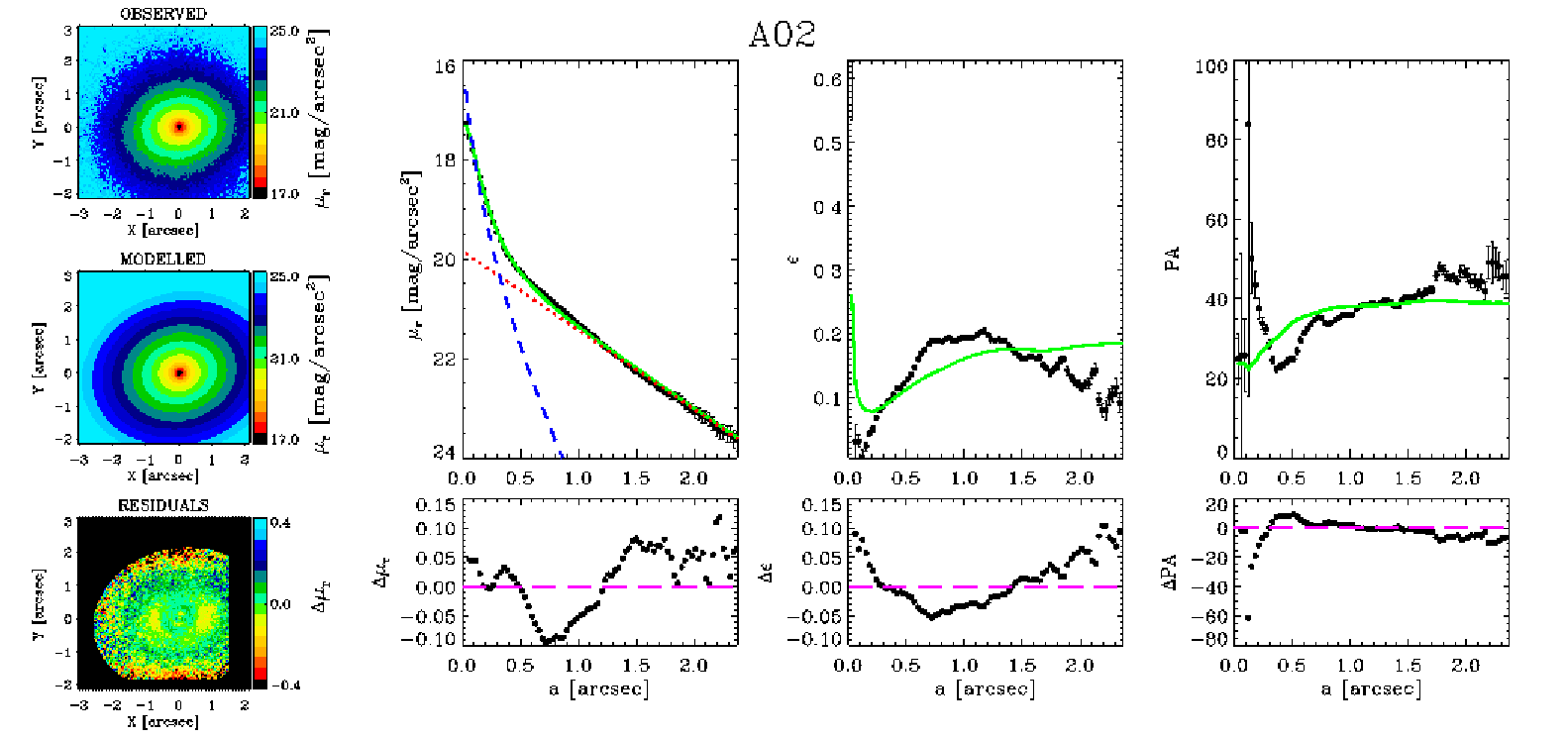}
\caption{As in Fig. \ref{fig:fit_S01} but for galaxy A02 (\sedisc\ model).}
\label{fig:fit_A02}
\end{figure*}

\begin{figure*}
\centering
\includegraphics[width=\textwidth, angle=0]{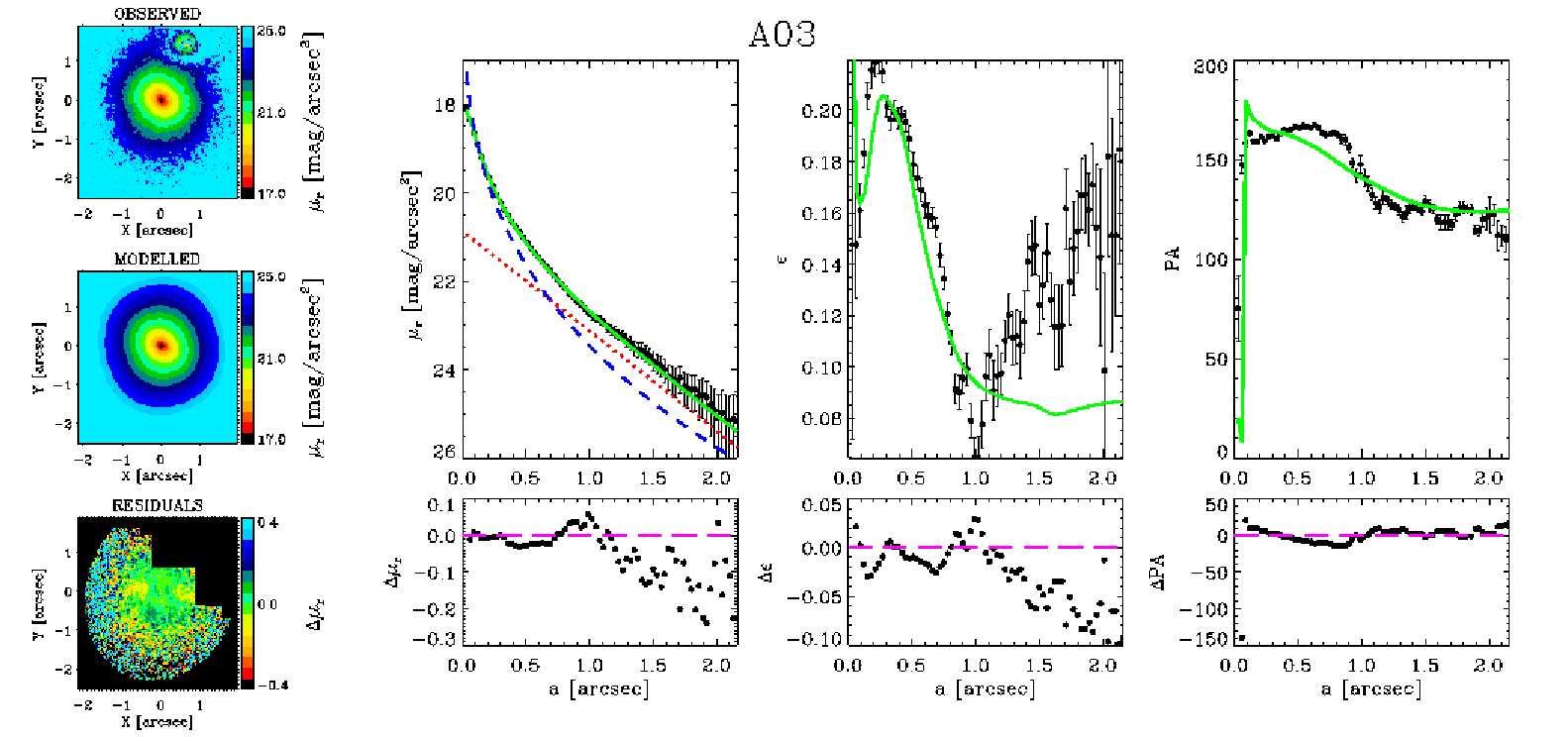}
\caption{As in Fig. \ref{fig:fit_S01} but for galaxy A03 (\sedisc\ model).}
\label{fig:fit_A03}
\end{figure*}

\clearpage

\begin{figure*}
\centering
\includegraphics[width=\textwidth, angle=0]{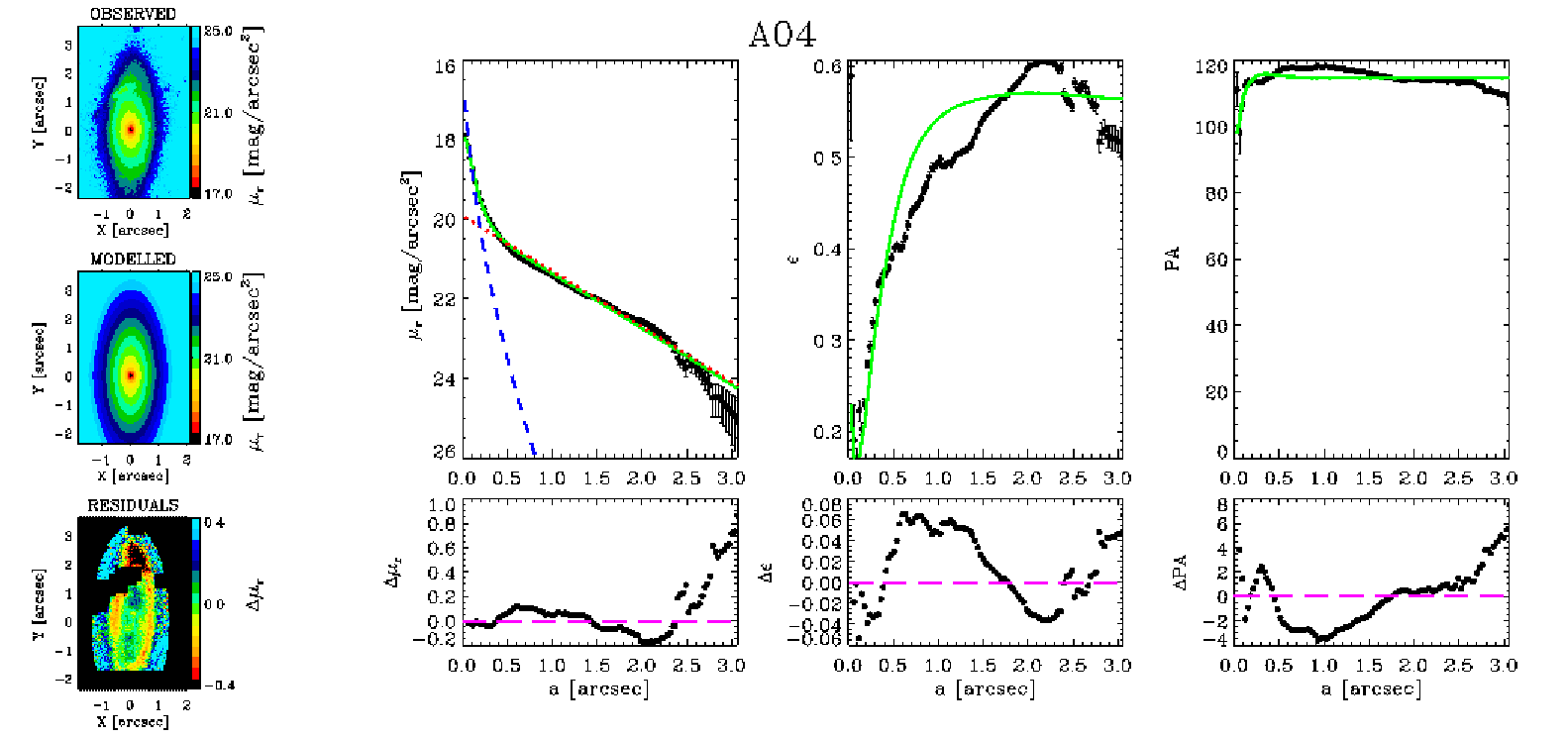}
\caption{As in Fig. \ref{fig:fit_S01} but for galaxy A04 (\sedisc\ model).}
\label{fig:fit_A04}
\end{figure*}

\begin{figure*}
\centering
\includegraphics[width=\textwidth, angle=0]{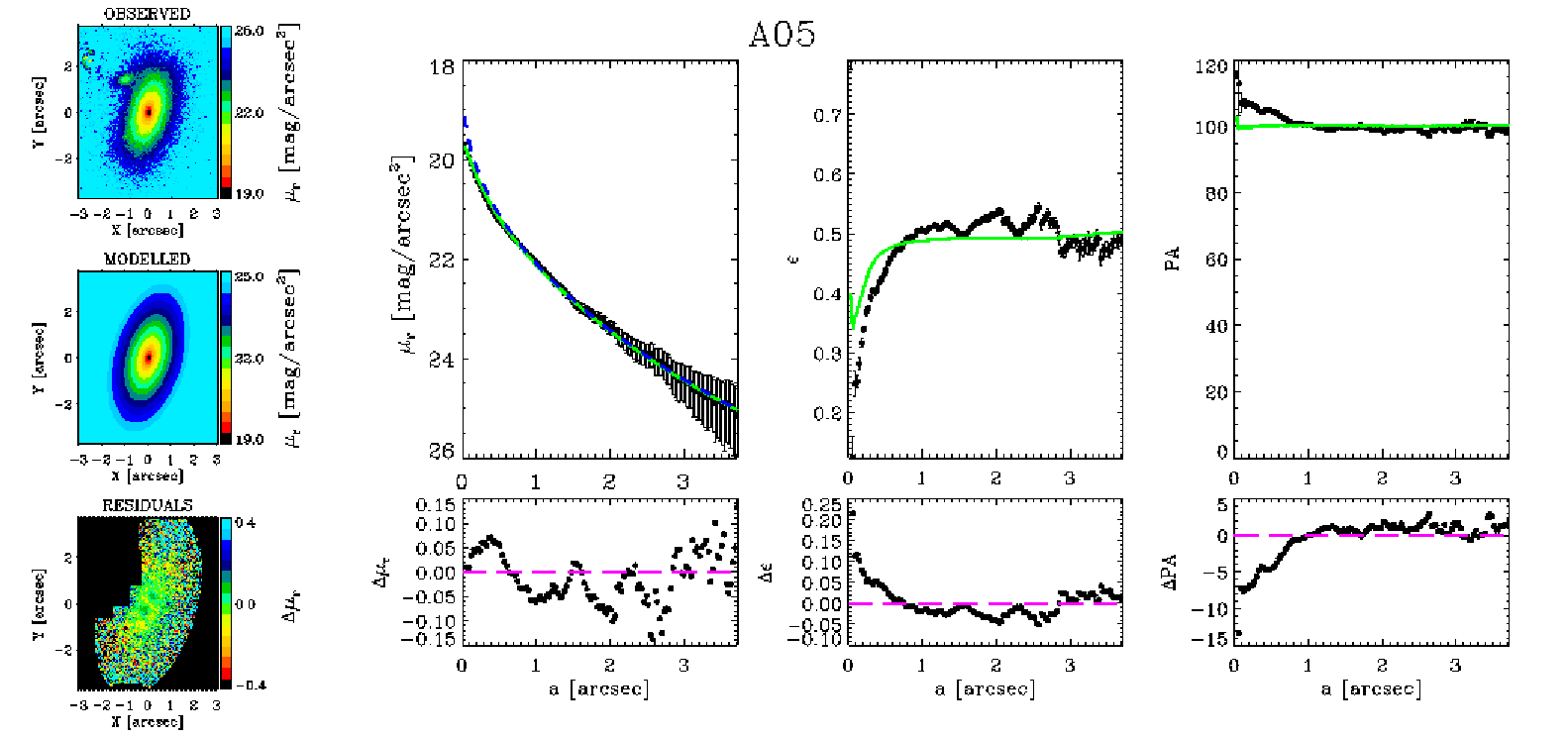}
\caption{As in Fig. \ref{fig:fit_S01} but for galaxy A05 fitted with a \sersic\ model).}
\label{fig:fit_A05}
\end{figure*}

\clearpage

\begin{figure*}
\centering
\includegraphics[width=\textwidth, angle=0]{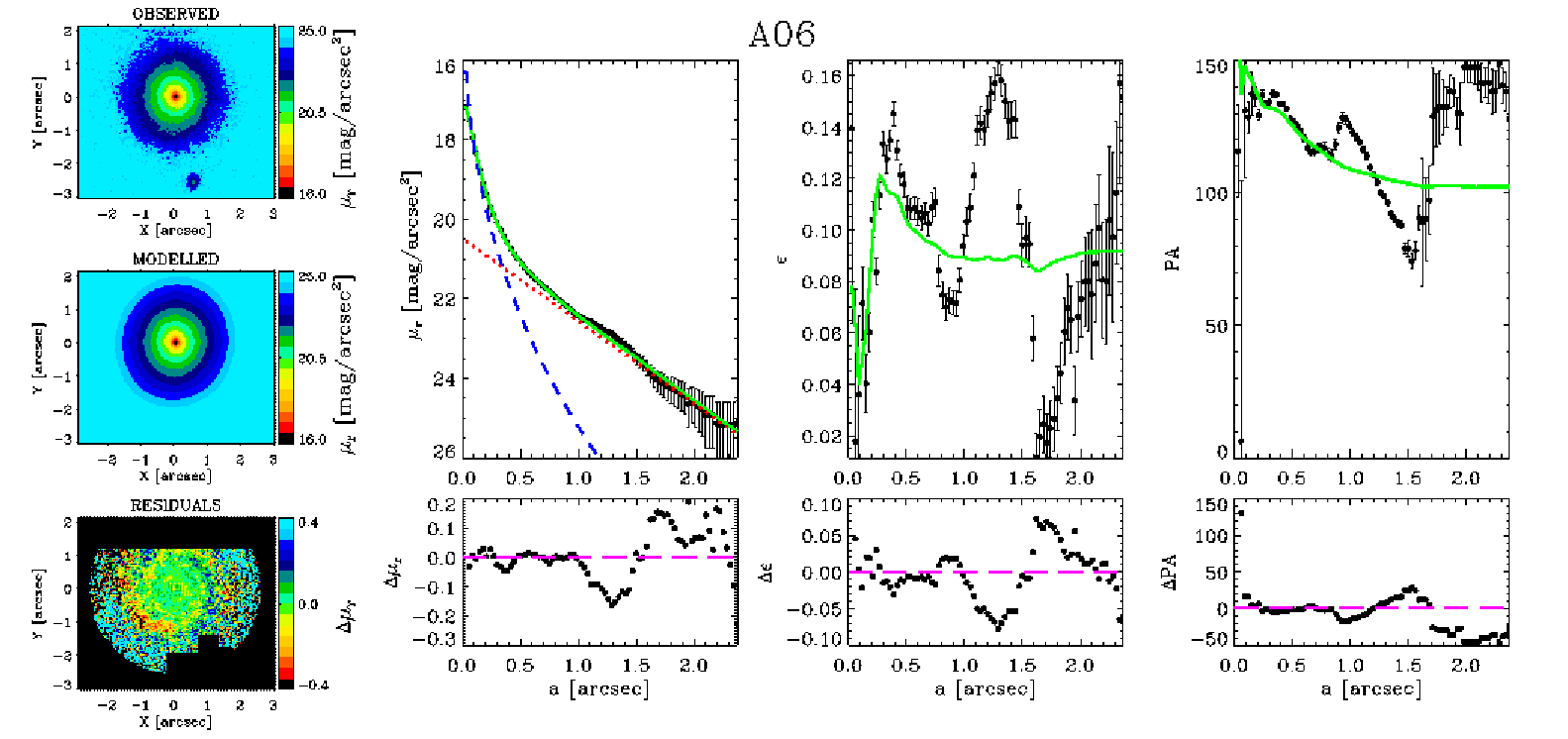}
\caption{As in Fig. \ref{fig:fit_S01} but for galaxy A06 (\sedisc\ model).}
\label{fig:fit_A06}
\end{figure*}

\begin{figure*}
\centering
\includegraphics[width=\textwidth, angle=0]{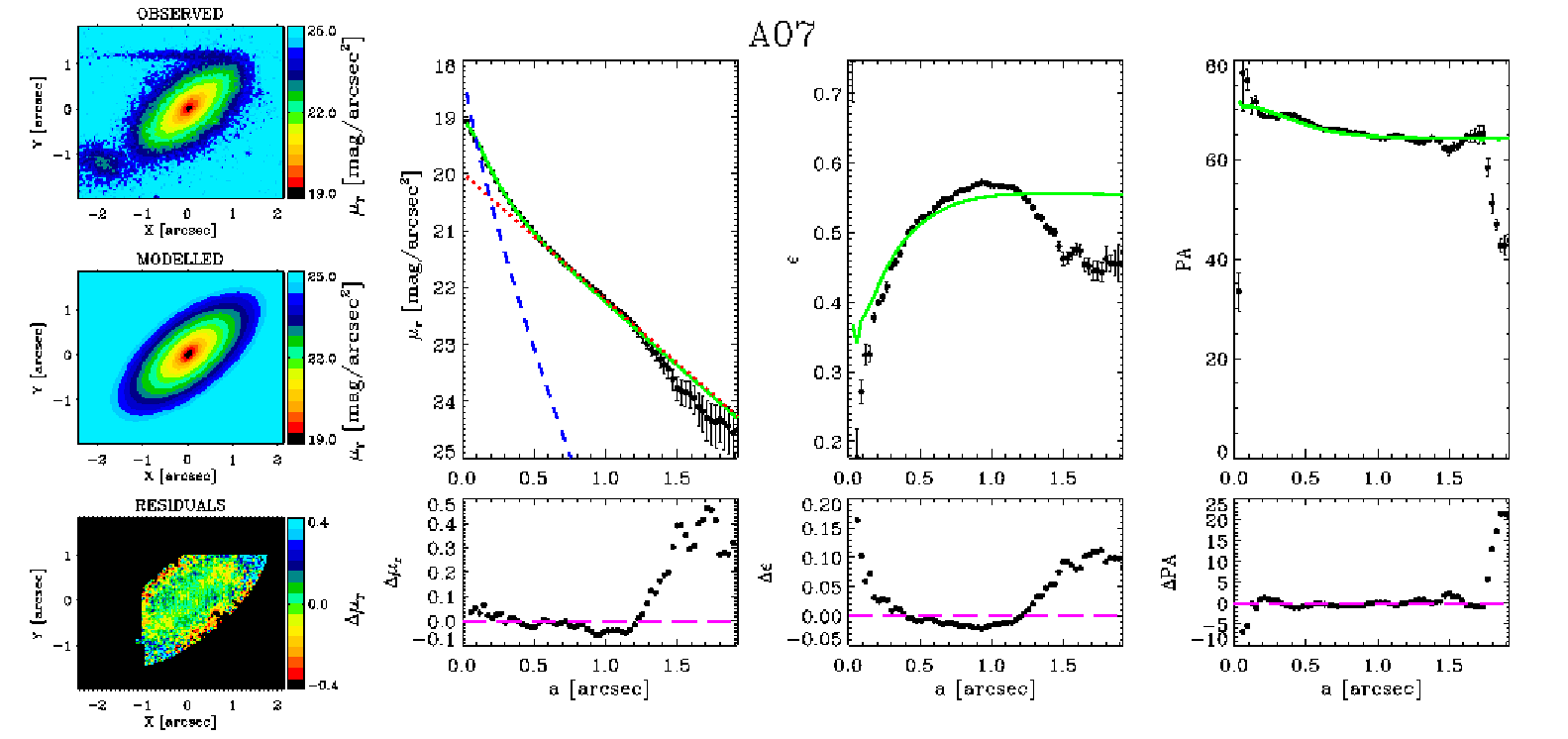}
\caption{As in Fig. \ref{fig:fit_S01} but for galaxy A07 (\sedisc\ model).}
\label{fig:fit_A07}
\end{figure*}

\clearpage

\begin{figure*}
\centering
\includegraphics[width=\textwidth, angle=0]{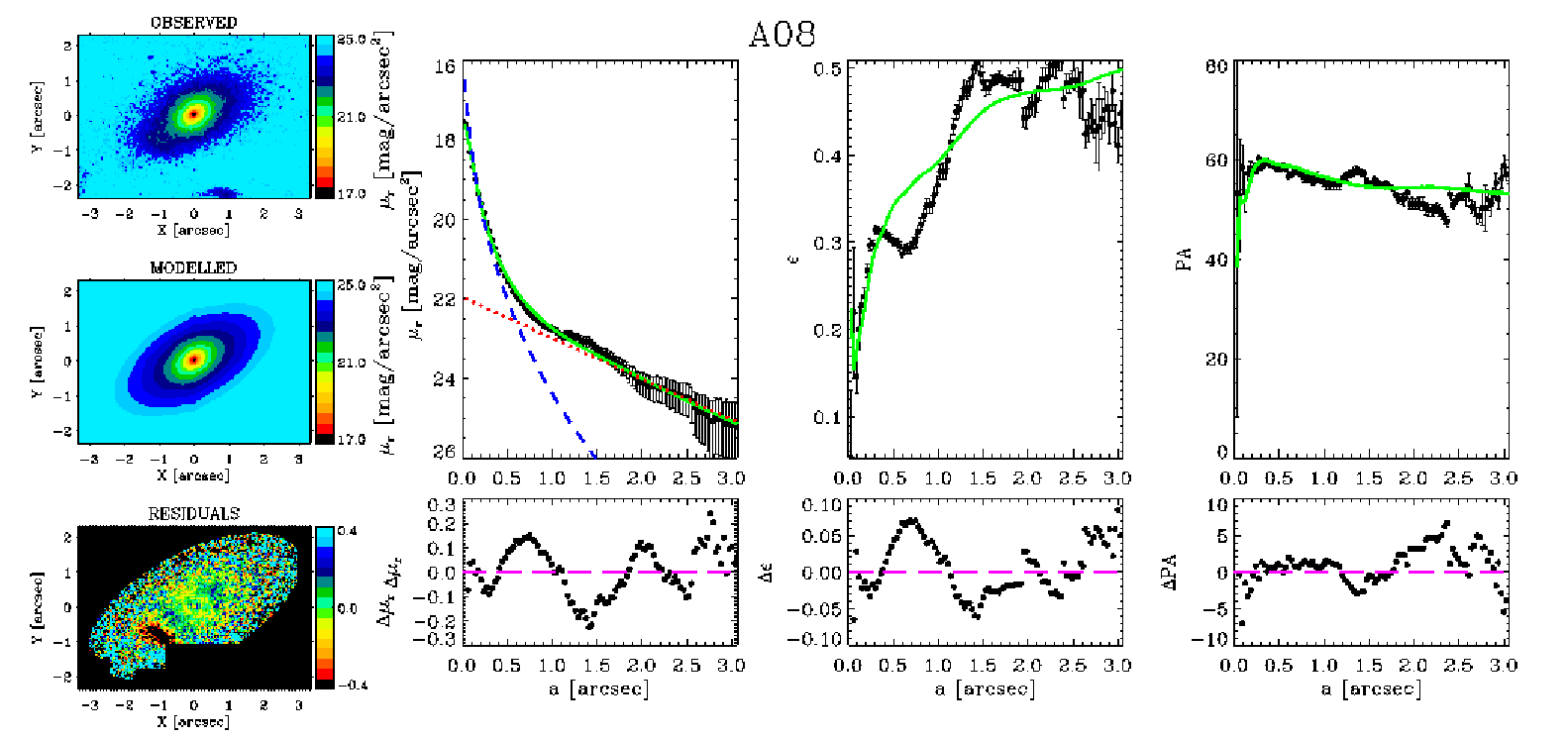}
\caption{As in Fig. \ref{fig:fit_S01} but for galaxy A08 (\sedisc\ model).}
\label{fig:fit_A08}
\end{figure*}

\begin{figure*}
\centering
\includegraphics[width=\textwidth, angle=0]{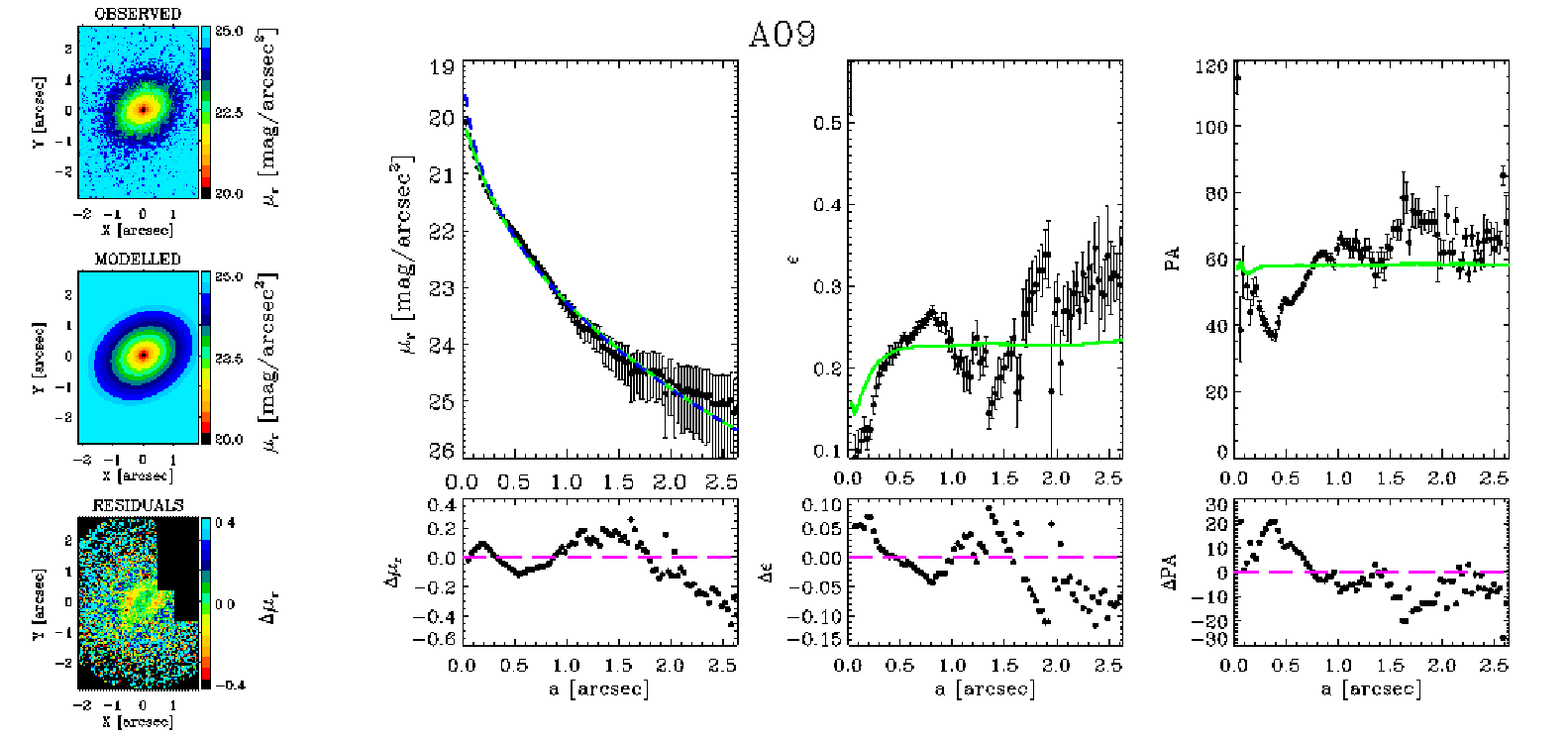}
\caption{As in Fig. \ref{fig:fit_S01} but for galaxy A09 fitted with a \sersic\ model.}
\label{fig:fit_A09}
\end{figure*}

\clearpage

\begin{figure*}
\centering
\includegraphics[width=\textwidth, angle=0]{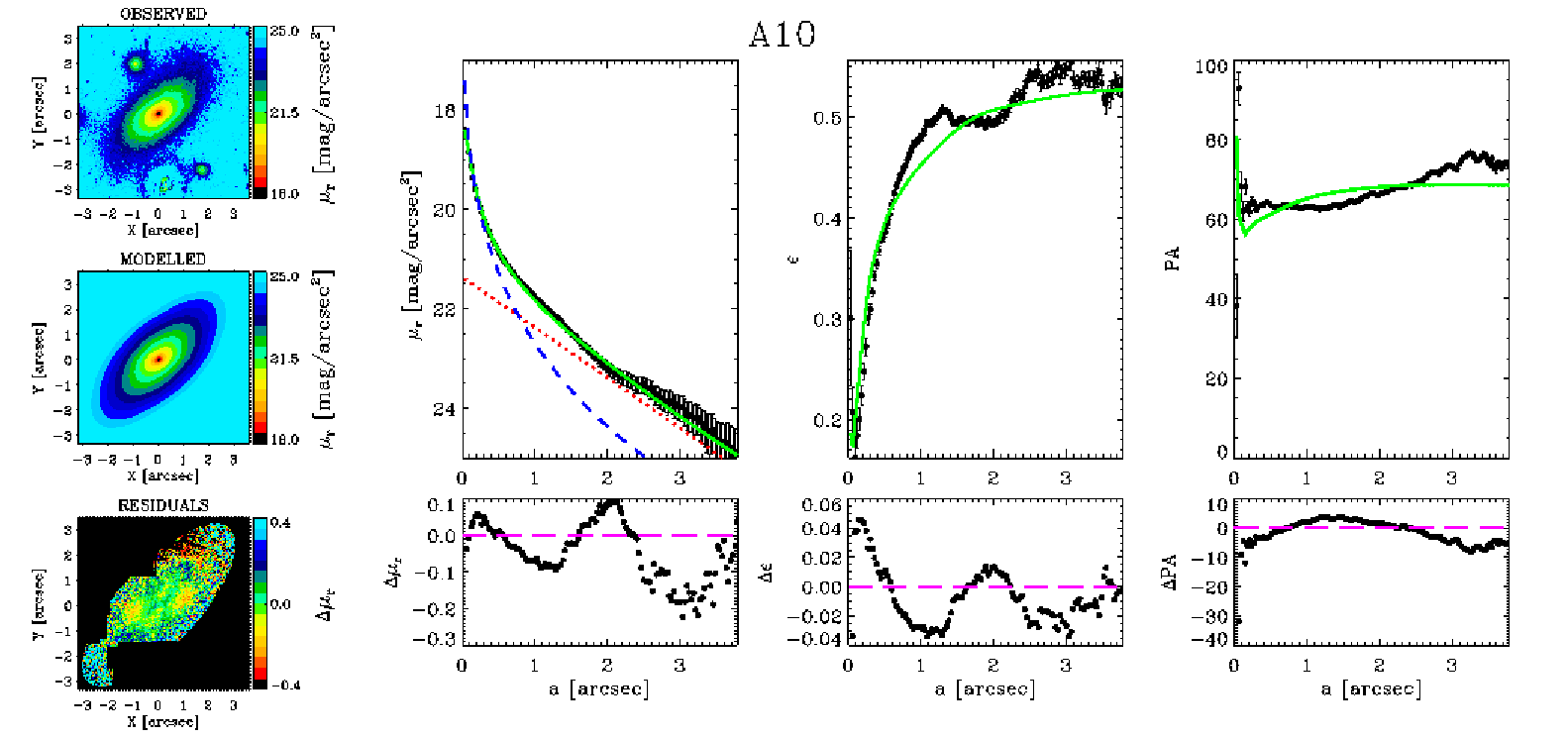}
\caption{As in Fig. \ref{fig:fit_S01} but for galaxy A10 (\sedisc\ model).}
\label{fig:fit_A10}
\end{figure*}

\begin{figure*}
\centering
\includegraphics[width=\textwidth, angle=0]{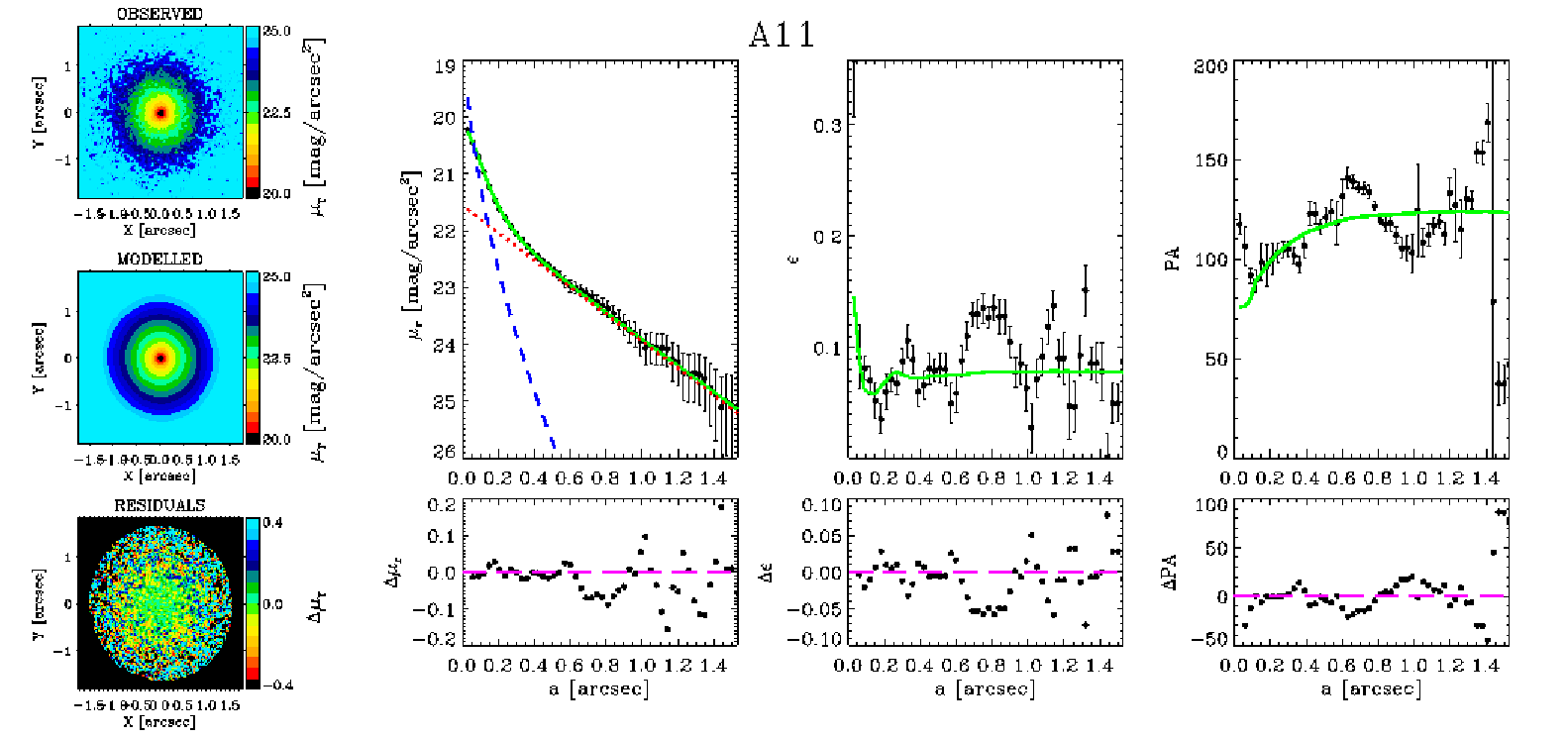}
\caption{As in Fig. \ref{fig:fit_S01} but for galaxy A11 (\sedisc\ model).}
\label{fig:fit_A11}
\end{figure*}

\begin{landscape}
\renewcommand{\tabcolsep}{3pt}
\begin{table}
\caption{Structural photometric parameters of the sample galaxies.}
\label{tab:phot_par}
\begin{scriptsize}
\begin{tabular}{clcccccccccccccc}
\hline
\noalign{\smallskip}
\multicolumn{1}{c}{ID} &
\multicolumn{1}{c}{model} &
\multicolumn{1}{c}{mag} &
\multicolumn{1}{c}{$\mu_e$} &
\multicolumn{1}{c}{$r_e$} &
\multicolumn{1}{c}{$n$} &
\multicolumn{1}{c}{$q_{\rm bulge}$} &
\multicolumn{1}{c}{$PA_{\rm bulge}$} &
\multicolumn{1}{c}{$\mu_{\rm 0,disc}$} &
\multicolumn{1}{c}{$h$} &
\multicolumn{1}{c}{$q_{\rm disc}$} &
\multicolumn{1}{c}{PA$_{\rm disc}$} &
\multicolumn{1}{c}{$\mu_{\rm 0,bar}$} &
\multicolumn{1}{c}{$r_{\rm bar}$} &
\multicolumn{1}{c}{$q_{\rm bar}$} &
\multicolumn{1}{c}{PA$_{\rm bar}$} \\
\multicolumn{1}{c}{} &
\multicolumn{1}{c}{} &
\multicolumn{1}{c}{(mag)} &
\multicolumn{1}{c}{$\left ( \frac{\rm mag}{\rm arcsec^2}\right )$} &
\multicolumn{1}{c}{(arcsec)} &
\multicolumn{1}{c}{} &
\multicolumn{1}{c}{} &
\multicolumn{1}{c}{($\degr$)} &
\multicolumn{1}{c}{$\left ( \frac{\rm mag}{\rm arcsec^2}\right )$} &
\multicolumn{1}{c}{(arcsec)} &
\multicolumn{1}{c}{} &
\multicolumn{1}{c}{($\degr$)} &
\multicolumn{1}{c}{$\left ( \frac{\rm mag}{\rm arcsec^2}\right )$} &
\multicolumn{1}{c}{(arcsec)} &
\multicolumn{1}{c}{} &
\multicolumn{1}{c}{($\degr$)} \\
\multicolumn{1}{c}{(1)} & 
\multicolumn{1}{c}{(2)} &   
\multicolumn{1}{c}{(3)} & 
\multicolumn{1}{c}{(4)} &   
\multicolumn{1}{c}{(5)} & 
\multicolumn{1}{c}{(6)} &
\multicolumn{1}{c}{(7)} & 
\multicolumn{1}{c}{(8)} &   
\multicolumn{1}{c}{(9)} & 
\multicolumn{1}{c}{(10)} &   
\multicolumn{1}{c}{(11)} & 
\multicolumn{1}{c}{(12)} &
\multicolumn{1}{c}{(13)} & 
\multicolumn{1}{c}{(14)} &
\multicolumn{1}{c}{(15)} &
\multicolumn{1}{c}{(16)} \\
\noalign{\smallskip}   
\hline
\noalign{\smallskip}       
\multicolumn{16}{c}{Spectroscopic sample}\\
\noalign{\smallskip}
\hline
%
 S01  & \sedisc &  19.01  &  $19.83\pm0.12$ & $0.22\pm0.017$ &  $ 3.59\pm0.12$ &  $0.71\pm0.003$ &  $  5.45\pm0.38$  &  $20.10\pm0.06$  &   $0.90\pm0.019$ &   $0.34\pm0.009$ &  $  3.18\pm0.27$ &   $...          $  &   $...          $ &   $ ...          $ & $ ...            $  \\
 S02  & \sedisc &  18.77  &  $18.22\pm0.12$ & $0.13\pm0.010$ &  $ 1.28\pm0.04$ &  $0.76\pm0.003$ &  $147.28\pm0.38$  &  $19.36\pm0.06$  &   $0.73\pm0.015$ &   $0.37\pm0.009$ &  $160.10\pm0.27$ &   $...          $  &   $...          $ &   $ ...          $ & $ ...            $  \\
 S03  & \sedibar&  17.54  &  $20.55\pm0.12$ & $0.50\pm0.037$ &  $ 1.45\pm0.07$ &  $0.73\pm0.002$ &  $117.68\pm0.28$  &  $21.59\pm0.05$  &   $2.26\pm0.033$ &   $0.95\pm0.006$ &  $135.12\pm0.47$ &   $21.59\pm 0.05$  &   $4.62\pm 0.069$ &   $ 0.26\pm 0.006$ & $ 101.82\pm 0.47 $  \\
 S04  & \sersic &  18.88  &  $24.92\pm0.05$ & $2.94\pm0.069$ &  $10.21\pm0.11$ &  $0.85\pm0.001$ &  $ 20.64\pm0.12$  &  $...         $  &   $...         $ &   $...         $ &  $...          $ &   $...          $  &   $...          $ &   $ ...          $ & $ ...            $  \\
 S05  & \sersic &  18.71  &  $21.62\pm0.05$ & $0.92\pm0.022$ &  $ 3.93\pm0.04$ &  $0.76\pm0.001$ &  $ 36.11\pm0.12$  &  $...         $  &   $...         $ &   $...         $ &  $...          $ &   $...          $  &   $...          $ &   $ ...          $ & $ ...            $  \\
 S06  & \sersic &  18.74  &  $20.46\pm0.05$ & $0.62\pm0.015$ &  $ 3.06\pm0.03$ &  $0.64\pm0.001$ &  $ 81.61\pm0.12$  &  $...         $  &   $...         $ &   $...         $ &  $...          $ &   $...          $  &   $...          $ &   $ ...          $ & $ ...            $  \\
 S07  & \sedisc &  20.10  &  $21.33\pm0.09$ & $0.36\pm0.020$ &  $ 1.11\pm0.04$ &  $0.38\pm0.004$ &  $129.93\pm0.47$  &  $21.37\pm0.05$  &   $1.13\pm0.021$ &   $0.32\pm0.008$ &  $129.38\pm0.51$ &   $...          $  &   $...          $ &   $ ...          $ & $ ...            $  \\
 S08  & \sedisc &  18.79  &  $20.31\pm0.12$ & $0.36\pm0.027$ &  $ 3.21\pm0.11$ &  $0.78\pm0.003$ &  $104.17\pm0.38$  &  $21.63\pm0.06$  &   $1.10\pm0.023$ &   $0.88\pm0.009$ &  $153.04\pm0.27$ &   $...          $  &   $...          $ &   $ ...          $ & $ ...            $  \\
 S09  & \sedisc &  18.53  &  $21.37\pm0.12$ & $0.64\pm0.049$ &  $ 5.62\pm0.19$ &  $0.79\pm0.003$ &  $ 63.50\pm0.38$  &  $22.61\pm0.06$  &   $1.73\pm0.036$ &   $0.86\pm0.009$ &  $147.45\pm0.27$ &   $...          $  &   $...          $ &   $ ...          $ & $ ...            $  \\
 S10  & \sedisc &  18.11  &  $18.95\pm0.10$ & $0.24\pm0.013$ &  $ 2.05\pm0.06$ &  $0.97\pm0.002$ &  $132.62\pm0.28$  &  $20.52\pm0.05$  &   $0.96\pm0.014$ &   $0.93\pm0.006$ &  $126.46\pm0.47$ &   $...          $  &   $...          $ &   $ ...          $ & $ ...            $  \\
 S11  & \sedisc &  19.23  &  $19.17\pm0.07$ & $0.13\pm0.004$ &  $ 3.05\pm0.12$ &  $0.87\pm0.004$ &  $129.27\pm0.57$  &  $19.49\pm0.01$  &   $0.67\pm0.002$ &   $0.33\pm0.001$ &  $119.78\pm0.12$ &   $...          $  &   $...          $ &   $ ...          $ & $ ...            $  \\
 S12  & \sedibar&  18.28  &  $19.76\pm0.12$ & $0.26\pm0.020$ &  $ 2.18\pm0.11$ &  $0.91\pm0.002$ &  $146.62\pm0.28$  &  $21.50\pm0.05$  &   $1.43\pm0.021$ &   $0.96\pm0.006$ &  $141.21\pm0.47$ &   $22.45\pm 0.05$  &   $3.78\pm 0.056$ &   $ 0.35\pm 0.006$ & $  87.04\pm 0.47 $  \\
 S13  & \sedisc &  18.65  &  $20.18\pm0.12$ & $0.39\pm0.029$ &  $ 2.71\pm0.09$ &  $0.54\pm0.003$ &  $129.33\pm0.38$  &  $20.35\pm0.06$  &   $0.92\pm0.019$ &   $0.57\pm0.009$ &  $141.66\pm0.27$ &   $...          $  &   $...          $ &   $ ...          $ & $ ...            $  \\
 S14  & \sedisc &  18.88  &  $19.33\pm0.12$ & $0.23\pm0.018$ &  $ 2.44\pm0.08$ &  $0.89\pm0.003$ &  $ 93.02\pm0.38$  &  $21.81\pm0.06$  &   $1.08\pm0.023$ &   $0.88\pm0.009$ &  $102.83\pm0.27$ &   $...          $  &   $...          $ &   $ ...          $ & $ ...            $  \\
 S15  & \sedibar&  19.65  &  $18.95\pm0.11$ & $0.09\pm0.007$ &  $ 1.52\pm0.09$ &  $0.76\pm0.004$ &  $ 15.53\pm0.47$  &  $20.85\pm0.05$  &   $0.76\pm0.014$ &   $0.60\pm0.008$ &  $ 25.32\pm0.52$ &   $21.29\pm 0.05$  &   $1.32\pm 0.025$ &   $ 0.32\pm 0.008$ & $  43.58\pm 0.52 $  \\
 S16  & \sedisc &  18.39  &  $20.50\pm0.10$ & $0.47\pm0.026$ &  $ 2.74\pm0.08$ &  $0.75\pm0.002$ &  $116.49\pm0.28$  &  $21.35\pm0.05$  &   $1.23\pm0.018$ &   $0.87\pm0.006$ &  $ 35.48\pm0.47$ &   $...          $  &   $...          $ &   $ ...          $ & $ ...            $  \\
 S18  & \sedisc &  15.82  &  $21.63\pm0.00$ & $2.52\pm0.014$ &  $ 0.96\pm0.01$ &  $0.83\pm0.001$ &  $157.06\pm0.17$  &  $22.40\pm0.03$  &   $7.70\pm0.169$ &   $0.81\pm0.002$ &  $152.35\pm0.26$ &   $...          $  &   $...          $ &   $ ...          $ & $ ...            $  \\
 S19  & \sedisc &  19.50  &  $19.12\pm0.12$ & $0.15\pm0.011$ &  $ 2.08\pm0.07$ &  $0.87\pm0.003$ &  $ 79.39\pm0.38$  &  $20.81\pm0.06$  &   $0.60\pm0.013$ &   $0.81\pm0.009$ &  $ 98.27\pm0.27$ &   $...          $  &   $...          $ &   $ ...          $ & $ ...            $  \\
 S20  & \sedibar&  19.75  &  $18.80\pm0.11$ & $0.09\pm0.006$ &  $ 1.07\pm0.06$ &  $0.70\pm0.004$ &  $  7.80\pm0.47$  &  $20.38\pm0.05$  &   $0.63\pm0.012$ &   $0.52\pm0.008$ &  $  5.31\pm0.52$ &   $21.60\pm 0.05$  &   $1.20\pm 0.023$ &   $ 0.50\pm 0.008$ & $ 136.78\pm 0.52 $  \\
 S21  & \sedibar&  18.42  &  $19.11\pm0.12$ & $0.14\pm0.011$ &  $ 1.27\pm0.06$ &  $0.82\pm0.002$ &  $154.07\pm0.28$  &  $20.36\pm0.05$  &   $1.04\pm0.015$ &   $0.67\pm0.006$ &  $ 46.45\pm0.47$ &   $21.83\pm 0.05$  &   $3.09\pm 0.046$ &   $ 0.31\pm 0.006$ & $  74.02\pm 0.47 $  \\
 S22  & \sedisc &  18.63  &  $19.66\pm0.12$ & $0.30\pm0.023$ &  $ 2.61\pm0.09$ &  $0.91\pm0.003$ &  $ 16.50\pm0.38$  &  $21.48\pm0.06$  &   $1.15\pm0.024$ &   $0.69\pm0.009$ &  $ 21.69\pm0.27$ &   $...          $  &   $...          $ &   $ ...          $ & $ ...            $  \\
 S23  & \sedisc &  19.55  &  $19.66\pm0.09$ & $0.26\pm0.014$ &  $ 2.15\pm0.07$ &  $0.60\pm0.004$ &  $ 60.44\pm0.47$  &  $21.04\pm0.05$  &   $0.67\pm0.013$ &   $0.54\pm0.008$ &  $ 29.34\pm0.51$ &   $...          $  &   $...          $ &   $ ...          $ & $ ...            $  \\
 S24  & \sedisc &  18.38  &  $19.82\pm0.10$ & $0.29\pm0.016$ &  $ 2.01\pm0.06$ &  $0.93\pm0.002$ &  $121.71\pm0.28$  &  $20.34\pm0.05$  &   $0.85\pm0.013$ &   $0.87\pm0.006$ &  $122.38\pm0.47$ &   $...          $  &   $...          $ &   $ ...          $ & $ ...            $  \\
 S25  & \sedibar&  18.82  &  $21.42\pm0.15$ & $0.50\pm0.051$ &  $ 5.32\pm0.31$ &  $0.92\pm0.003$ &  $ 13.49\pm0.38$  &  $21.78\pm0.06$  &   $1.09\pm0.023$ &   $0.80\pm0.009$ &  $159.84\pm0.27$ &   $21.06\pm 0.06$  &   $1.17\pm 0.025$ &   $ 0.44\pm 0.009$ & $  25.24\pm 0.27 $  \\
 S26  & \sedisc &  18.81  &  $19.78\pm0.12$ & $0.33\pm0.025$ &  $ 2.61\pm0.09$ &  $0.80\pm0.003$ &  $ 52.62\pm0.38$  &  $20.65\pm0.06$  &   $0.69\pm0.014$ &   $0.66\pm0.009$ &  $ 54.01\pm0.27$ &   $...          $  &   $...          $ &   $ ...          $ & $ ...            $  \\
 S27  & \sersic &  17.72  &  $24.27\pm0.04$ & $4.03\pm0.092$ &  $ 6.79\pm0.06$ &  $0.88\pm0.001$ &  $ 66.02\pm0.09$  &  $...         $  &   $...         $ &   $...         $ &  $...          $ &   $...          $  &   $...          $ &   $ ...          $ & $ ...            $  \\
 S28  & \sersic &  17.29  &  $21.28\pm0.03$ & $1.72\pm0.026$ &  $ 2.13\pm0.01$ &  $0.80\pm0.000$ &  $144.20\pm0.06$  &  $...         $  &   $...         $ &   $...         $ &  $...          $ &   $...          $  &   $...          $ &   $ ...          $ & $ ...            $  \\
 S29  & \sedisc &  19.48  &  $19.46\pm0.12$ & $0.11\pm0.008$ &  $ 2.57\pm0.09$ &  $0.86\pm0.003$ &  $106.13\pm0.38$  &  $20.11\pm0.06$  &   $0.69\pm0.015$ &   $0.48\pm0.009$ &  $ 24.19\pm0.27$ &   $...          $  &   $...          $ &   $ ...          $ & $ ...            $  \\
 S30  & \sedisc &  18.74  &  $20.90\pm0.12$ & $0.58\pm0.044$ &  $ 3.81\pm0.13$ &  $0.83\pm0.003$ &  $ 91.84\pm0.38$  &  $22.22\pm0.06$  &   $1.02\pm0.021$ &   $0.65\pm0.009$ &  $ 82.89\pm0.27$ &   $...          $  &   $...          $ &   $ ...          $ & $ ...            $  \\
 S31  & \sersic &  18.38  &  $22.17\pm0.04$ & $1.38\pm0.032$ &  $ 4.36\pm0.04$ &  $0.73\pm0.001$ &  $174.35\pm0.09$  &  $...         $  &   $...         $ &   $...         $ &  $...          $ &   $...          $  &   $...          $ &   $ ...          $ & $ ...            $  \\
 S32  & \sedisc &  19.39  &  $19.70\pm0.12$ & $0.22\pm0.017$ &  $ 1.13\pm0.04$ &  $0.77\pm0.003$ &  $ 94.29\pm0.38$  &  $21.41\pm0.06$  &   $1.09\pm0.023$ &   $0.56\pm0.009$ &  $ 75.04\pm0.27$ &   $...          $  &   $...          $ &   $ ...          $ & $ ...            $  \\
 S33  & \sedisc &  18.03  &  $19.20\pm0.10$ & $0.27\pm0.015$ &  $ 2.01\pm0.06$ &  $0.92\pm0.002$ &  $160.54\pm0.28$  &  $20.63\pm0.05$  &   $1.10\pm0.016$ &   $0.92\pm0.006$ &  $ 99.91\pm0.47$ &   $...          $  &   $...          $ &   $ ...          $ & $ ...            $  \\
 S34  & \sedisc &  16.77  &  $21.15\pm0.06$ & $1.57\pm0.057$ &  $ 3.73\pm0.06$ &  $0.73\pm0.001$ &  $ 97.05\pm0.19$  &  $21.98\pm0.06$  &   $3.54\pm0.082$ &   $0.47\pm0.009$ &  $ 89.18\pm0.36$ &   $...          $  &   $...          $ &   $ ...          $ & $ ...            $  \\
 S35  & \sedisc &  20.50  &  $21.52\pm0.09$ & $0.35\pm0.019$ &  $ 3.41\pm0.12$ &  $0.51\pm0.004$ &  $ 75.34\pm0.47$  &  $21.42\pm0.05$  &   $0.52\pm0.010$ &   $0.66\pm0.008$ &  $ 69.32\pm0.51$ &   $...          $  &   $...          $ &   $ ...          $ & $ ...            $  \\
\noalign{\smallskip}       
\hline
\end{tabular} 
\end{scriptsize}
\end{table}    
\end{landscape}

\begin{landscape}
\renewcommand{\tabcolsep}{3pt}
\begin{table}
\contcaption{}
\begin{scriptsize}
\begin{tabular}{clcccccccccccccc}
\hline 
\multicolumn{1}{c}{(1)} & 
\multicolumn{1}{c}{(2)} &   
\multicolumn{1}{c}{(3)} & 
\multicolumn{1}{c}{(4)} &   
\multicolumn{1}{c}{(5)} & 
\multicolumn{1}{c}{(6)} &
\multicolumn{1}{c}{(7)} & 
\multicolumn{1}{c}{(8)} &   
\multicolumn{1}{c}{(9)} & 
\multicolumn{1}{c}{(10)} &   
\multicolumn{1}{c}{(11)} & 
\multicolumn{1}{c}{(12)} &
\multicolumn{1}{c}{(13)} & 
\multicolumn{1}{c}{(14)} &
\multicolumn{1}{c}{(15)} &
\multicolumn{1}{c}{(16)} \\
\noalign{\smallskip}   
\hline
\noalign{\smallskip}       
 S36  & \sedisc &  16.90  &  $21.35\pm0.06$ & $1.41\pm0.051$ &  $ 3.89\pm0.07$ &  $0.77\pm0.001$ &  $ 58.05\pm0.19$  &  $22.60\pm0.06$  &   $5.23\pm0.122$ &   $0.47\pm0.009$ &  $ 49.10\pm0.36$ &    ...             &   $...          $ &   $ ...          $ & $ ...            $  \\
 S37  & \sedibar&  18.19  &  $21.74\pm0.12$ & $0.41\pm0.030$ &  $ 3.71\pm0.18$ &  $0.70\pm0.002$ &  $152.99\pm0.28$  &  $20.63\pm0.05$  &   $1.76\pm0.026$ &   $0.32\pm0.006$ &  $107.62\pm0.47$ &   $19.76\pm 0.05$  &   $2.06\pm 0.031$ &   $ 0.28\pm 0.006$ & $ 125.65\pm 0.47 $  \\
 S38  & \sedisc &  19.47  &  $20.91\pm0.12$ & $0.39\pm0.030$ &  $ 3.05\pm0.11$ &  $0.72\pm0.003$ &  $107.05\pm0.38$  &  $20.43\pm0.06$  &   $0.86\pm0.018$ &   $0.21\pm0.009$ &  $117.22\pm0.27$ &    ...             &   $...          $ &   $ ...          $ & $ ...            $  \\
 S39  & \sedibar&  18.74  &  $20.26\pm0.15$ & $0.44\pm0.045$ &  $ 2.25\pm0.13$ &  $0.66\pm0.003$ &  $177.61\pm0.38$  &  $21.60\pm0.06$  &   $1.19\pm0.025$ &   $0.65\pm0.009$ &  $171.71\pm0.27$ &   $20.63\pm 0.06$  &   $0.86\pm 0.018$ &   $ 0.44\pm 0.009$ & $  63.01\pm 0.27 $  \\
 S40  & \sedisc &  19.62  &  $18.69\pm0.02$ & $0.11\pm0.001$ &  $ 1.97\pm0.03$ &  $0.70\pm0.003$ &  $106.99\pm0.47$  &  $20.40\pm0.01$  &   $0.58\pm0.003$ &   $0.64\pm0.002$ &  $103.54\pm0.24$ &   ...              &   $...          $ &   $ ...          $ & $ ...            $  \\
 S41  & \sedisc &  18.27  &  $19.81\pm0.10$ & $0.36\pm0.020$ &  $ 1.93\pm0.06$ &  $0.68\pm0.002$ &  $150.16\pm0.28$  &  $19.95\pm0.05$  &   $1.21\pm0.018$ &   $0.34\pm0.006$ &  $145.07\pm0.47$ &   ...              &   $...          $ &   $ ...          $ & $ ...            $  \\
 S42  & \sedisc &  18.52  &  $17.01\pm0.07$ & $0.04\pm0.001$ &  $ 9.93\pm0.39$ &  $0.35\pm0.004$ &  $ 78.42\pm0.57$  &  $20.39\pm0.01$  &   $0.98\pm0.003$ &   $0.86\pm0.001$ &  $110.87\pm0.12$ &   ...              &   $...          $ &   $ ...          $ & $ ...            $  \\
 S43  & \sedibar&  19.68  &  $18.70\pm0.11$ & $0.08\pm0.006$ &  $ 0.50\pm0.03$ &  $0.75\pm0.004$ &  $ 99.52\pm0.47$  &  $20.55\pm0.05$  &   $0.56\pm0.011$ &   $0.96\pm0.008$ &  $ 43.90\pm0.52$ &   $21.63\pm 0.05$  &   $1.18\pm 0.022$ &   $ 0.25\pm 0.008$ & $  65.39\pm 0.52 $  \\
 S44  & \sedisc &  19.06  &  $18.34\pm0.12$ & $0.13\pm0.010$ &  $ 1.69\pm0.06$ &  $0.78\pm0.003$ &  $126.51\pm0.38$  &  $19.96\pm0.06$  &   $0.61\pm0.013$ &   $0.61\pm0.009$ &  $149.46\pm0.27$ &    ...             &   $...          $ &   $ ...          $ & $ ...            $  \\
 S45  & \sedisc &  19.14  &  $21.10\pm0.12$ & $0.41\pm0.031$ &  $ 2.74\pm0.09$ &  $0.69\pm0.003$ &  $ 77.11\pm0.38$  &  $20.47\pm0.06$  &   $1.05\pm0.022$ &   $0.32\pm0.009$ &  $ 84.92\pm0.27$ &    ...             &   $...          $ &   $ ...          $ & $ ...            $  \\
 S46  & \sedisc &  18.68  &  $18.86\pm0.12$ & $0.24\pm0.019$ &  $ 1.63\pm0.06$ &  $0.62\pm0.003$ &  $123.56\pm0.38$  &  $20.53\pm0.06$  &   $0.94\pm0.020$ &   $0.53\pm0.009$ &  $121.53\pm0.27$ &    ...             &   $...          $ &   $ ...          $ & $ ...            $  \\
 S47  & \sedisc &  19.58  &  $21.90\pm0.09$ & $0.49\pm0.026$ &  $ 4.07\pm0.14$ &  $0.87\pm0.004$ &  $134.40\pm0.47$  &  $21.79\pm0.05$  &   $0.76\pm0.014$ &   $0.94\pm0.008$ &  $ 67.34\pm0.51$ &    ...             &   $...          $ &   $ ...          $ & $ ...            $  \\
 S48  & \sedibar&  19.14  &  $20.79\pm0.15$ & $0.36\pm0.037$ &  $ 4.12\pm0.24$ &  $0.89\pm0.003$ &  $114.25\pm0.38$  &  $22.00\pm0.06$  &   $1.02\pm0.021$ &   $0.79\pm0.009$ &  $133.57\pm0.27$ &   $21.96\pm 0.06$  &   $1.09\pm 0.023$ &   $ 0.64\pm 0.009$ & $  85.30\pm 0.27 $  \\
 S49  & \sedisc &  17.39  &  $19.98\pm0.06$ & $0.50\pm0.018$ &  $ 2.72\pm0.05$ &  $0.95\pm0.001$ &  $  0.73\pm0.19$  &  $20.77\pm0.06$  &   $1.55\pm0.036$ &   $0.88\pm0.009$ &  $141.83\pm0.36$ &    ...             &   $...          $ &   $ ...          $ & $ ...            $  \\
 S50  & \sedibar&  19.43  &  $18.92\pm0.15$ & $0.10\pm0.011$ &  $ 1.98\pm0.12$ &  $0.93\pm0.003$ &  $ 60.03\pm0.38$  &  $20.78\pm0.06$  &   $0.67\pm0.014$ &   $0.77\pm0.009$ &  $  9.30\pm0.27$ &   $21.59\pm 0.06$  &   $1.26\pm 0.026$ &   $ 0.61\pm 0.009$ & $  38.02\pm 0.27 $  \\
 S51  & \sedisc &  19.21  &  $22.11\pm0.12$ & $0.66\pm0.050$ &  $ 7.62\pm0.26$ &  $0.82\pm0.003$ &  $ 33.77\pm0.38$  &  $22.63\pm0.06$  &   $1.20\pm0.025$ &   $0.62\pm0.009$ &  $115.75\pm0.27$ &    ...             &   $...          $ &   $ ...          $ & $ ...            $  \\
 S52  & \sedisc &  19.11  &  $17.75\pm0.07$ & $0.08\pm0.003$ &  $ 1.57\pm0.06$ &  $0.93\pm0.004$ &  $ 93.39\pm0.57$  &  $19.52\pm0.01$  &   $0.41\pm0.001$ &   $0.93\pm0.001$ &  $ 75.64\pm0.12$ &    ...             &   $...          $ &   $ ...          $ & $ ...            $  \\
 S53  & \sedisc &  17.27  &  $20.87\pm0.06$ & $0.68\pm0.025$ &  $ 3.97\pm0.07$ &  $0.82\pm0.001$ &  $ 51.20\pm0.19$  &  $20.71\pm0.06$  &   $1.87\pm0.044$ &   $0.74\pm0.009$ &  $  1.38\pm0.36$ &    ...             &   $...          $ &   $ ...          $ & $ ...            $  \\
 S54  & \sedisc &  17.76  &  $20.03\pm0.10$ & $0.50\pm0.028$ &  $ 1.49\pm0.04$ &  $0.80\pm0.002$ &  $164.95\pm0.28$  &  $21.13\pm0.05$  &   $1.59\pm0.024$ &   $0.91\pm0.006$ &  $130.19\pm0.47$ &    ...             &   $...          $ &   $ ...          $ & $ ...            $  \\
\noalign{\smallskip}   
\hline
\noalign{\smallskip}       
\multicolumn{16}{c}{Ancillary sample}\\
\noalign{\smallskip}
\hline
\noalign{\smallskip}       
A01 & \sedisc &    $17.90$  &  $20.28\pm0.10$  &   $0.22\pm0.012$  &   $0.80\pm0.02$ &    $1.00\pm0.002$ &  $115.12\pm0.28$ &   $20.00\pm0.05$  &   $1.05\pm0.016$  &   $0.94\pm0.006$ &  $ 40.83\pm0.47$ & ...            &...            &...          &...             \\
A02 & \sedisc &    $18.51$  &  $18.82\pm0.12$  &   $0.17\pm0.013$  &   $1.84\pm0.06$ &    $0.91\pm0.003$ &  $ 25.88\pm0.38$ &   $19.84\pm0.06$  &   $0.68\pm0.014$  &   $0.81\pm0.009$ &  $ 39.06\pm0.27$ & ...            &...            &...          &...             \\
A03 & \sedisc &    $19.70$  &  $20.69\pm0.09$  &   $0.33\pm0.018$  &   $3.06\pm0.11$ &    $0.70\pm0.004$ &  $164.59\pm0.47$ &   $20.87\pm0.05$  &   $0.48\pm0.009$  &   $0.82\pm0.008$ &  $ 96.97\pm0.51$ & ...            &...            &...          &...             \\
A04 & \sedisc &    $19.22$  &  $18.84\pm0.07$  &   $0.11\pm0.003$  &   $1.84\pm0.07$ &    $0.67\pm0.004$ &  $119.12\pm0.57$ &   $19.91\pm0.01$  &   $0.78\pm0.003$  &   $0.40\pm0.001$ &  $114.58\pm0.12$ & ...            &...            &...          &...             \\
A05 & \sersic &    $19.62$  &  $22.92\pm0.02$  &   $1.57\pm0.013$  &   $2.29\pm0.02$ &    $0.49\pm0.001$ &  $100.25\pm0.15$ &   $...         $  &   $...         $  &   $...         $ &  $...          $ & ...            &...            &...          &...             \\
A06 & \sedisc &    $19.33$  &  $18.41\pm0.07$  &   $0.12\pm0.004$  &   $2.66\pm0.10$ &    $0.81\pm0.004$ &  $133.35\pm0.57$ &   $20.48\pm0.01$  &   $0.52\pm0.002$  &   $0.90\pm0.001$ &  $101.06\pm0.12$ & ...            &...            &...          &...             \\
A07 & \sedisc &    $20.28$  &  $20.44\pm0.09$  &   $0.20\pm0.011$  &   $1.29\pm0.04$ &    $0.46\pm0.004$ &  $ 72.81\pm0.47$ &   $19.98\pm0.05$  &   $0.49\pm0.009$  &   $0.41\pm0.008$ &  $ 64.09\pm0.51$ & ...            &...            &...          &...             \\
A08 & \sedisc &    $19.80$  &  $19.10\pm0.09$  &   $0.16\pm0.009$  &   $2.96\pm0.10$ &    $0.61\pm0.004$ &  $ 60.99\pm0.47$ &   $21.96\pm0.05$  &   $1.06\pm0.020$  &   $0.46\pm0.008$ &  $ 53.08\pm0.51$ & ...            &...            &...          &...             \\
A09 & \sersic &    $20.37$  &  $23.58\pm0.02$  &   $1.18\pm0.010$  &   $2.58\pm0.02$ &    $0.76\pm0.001$ &  $ 58.05\pm0.15$ &   $...         $  &   $...         $  &   $...         $ &  $...          $ & ...            &...            &...          &...             \\
A10 & \sedisc &    $19.31$  &  $22.14\pm0.12$  &   $0.77\pm0.059$  &   $4.19\pm0.14$ &    $0.58\pm0.003$ &  $ 58.38\pm0.38$ &   $21.36\pm0.06$  &   $1.07\pm0.023$  &   $0.38\pm0.009$ &  $ 72.36\pm0.27$ & ...            &...            &...          &...             \\
A11 & \sedisc &    $21.19$  &  $21.62\pm0.19$  &   $0.13\pm0.017$  &   $1.84\pm0.22$ &    $0.88\pm0.013$ &  $ 99.74\pm3.63$ &   $21.56\pm0.06$  &   $0.46\pm0.010$  &   $0.92\pm0.006$ &  $124.17\pm1.80$ & ...            &...            &...          &...             \\
\noalign{\smallskip}       
\hline
\end{tabular}
{\em Note.}  Best-fit observed parameters of the sample galaxies
resulting from the photometric decomposition.  Col. (1): galaxy ID.
Col. (2): fit-type according to morphological classification (see
Table \ref{tab:morph_type}).  Col. (3): total magnitude.
Col. (4)-(8): bulge parameters, i.e., effective surface brightness
$\mu_e$ and radius $r_e$, S\'ersic index $n$, axis ratio $q_{\rm
  bulge}$, and position angle PA$_{\rm bulge}$ Col. (9)-(12): disc
parameters, i.e., central surface brightness $\mu_0$, scale length
$h$, axis ratio $q_{\rm disc}$, and position angle PA$_{\rm disc}$.
Col. (13)-(16): bar parameters, i.e., central surface brightness
$\mu_{\rm 0,bar}$, bar radius $r_{\rm bar}$, axis ratio $q_{\rm bar}$,
and position angle PA$_{\rm bar}$. The PA are measured
counterclockwise from North to East.
\end{scriptsize}
\end{table}
\end{landscape}

\begin{table*}
\begin{minipage}{126mm}
\caption{\small{Structural photometric parameters of the ETGs spectroscopic sample fitted with a \devauc\ model.}}
\label{tab:devauc_par}
\begin{scriptsize}
\begin{center}
\begin{tabular}{rrrrrr}
\hline
\multicolumn{1}{c}{Galaxy} &
\multicolumn{1}{c}{mag} &
\multicolumn{1}{c}{$\mu_e$} &
\multicolumn{1}{c}{$r_e$} &
\multicolumn{1}{c}{$q_{\rm bulge}$} &
\multicolumn{1}{c}{PA$_{\rm bulge}$} \\
\multicolumn{1}{c}{ID} &
\multicolumn{1}{c}{(mag)} &
\multicolumn{1}{c}{(mag/arcsec$^{2}$)} &
\multicolumn{1}{c}{(arcsec)} &
\multicolumn{1}{c}{} &
\multicolumn{1}{c}{($\degr$)} \\
\multicolumn{1}{c}{(1)} &
\multicolumn{1}{c}{(2)} &
\multicolumn{1}{c}{(3)} &
\multicolumn{1}{c}{(4)} &
\multicolumn{1}{c}{(5)} &
\multicolumn{1}{c}{(6)} \\
\hline
 S02  &     18.59  &  $21.04\pm     0.01$ &    $ 0.98\pm    0.007$ &     $0.44\pm    0.001$ & $  159.34\pm     0.11$\\   
 S04  &     19.57  &  $21.79\pm     0.01$ &    $ 0.63\pm    0.005$ &     $0.86\pm    0.002$ & $   19.01\pm     0.14$\\    
 S05  &     18.71  &  $21.65\pm     0.01$ &    $ 0.94\pm    0.006$ &     $0.76\pm    0.001$ & $   36.11\pm     0.11$\\    
 S06  &     18.65  &  $20.85\pm     0.01$ &    $ 0.72\pm    0.005$ &     $0.64\pm    0.001$ & $   81.89\pm     0.11$\\    
 S07  &     19.59  &  $24.24\pm     0.01$ &    $ 3.10\pm    0.027$ &     $0.33\pm    0.002$ & $  129.97\pm     0.14$\\    
 S08  &     18.76  &  $21.80\pm     0.01$ &    $ 0.92\pm    0.006$ &     $0.85\pm    0.001$ & $  109.79\pm     0.11$\\    
 S09  &     18.69  &  $21.88\pm     0.01$ &    $ 0.99\pm    0.007$ &     $0.85\pm    0.001$ & $   63.68\pm     0.11$\\    
 S10  &     18.02  &  $21.16\pm     0.01$ &    $ 0.92\pm    0.005$ &     $0.95\pm    0.001$ & $  125.60\pm     0.06$\\    
 S11  &     19.02  &  $21.44\pm     0.01$ &    $ 0.99\pm    0.007$ &     $0.41\pm    0.001$ & $  119.99\pm     0.11$\\    
 S14  &     18.95  &  $20.79\pm     0.01$ &    $ 0.52\pm    0.003$ &     $0.90\pm    0.001$ & $   93.01\pm     0.11$\\    
 S16  &     18.24  &  $22.27\pm     0.01$ &    $ 1.42\pm    0.008$ &     $0.89\pm    0.001$ & $  111.74\pm     0.06$\\    
 S17  &     19.12  &  $20.12\pm     0.01$ &    $ 0.51\pm    0.003$ &     $0.43\pm    0.001$ & $   59.14\pm     0.11$\\    
 S18  &     15.23  &  $24.92\pm     0.19$ &    $20.05\pm    2.033$ &     $0.82\pm    0.001$ & $  153.36\pm     0.08$\\    
 S19  &     19.43  &  $21.17\pm     0.01$ &    $ 0.51\pm    0.003$ &     $0.85\pm    0.001$ & $   90.00\pm     0.11$\\    
 S20  &     19.42  &  $22.64\pm     0.01$ &    $ 1.17\pm    0.008$ &     $0.62\pm    0.001$ & $  178.19\pm     0.11$\\    
 S22  &     18.61  &  $21.03\pm     0.01$ &    $ 0.70\pm    0.005$ &     $0.84\pm    0.001$ & $   17.20\pm     0.11$\\    
 S23  &     19.42  &  $21.07\pm     0.01$ &    $ 0.57\pm    0.004$ &     $0.63\pm    0.001$ & $   49.23\pm     0.11$\\    
 S24  &     18.14  &  $21.99\pm     0.01$ &    $ 1.30\pm    0.008$ &     $0.90\pm    0.001$ & $  115.12\pm     0.06$\\    
 S25  &     18.83  &  $21.80\pm     0.01$ &    $ 0.90\pm    0.006$ &     $0.84\pm    0.001$ & $   12.57\pm     0.11$\\    
 S26  &     18.64  &  $20.98\pm     0.01$ &    $ 0.71\pm    0.005$ &     $0.75\pm    0.001$ & $   53.86\pm     0.11$\\    
 S27  &     18.17  &  $22.55\pm     0.01$ &    $ 1.68\pm    0.010$ &     $0.89\pm    0.001$ & $   69.72\pm     0.06$\\    
 S28  &     17.00  &  $22.50\pm     0.01$ &    $ 2.97\pm    0.020$ &     $0.79\pm    0.001$ & $  143.86\pm     0.05$\\    
 S30  &     18.68  &  $21.31\pm     0.01$ &    $ 0.78\pm    0.005$ &     $0.80\pm    0.001$ & $   89.05\pm     0.11$\\    
 S31  &     18.43  &  $21.99\pm     0.01$ &    $ 1.26\pm    0.008$ &     $0.73\pm    0.001$ & $  174.21\pm     0.06$\\    
 S34  &     16.69  &  $21.91\pm     0.01$ &    $ 2.83\pm    0.019$ &     $0.67\pm    0.001$ & $   94.05\pm     0.05$\\    
 S35  &     20.26  &  $22.73\pm     0.01$ &    $ 0.86\pm    0.007$ &     $0.57\pm    0.002$ & $   73.32\pm     0.14$\\    
 S36  &     17.00  &  $22.24\pm     0.01$ &    $ 2.73\pm    0.018$ &     $0.73\pm    0.001$ & $   54.81\pm     0.05$\\    
 S38  &     19.29  &  $21.79\pm     0.01$ &    $ 1.00\pm    0.007$ &     $0.44\pm    0.001$ & $  115.63\pm     0.11$\\    
 S39  &     18.64  &  $21.56\pm     0.01$ &    $ 0.98\pm    0.007$ &     $0.68\pm    0.001$ & $  177.53\pm     0.11$\\    
 S40  &     19.54  &  $21.34\pm     0.01$ &    $ 0.59\pm    0.005$ &     $0.67\pm    0.002$ & $  104.43\pm     0.14$\\    
 S41  &     17.92  &  $22.01\pm     0.01$ &    $ 2.03\pm    0.012$ &     $0.47\pm    0.001$ & $  145.60\pm     0.06$\\    
 S44  &     18.98  &  $20.74\pm     0.01$ &    $ 0.57\pm    0.004$ &     $0.69\pm    0.001$ & $  143.60\pm     0.11$\\    
 S45  &     18.83  &  $22.66\pm     0.01$ &    $ 1.89\pm    0.013$ &     $0.43\pm    0.001$ & $   83.90\pm     0.11$\\    
 S46  &     18.60  &  $20.86\pm     0.01$ &    $ 0.78\pm    0.005$ &     $0.57\pm    0.001$ & $  122.33\pm     0.11$\\    
 S47  &     19.40  &  $22.87\pm     0.01$ &    $ 1.07\pm    0.007$ &     $0.94\pm    0.001$ & $  127.17\pm     0.11$\\    
 S48  &     19.11  &  $21.66\pm     0.01$ &    $ 0.73\pm    0.005$ &     $0.87\pm    0.001$ & $  114.89\pm     0.11$\\    
 S49  &     17.21  &  $21.98\pm     0.01$ &    $ 1.95\pm    0.013$ &     $0.93\pm    0.001$ & $  157.07\pm     0.05$\\    
 S51  &     19.40  &  $21.74\pm     0.01$ &    $ 0.62\pm    0.004$ &     $1.00\pm    0.001$ & $   41.66\pm     0.11$\\    
 S52  &     19.01  &  $20.62\pm     0.01$ &    $ 0.46\pm    0.003$ &     $0.94\pm    0.001$ & $   79.36\pm     0.11$\\    
 S53  &     16.95  &  $22.90\pm     0.01$ &    $ 3.55\pm    0.024$ &     $0.83\pm    0.001$ & $   17.19\pm     0.05$\\    
 S54  &     17.55  &  $22.54\pm     0.01$ &    $ 2.25\pm    0.013$ &     $0.87\pm    0.001$ & $  158.90\pm     0.06$\\    
\hline
\end{tabular}
\end{center}
{\em Note.}   Best-fit observed \devauc\ parameters.  Col. (1): galaxy ID.
Col. (2): total magnitude.
Col. (3): effective surface brightness. 
Col. (4): effective radius.
Col. (5): axis ratio.
Col. (6): position angle, measured counterclockwise from North to East.
\end{scriptsize}
\end{minipage}
\end{table*}                                     

\begin{table*}
\begin{minipage}{126mm}
\caption{\small{Structural photometric parameters of the ETGs spectroscopic sample fitted with a \sersic\ model}}
\label{tab:sersic_par}
\begin{scriptsize}
\begin{center}
\begin{tabular}{rrrrrrr}
\hline
\multicolumn{1}{c}{Galaxy} &
\multicolumn{1}{c}{mag} &
\multicolumn{1}{c}{$\mu_e$} &
\multicolumn{1}{c}{$r_e$} &
\multicolumn{1}{c}{$n$} &
\multicolumn{1}{c}{$q_{\rm bulge}$} &
\multicolumn{1}{c}{PA$_{\rm bulge}$} \\
\multicolumn{1}{c}{ID} &
\multicolumn{1}{c}{(mag)} &
\multicolumn{1}{c}{(mag/arcsec$^{2}$)} &
\multicolumn{1}{c}{(arcsec)} &
\multicolumn{1}{c}{} &
\multicolumn{1}{c}{} &
\multicolumn{1}{c}{($\degr$)} \\
\multicolumn{1}{c}{(1)} &
\multicolumn{1}{c}{(2)} &
\multicolumn{1}{c}{(3)} &
\multicolumn{1}{c}{(4)} &
\multicolumn{1}{c}{(5)} &
\multicolumn{1}{c}{(6)} &
\multicolumn{1}{c}{(7)} \\
\hline
 S02 &       18.56&    $21.16\pm0.05$&     $1.03\pm0.024$&   $  4.27\pm0.04$&     $0.44\pm0.001$&   $159.35\pm0.12$\\    
 S04 &       18.88&    $24.92\pm0.05$&     $2.94\pm0.069$&   $ 10.21\pm0.11$&     $0.85\pm0.001$&   $ 20.64\pm0.12$\\    
 S05 &       18.71&    $21.62\pm0.05$&     $0.92\pm0.022$&   $  3.93\pm0.04$&     $0.76\pm0.001$&   $ 36.11\pm0.12$\\    
 S06 &       18.74&    $20.46\pm0.05$&     $0.62\pm0.015$&   $  3.06\pm0.03$&     $0.64\pm0.001$&   $ 81.61\pm0.12$\\    
 S07 &       19.87&    $23.33\pm0.02$&     $1.94\pm0.016$&   $  2.78\pm0.02$&     $0.34\pm0.001$&   $129.57\pm0.15$\\    
 S08 &       18.38&    $23.32\pm0.04$&     $1.96\pm0.045$&   $  6.73\pm0.06$&     $0.84\pm0.001$&   $110.97\pm0.09$\\    
 S09 &       17.86&    $25.56\pm0.04$&     $6.09\pm0.140$&   $ 11.57\pm0.11$&     $0.85\pm0.001$&   $ 64.64\pm0.09$\\    
 S10 &       17.92&    $21.60\pm0.04$&     $1.12\pm0.026$&   $  4.97\pm0.05$&     $0.95\pm0.001$&   $126.45\pm0.09$\\    
 S11 &       19.00&    $21.59\pm0.05$&     $1.07\pm0.025$&   $  4.16\pm0.04$&     $0.41\pm0.001$&   $119.98\pm0.12$\\    
 S14 &       18.80&    $21.44\pm0.05$&     $0.69\pm0.016$&   $  5.51\pm0.06$&     $0.90\pm0.001$&   $ 93.80\pm0.12$\\    
 S16 &       18.14&    $22.65\pm0.04$&     $1.71\pm0.039$&   $  4.71\pm0.04$&     $0.89\pm0.001$&   $112.26\pm0.09$\\    
 S17 &       19.16&    $19.89\pm0.05$&     $0.47\pm0.011$&   $  3.31\pm0.03$&     $0.43\pm0.001$&   $ 59.16\pm0.12$\\    
 S18 &       16.06&    $22.74\pm0.02$&     $6.01\pm0.083$&   $  1.86\pm0.02$&     $0.83\pm0.001$&   $157.18\pm0.07$\\    
 S19 &       19.30&    $21.71\pm0.05$&     $0.65\pm0.015$&   $  5.26\pm0.05$&     $0.84\pm0.001$&   $ 90.62\pm0.12$\\    
 S20 &       19.44&    $22.58\pm0.05$&     $1.14\pm0.027$&   $  3.91\pm0.04$&     $0.62\pm0.001$&   $178.20\pm0.12$\\    
 S22 &       18.52&    $21.39\pm0.05$&     $0.82\pm0.019$&   $  4.72\pm0.05$&     $0.84\pm0.001$&   $ 18.18\pm0.12$\\    
 S23 &       19.48&    $20.78\pm0.05$&     $0.50\pm0.012$&   $  3.33\pm0.03$&     $0.63\pm0.001$&   $ 49.26\pm0.12$\\    
 S24 &       18.23&    $21.65\pm0.04$&     $1.11\pm0.025$&   $  3.41\pm0.03$&     $0.90\pm0.001$&   $123.29\pm0.09$\\    
 S25 &       18.53&    $23.02\pm0.05$&     $1.62\pm0.038$&   $  6.31\pm0.06$&     $0.84\pm0.001$&   $ 11.51\pm0.12$\\    
 S26 &       18.72&    $20.64\pm0.05$&     $0.61\pm0.014$&   $  3.30\pm0.03$&     $0.75\pm0.001$&   $ 53.24\pm0.12$\\    
 S27 &       17.72&    $24.27\pm0.04$&     $4.03\pm0.092$&   $  6.79\pm0.06$&     $0.88\pm0.001$&   $ 66.02\pm0.09$\\    
 S28 &       17.29&    $21.28\pm0.03$&     $1.72\pm0.026$&   $  2.13\pm0.01$&     $0.80\pm0.001$&   $144.20\pm0.06$\\    
 S30 &       18.65&    $21.47\pm0.05$&     $0.84\pm0.020$&   $  4.29\pm0.04$&     $0.80\pm0.001$&   $ 89.22\pm0.12$\\    
 S31 &       18.38&    $22.17\pm0.04$&     $1.38\pm0.032$&   $  4.36\pm0.04$&     $0.73\pm0.001$&   $174.35\pm0.09$\\    
 S34 &       16.58&    $22.37\pm0.03$&     $3.54\pm0.053$&   $  4.86\pm0.03$&     $0.67\pm0.001$&   $ 94.19\pm0.06$\\    
 S35 &       20.30&    $22.57\pm0.02$&     $0.80\pm0.007$&   $  3.72\pm0.03$&     $0.58\pm0.001$&   $ 73.32\pm0.15$\\    
 S36 &       16.62&    $23.71\pm0.03$&     $5.76\pm0.087$&   $  6.53\pm0.04$&     $0.72\pm0.000$&   $ 54.64\pm0.06$\\    
 S38 &       19.42&    $21.27\pm0.05$&     $0.80\pm0.019$&   $  2.93\pm0.03$&     $0.44\pm0.001$&   $115.59\pm0.12$\\    
 S39 &       18.70&    $21.27\pm0.05$&     $0.86\pm0.020$&   $  3.41\pm0.04$&     $0.68\pm0.001$&   $177.77\pm0.12$\\    
 S40 &       19.35&    $22.23\pm0.05$&     $0.89\pm0.021$&   $  5.65\pm0.06$&     $0.66\pm0.001$&   $104.85\pm0.12$\\    
 S41 &       18.05&    $21.51\pm0.04$&     $1.59\pm0.036$&   $  3.25\pm0.03$&     $0.46\pm0.001$&   $145.57\pm0.09$\\    
 S44 &       18.87&    $21.20\pm0.05$&     $0.70\pm0.017$&   $  5.07\pm0.05$&     $0.68\pm0.001$&   $144.44\pm0.12$\\    
 S45 &       19.02&    $21.95\pm0.05$&     $1.35\pm0.032$&   $  2.76\pm0.03$&     $0.43\pm0.001$&   $ 83.73\pm0.12$\\    
 S46 &       18.60&    $20.87\pm0.05$&     $0.79\pm0.019$&   $  4.03\pm0.04$&     $0.57\pm0.001$&   $122.33\pm0.12$\\    
 S47 &       19.21&    $23.63\pm0.05$&     $1.57\pm0.037$&   $  5.19\pm0.05$&     $0.94\pm0.001$&   $129.38\pm0.12$\\    
 S48 &       18.93&    $22.45\pm0.05$&     $1.05\pm0.025$&   $  5.65\pm0.06$&     $0.87\pm0.001$&   $115.44\pm0.12$\\    
 S49 &       16.94&    $22.99\pm0.03$&     $3.25\pm0.049$&   $  5.63\pm0.04$&     $0.93\pm0.001$&   $154.97\pm0.06$\\    
 S51 &       18.68&    $25.12\pm0.05$&     $3.26\pm0.077$&   $ 11.79\pm0.12$&     $0.92\pm0.001$&   $ 42.38\pm0.12$\\    
 S52 &       18.91&    $21.21\pm0.05$&     $0.59\pm0.014$&   $  5.11\pm0.05$&     $0.93\pm0.001$&   $ 79.88\pm0.12$\\    
 S53 &       16.55&    $24.37\pm0.03$&     $7.58\pm0.114$&   $  6.26\pm0.04$&     $0.83\pm0.001$&   $ 18.11\pm0.06$\\    
 S54 &       17.65&    $22.17\pm0.04$&     $1.87\pm0.043$&   $  3.49\pm0.03$&     $0.87\pm0.001$&   $160.09\pm0.09$\\        
\hline
\end{tabular}
\end{center}
{\em Note.}   Best-fit observed \sersic\ parameters.  Col. (1): galaxy ID.
Col. (2): total magnitude.
Col. (3): effective surface brightness. 
Col. (4): effective radius.
Col. (5): S\'ersic index.
Col. (6): axis ratio.
Col. (7): position angle, measured counterclockwise from North to East.
\end{scriptsize}
\end{minipage}
\end{table*}

\bsp	
\label{lastpage}
\end{document}